\documentclass[10pt,a4paper,reqno]{amsart}
\usepackage[foot]{amsaddr}
\usepackage{a4wide}
\usepackage{amsmath,amsthm,amssymb, amsfonts, amsbsy, mathrsfs}
\usepackage{bbm}
\usepackage{enumerate}
\usepackage{xparse}
\usepackage{bm}
\usepackage{filecontents}
\usepackage{tikz}
\usetikzlibrary{calc}
\usetikzlibrary{decorations.markings}
\usepackage{subcaption}
\usepackage{hyperref}
\usepackage{cite}
\usepackage{appendix}
\usepackage{booktabs}
\usepackage{graphicx}
\usepackage[portrait,margin=3cm]{geometry}
\usepackage{color, colortbl}
\definecolor{darkgrey}{gray}{0.35}
\definecolor{grey}{gray}{0.5}
\definecolor{lightgrey}{gray}{0.8}
\definecolor{ballblue}{rgb}{0, 0.5,0.5}
\definecolor{lightballblue}{rgb}{0, 0.8,0.8}
\definecolor{dbblue}{rgb}{0, 0.4,0.4}

\newcommand{\Ev}{\mathbb{E}}
\newcommand*{\be}{\begin{equation}}
\newcommand*{\ee}{\end{equation}}
\newcommand*{\ba}{\begin{aligned}}
	\newcommand*{\ea}{\end{aligned}}
\newcommand*{\barr}{\begin{array}{c}}
	\newcommand*{\earr}{\end{array}}

\newcommand{\wit}{\widetilde}

\newcommand*{\dd}{\mathrm{d}}

	\newcommand{\Z}{\mathbb{Z}}

\newcommand{\E}{\mathbb{E}}

\newcommand{\R}{\mathbb{R}}

\newcommand{\eps}{\varepsilon}
\renewcommand{\epsilon}{\eps}

\newtheorem{theorem}{Theorem}[section]

\newtheorem{definition}[theorem]{Definition}

\newtheorem*{rep@theorem}{\rep@title}
\newcommand{\newreptheorem}[2]{%
	\newenvironment{rep#1}[1]{%
		\def\rep@title{#2 \ref{##1} (restated)}%
		\begin{rep@theorem}}%
		{\end{rep@theorem}}}
\newreptheorem{theorem}{Theorem}
\newreptheorem{lemma}{Lemma}
\graphicspath{{images/}}
\title[Epidemics on spatial networks]{Not all interventions are equal for the height of the second peak}
\author{Tim Hulshof$^\star$}
\address{$\star$Bureau WO, Eindhoven}
\email{timhulshof@bureauwo.nl}
\author{Joost Jorritsma$^\dagger$}
\author{J\'ulia Komj\'athy$^\dagger$}
\address{$\dagger$ Department of Mathematics and
	    Computer Science, Eindhoven University of Technology, P.O.\ Box 513,
	    5600 MB Eindhoven, The Netherlands.}
	    \email{j.jorritsma@tue.nl, j.komjathy@tue.nl}
\thanks{Acknowledgements. The work of JJ and JK is partly supported by the Netherlands Organisation for Scientific Research (NWO) through grant NWO 613.009.122.}
\begin{document}
\begin{abstract}
  In this paper we conduct a simulation study of the spread of an epidemic like COVID-19 with temporary immunity on finite spatial and non-spatial network models.
  In particular, we assume that an epidemic spreads stochastically on a scale-free network and that each infected individual in the network gains a \emph{temporary immunity} after its infectious period is over. After the temporary immunity period is over, the individual becomes susceptible to the virus again.

  When the underlying contact network is embedded in Euclidean geometry, we model three different intervention strategies that aim to control the spread of the epidemic: social distancing, restrictions on travel, and restrictions on maximal number of social contacts per node.

  Our first finding is that on a finite network, a long enough average immunity period leads to extinction of the pandemic after the first peak, analogous to the concept of ``herd immunity''. For each model, there is a critical average immunity length $L_c$ above which this happens.

  Our second finding is that all three interventions manage to flatten the first peak (the travel restrictions most efficiently), as well as decrease the critical immunity length $L_c$, but elongate the epidemic. However, when the average immunity length $L$ is shorter than $L_c$, the price for the flattened first peak is often a high second peak: for limiting the maximal number of contacts, the second peak can be as high as $1/3$ of the first peak, and   twice as high as it would be without intervention. Thirdly, interventions introduce oscillations into the system and the time to reach equilibrium is, for almost all scenarios, much longer. We conclude that network-based epidemic models can show a variety of behaviors that are not captured by the continuous compartmental models.
\end{abstract}
\maketitle

\section{Introduction}
When recovering from a disease grants temporary immunity against it, it can happen that an epidemic dies out locally, but survives elsewhere, returning at a later point in time. We observe a ``second peak''. A second peak can also happen when interventions are effectively applied to slow down the spread of a disease locally, but are then lifted. This phenomenon has a clear geometric component. Standard compartmental models for epidemic curves are inherently a-geometric, because they assume a perfectly mixed population. Agent-based epidemiological models allow for an embedding of the population in a geometric space to capture this effect more clearly.

In this paper we study this geometric effect empirically. We conduct a simulation study about the effect of temporary immunity on spreading processes on different agent-based models.  We also model the effect of three different interventions:
\begin{itemize}
\item[{ \bf A)}] social distancing,
\item[{\bf B)}] traveling restrictions, and
\item[{\bf C)}] limiting the number of social contacts.
\end{itemize}
 Our simulation results may serve as a qualitative indication of possible outcomes of epidemic spread and intervention strategies for highly infectious diseases, such as the current COVID-19 pandemic. It also gives qualitative predictions about the ways an epidemic might (or might not) return, depending on the average duration and level of temporary immunity.
The focus is twofold: first, on understanding how the underlying space affects the outcome of the simulations, and second, to see and compare the effect of the three intervention strategies on the pandemic. For the first goal, we compare four scenarios with underlying geometries:
\begin{itemize}
\item {\bf Scenario 1:} The underlying network is ``purely geometric'', ignoring long-range connections. For this scenario we use the square nearest-neighbor torus $\mathbb{Z}^2_n$ as the base graph.
\item {\bf Scenario 2:} The underlying network is a ``mean-field network'', ignoring the spatial component. For this scenario we use the \emph{configuration model,} which can mimic the local statistical properties of real human contact networks, such as degree distributions.
\item {\bf Scenario 3:} The underlying network is a mixture between a geometric and mean-field network. For this we use \emph{Geometric Inhomogeneous Random Graphs}, which possess geometric features and can match the statistical properties of real human contact networks.
  \item {\bf Scenario 4:} The epidemic is modeled in a ``mean-field continuous space'', ignoring the spatial component \emph{and} approximating the discrete population by a continuum. For this we use \emph{systems of ordinary differential equations} (also called: a compartmental model), which are currently the most popular tool for epidemic curve modeling.
\end{itemize}
Our aim is to compare these models, and see what the effect is of considering more realistic representations of the underlying space and spreading mechanisms. We, however, emphasize that our paper provides qualitative estimates, not quantitative ones. As a result, the study here might underpin or support certain intervention strategies more than others, but we refrain from (and see no justification for) using these simplified models to make numerical predictions with respect to the current COVID-19 outbreak.
 \medskip
\begin{figure}[t]\begin{center}
\begin{subfigure}{0.19\textwidth}
\centering
\includegraphics[width=\textwidth]{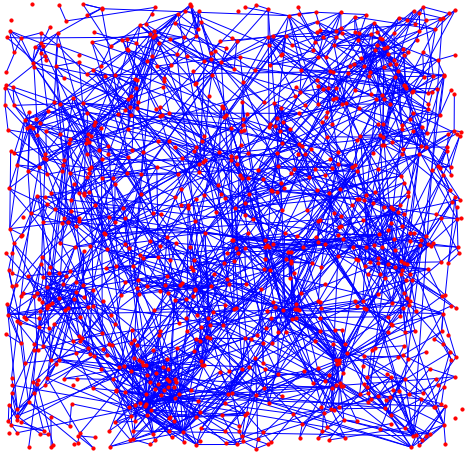}
\caption{No intervention}
\end{subfigure}
\begin{subfigure}{0.19\textwidth}
\centering
\includegraphics[width=\textwidth]{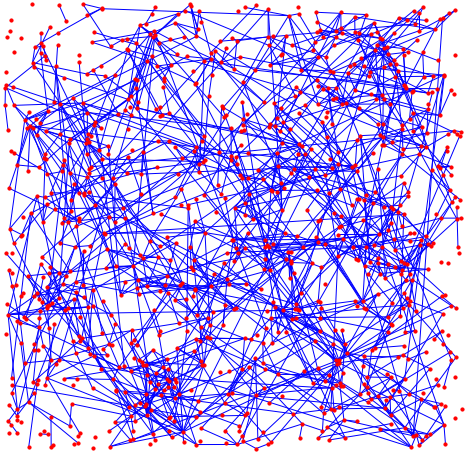}
\caption{Social dist.}
\end{subfigure}
\begin{subfigure}{0.19\textwidth}
\centering
\includegraphics[width=\textwidth]{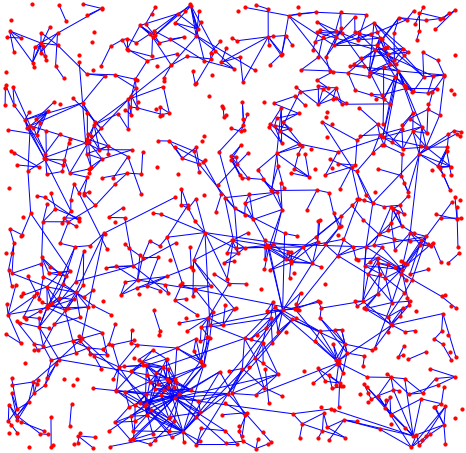}
\caption{Less travel (w)}
\end{subfigure}
\begin{subfigure}{0.19\textwidth}
\centering
\includegraphics[width=\textwidth]{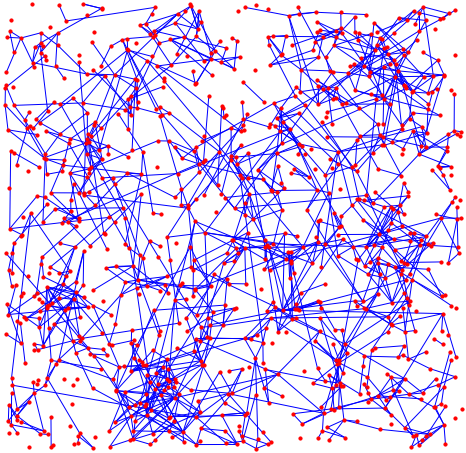}
\caption{Less travel (s)}
\end{subfigure}
\begin{subfigure}{0.19\textwidth}
\centering
\includegraphics[width=\textwidth]{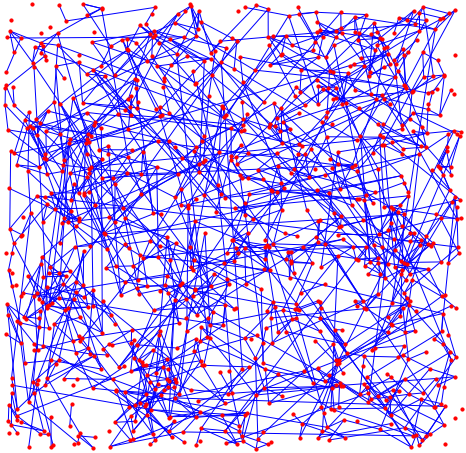}
\caption{Limiting degree}
\end{subfigure}
\end{center}
\caption{Visualization of the change in the underlying contact network under interventions for Scenario 3 (see Figure \ref{fig:girg2-interventions} below). By (w) and (s) we denote the weak and strong travel restrictions, respectively.  }\label{fig:networks-intro}
\end{figure}
\subsection{Epidemic spread with temporary immunity on a network}
\label{s:epi-girg}
We choose the simplest possible model that shows, qualitatively, the behavior that we would like to observe, namely, the effect of temporary immunity. This model can, however, be easily modified to accommodate other compartments, such as an incubation period (exposed state) (as e.g. in \cite{CovidKuchStroebFacebookConn2020}), deaths, asymptomatic cases, etc. For simplicity,  we describe the dynamics in \emph{discrete time}. A continuous time version is analogous and shows the same qualitative behavior. The model is similar to the one studied in \cite{SouTom10}, where the focus is on the critical regime.

\begin{figure}[ht]\begin{center}
  \includegraphics[width=0.85\columnwidth]{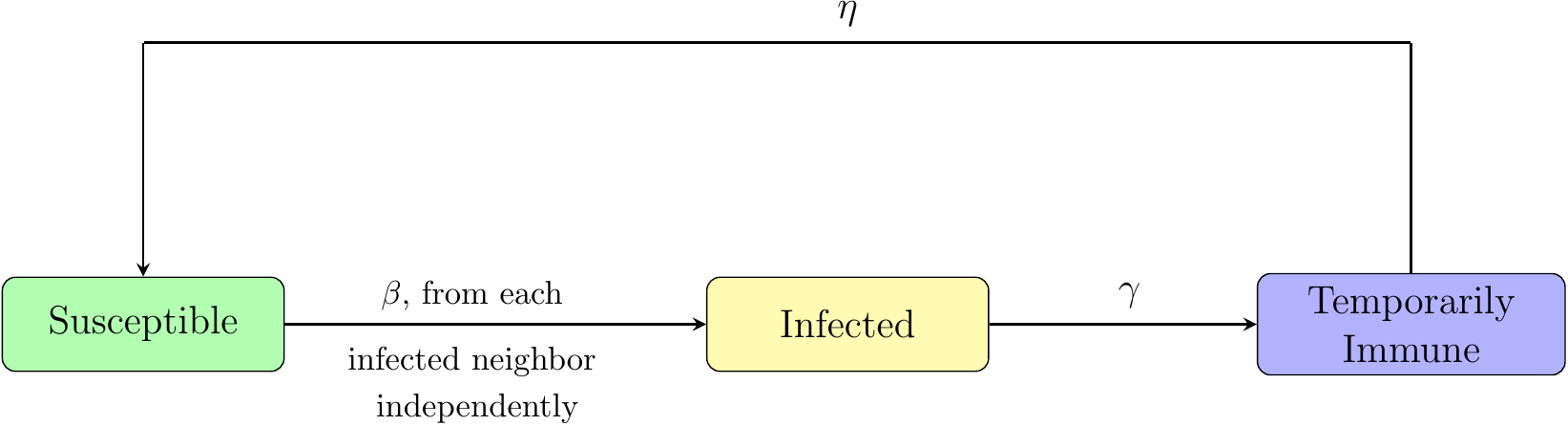}
  \caption{Schematic diagram of node states and their transition probabilities. A susceptible node $u$ may only become infected when it has at least one infected neighbor. Each of the infected neighbors infects $u$ with probability $\beta$, independently, in each step.}
\label{fig:comp1}
\end{center}\end{figure}

The spreading process changes at discrete time steps, $t=\{0, 1,2,\dots\}$, each time step corresponds to, say, a day.
We fix the network $G$ in advance. We think of nodes in the network as individuals.
Each node within the network can be in three possible states: \emph{susceptible (S), infected (I) or temporarily immune (T)}.
The neighbors of a node $u$ are nodes with a direct connection (also called link, or edge) to $u$. A connection may correspond to a friendship or an acquaintance, or simply a contact event.
The discrete time dynamics between the three states are described as follows (see also Figure \ref{fig:comp1}):
\begin{itemize}
\item{\bf Infecting:}
Each infected node, while being infectious, \emph{infects} each of its neighbors within the network with probability $\beta$ at every time-step. Infections to different neighbors happen independently.
\item{\bf Healing:} When infectious, each node \emph{heals} with probability $\gamma$ at every time-step, independently of other nodes, and independently of infecting other nodes. The average infectious period of an infected node is $1/\gamma$ time-steps. Upon healing, the node becomes temporarily immune. \item{\bf Losing immunity:}  Each temporarily immune node loses its immunity with probability $\eta$ at every time-step, independently of other nodes. The average immune period of a node is thus $1/\eta$ time-steps.  After losing immunity, the node becomes susceptible again.
\end{itemize}
We run these dynamics on three different types of networks, covering the above mentioned scenarios: the purely geometric square grid, the pure network configuration model, and the mixture Geometric Inhomogeneous Random Graph (GIRG) (see Definitions \ref{def:grid}, \ref{def:config}, and \ref{def:sclm}) below.

We adapt these dynamics to obtain a continuous S-I-T-S compartmental model (Scenario 4) in Section \ref{s:ODE}.
We choose the parameters of the underlying network models in Scenarios 1--3 so that the models have comparable average degree (and when possible, also comparable variance), and then vary the parameters $\beta, \gamma$ and $\eta$. In the remainder of this section we give a quick summary of our findings, then in Sections \ref{s:models} and \ref{s:interventions} we define the four scenarios and three intervention methods in more detail, and in Section \ref{s:results} we present our detailed simulation results.

\subsection{Phases of S-I-T-S on continuous compartmental models.} \label{s:phases-sits}
For Scenario 4, we observe three (commonly known) phases of the epidemic, that we briefly describe here for comparison. For the model description, see Section \ref{s:ODE}.
\begin{enumerate}
\item [1)] {\bf Subcritical and critical phase}: \emph{Immediate extinction}. Whenever the basic reproduction number, $R_0=\beta/\gamma\le 1$, the epidemic dies out  without producing a peak. For $\beta/\gamma<1$, it dies out quickly (logarithmic in the initial number of infected), while for $\beta/\gamma=1$ it dies out slowly (polynomial in the initial number of infected).  See Figure \ref{fig:ode3}.
\item[2)] {\bf Supercritical phase}: \emph{Peaks of decreasing magnitude towards a limiting stationary proportion}.
Whenever $\beta/\gamma> 1$,
the proportion of infected population stabilizes at $ \frac{\eta}{\eta+\gamma}\big(1-\frac{\gamma}{\beta}\big)$. There are several larger peaks before the equilibrium is reached, see Figures \ref{fig:ode1}--\ref{fig:ode3}. The incline and decline of peaks are exponential. See more in Section \ref{s:ODE}.
\end{enumerate}

\subsection{New supercritical phases in S-I-T-S epidemics on network models.}\label{s:phases-sits-graph}
The presence of the network structure, as well as the underlying space, combined with temporary immunity has a surprising effect on the qualitative behavior of the epidemic. More specifically, new phases arise in the phase diagram compared to those in Section \ref{s:phases-sits}. The duration of the immune period plays a more profound role than in the continuous compartmental model of Section \ref{s:ODE}, where the single supercritical  phase is entirely determined by the ratio $\beta/\gamma$ of the infection rate and the healing rate. For finite spatial network models, there is no well-defined definition of $R_0$: the average node-degree as well as the topology of the network influences the probability of entering each phase. See \cite[Figure 1]{SouTom10} for a phase diagram for Scenario 1.
For the S-I-T-S epidemic spread on all three network models (Scenarios 1, 2, and 3) we observe  the following three phases, with 2S below being new compared to Scenario 4:
\begin{itemize}
\item[1)] {\bf Subcritical and critical phase}: \emph{Immediate extinction.} The epidemic goes extinct almost immediately (for all runs), and the number of infected decays exponentially in this case. When $R_0\approx 1$, the epidemic may die out more slowly. We emphasize though, that the critical and subcritical phase do not separate so clearly on the stochastic S-I-T-S model on networks: due to the stochastic nature, the epidemic might die out quickly even when $R_0> 1$. For the extinction time (logarithmic vs polynomial of the network size), the initial number of infected individuals plays a more important role. This is, however, out of the scope of our current study and we do not pursue this direction further.
\item[2)] {\bf Supercritical phases}: \emph{Possible large outbreak.}  With positive probability, there is a large outbreak. For network-based S-I-T-S, the supercritical phase separates into two sub-phases:
\begin{itemize}
\item[2S)] {\bf Extinction after a Single peak}: When the duration of the immune period is long, the epidemic has a single peak, after which it immediately goes extinct. Heuristically, the reason for the existence of this new phase is that after the first epidemic peak, earlier infected nodes become immune and maintain their immunity long-enough to provide barriers in the network that the infection cannot pass through. See Figure \ref{fig:girg-cmtau33-1top}.
\item[2M)] {\bf Long-time survival, Many peaks:} When the duration of the immunity period is relatively short, the epidemic follows a (qualitatively) similar curve to the one observed in the ODE in Section \ref{s:ODE}. There is a first major outbreak, followed by smaller second, third, etc., peaks, decreasing in magnitude, and eventually settling on a \emph{(meta-stable) stationary proportion} of infected and immune population. See Figure \ref{fig:girg-cmtau33-1bottom}. We mention that this equilibrium is a meta-stable state, since the all-susceptible  configuration is an absorbing state. The time to reach that, however, is exponentially long in the network size, see \cite{ChaDur09}.
\end{itemize}
\end{itemize}
\begin{figure}[t]\begin{center}
		\includegraphics[width=0.6\textwidth]{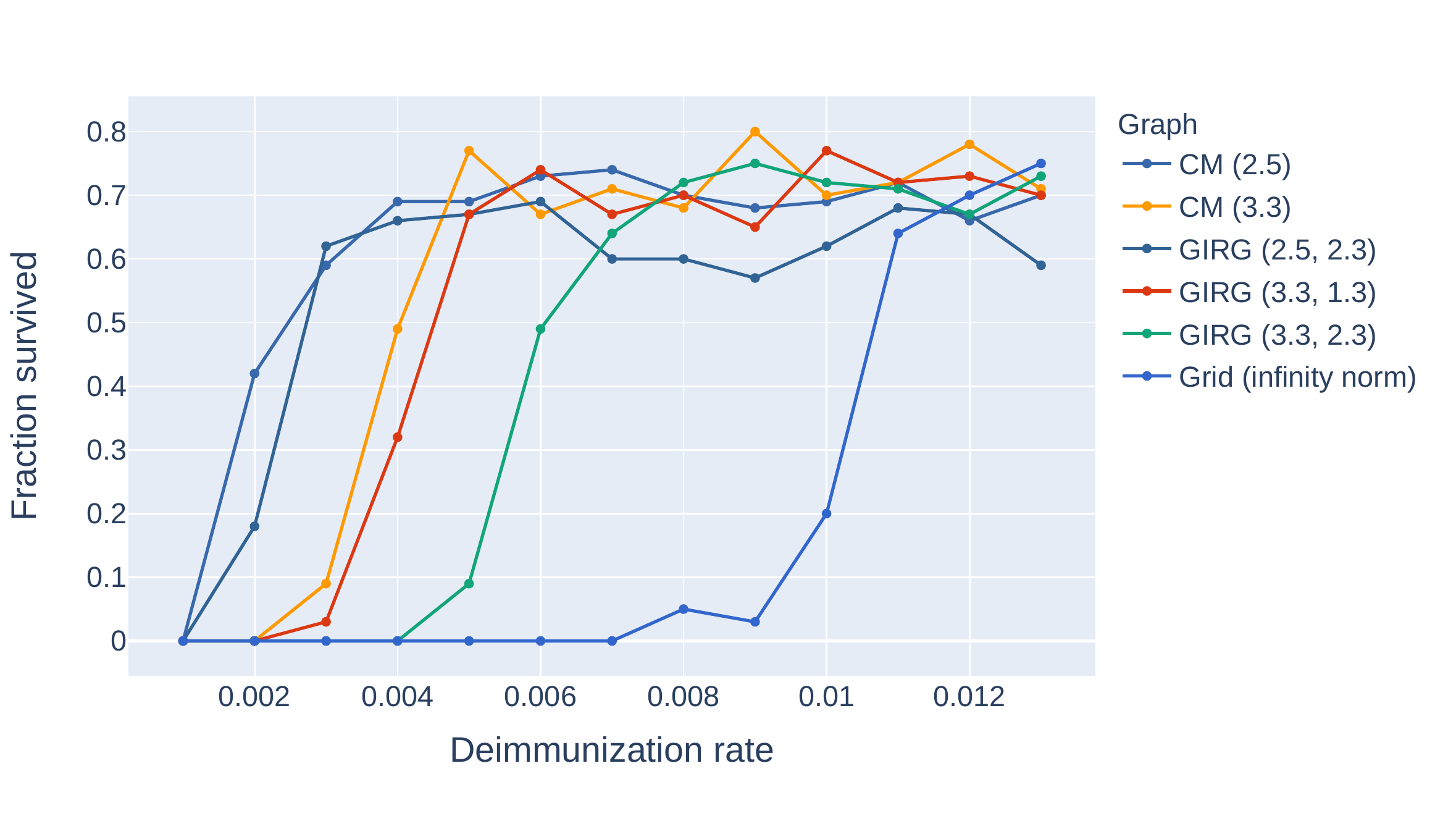}
		\caption{A comparison of the probability of phase 2M on eight different networks with the same average degree $8$ and number of nodes $N=160000$, as a function of the $\eta$.  Probabilities were computed using a $100$ runs for all parameter values ($\eta$) and models, with $\beta=0.225,  \gamma=0.2$.  Legend: CM, GIRG: Scenarios 2 and 3, with different parameters and Grid: Scenario 1.  For all networks we see a sharp transition at a critical value of $\eta$ where the system moves from phase 2P (single peak followed by extinction) to 2M (survival with multiple peaks) with overwhelming probability.
				} \label{fig:survival2}.
\end{center}\end{figure}

{\bf Stochastic transition between phases with a sharp threshold}: Due to the stochastic nature of the model, the phases 1, 2S, 2M are, (at least on a finite but large network) intertwined. This means that for a given fixed set of parameter $\beta, \gamma, \eta$, a single run of the S-I-T-S epidemic may enter any of the three phases, with different probabilities. However, the probability of entering a given phase undergoes a sharp transition in the parameter values: for fixed $\beta, \gamma$, a small change in $\eta$ results in a shift from almost always seeing phase 2S to almost always seeing phase 2M. This is called a \emph{sharp threshold phenomenon}, see \cite{Hugosharp, FridKal96}. The sharp threshold is defined as the value of $\eta_t(N)$ where the survival probability crosses $1/2$. The same theory suggests that the threshold $\eta_t(N)$ converges in the large network limit.
The probability of long survival (phase 2M) can be seen in Figure \ref{fig:survival2} for eight different models as a function of $\eta$, all with the \emph{same average degree} $\E[\deg (u)]=8$.
We conjecture that the long survival probabilities are monotone in $\eta$, see Figure \ref{fig:survival-500}, where the experiment is carried out with more runs for each value of $\eta$.

{\bf Height of the first peak:} The height of the first peak, in all models, is insensitive to $\eta$, just like for Scenario 4, see Figure \ref{fig:first-second-peak-g2}
A possible explanation for this insensitivity is that during the  initial spread, the S-I-T-S process is closely approximated by an S-I-R process, where temporarily immune nodes are simply removed.

{\bf Shape and location of the first peak:}
For Scenario 2 (mean-field network), the shape of the first and later peaks are similar to that of Scenario 4,  i.e., exponential incline followed by exponential decline. Scenario 1, the grid, however, has a \emph{linear} incline followed by linear decline, if there is a first peak, which is not true, when $\eta$ is large. See Section \ref{s:lattice}. The shape of the curve for Scenario 3 depends on the parameter values:  as long as there are many long-range connections, and hubs in the network (nodes with very high degree), the curve resembles that of Scenario 2.

When this is not the case, Scenario 3 starts to gradually resemble the epidemic curve of the grid (Scenario 1): linear incline followed by linear decline, see Figure \ref{fig:girg-cmtau33-1}. This is in accordance with theoretical results \eqref{eq:dist-girg} related to the average distance in the network being short (poly-logarithmic) when there are many long-connections  but long (polynomial) when these are scarce.
See \cite{CovidKuchStroebFacebookConn2020} and references therein for the role of long-connections.

{\bf Topology-dependent fluctuations after the first peak:} The fluctuations after the first peak are highly dependent on the network topology. For power-law networks  with infinite asymptotic variance, there is no second peak and the system reaches equilibrium very quickly after the first peak. For finite variance networks, several further peaks occur, see e.g. Figures \ref{fig:twogirg-noint}, \ref{fig:girg1-interventions}.  The transition is most profound when interventions change the topology, see Section \ref{s:short-interventions} below.

We mention that for Scenario 3 the way the second peak happens is entirely different from the first peak: the initial spread is local,  emanating from the source, (with some long connections causing non-local new infection centers), while the second peak is spread out, infections occur everywhere in space. This is quite natural, since during the initial wave, the epidemic fills the space, and thus the second peak happens roughly when the immunity arising from the first wave starts to wear off, and at that moment it does not have a well-defined source anymore.
This is why we never see the epidemic to die out after a second peak. It either dies out after the first peak, or many peaks occur and the system reaches equilibrium.

We conclude that the continuous ODE approximations often fail to capture the later behavior of the epidemic, which depends highly on the network topology.

\subsection{Comparison of intervention methods A) B) C) on Scenario 3.}\label{s:short-interventions}
We carry out the intervention methods on Scenario 3, the geometric scale-free network model.
We model intervention (A),  keeping physical distance by randomly removing connections from the network.

Intervention (B), allowing for less travel, is modeled in two different ways. Once, it is modeled by drawing for each present edge a random exponential threshold with a common mean $L$, and cutting the edge if it exceeds its threshold value, resulting in a connection-length distribution that decays exponentially above the threshold $L$. Secondly, it is also modeled by increasing the long-range parameter $\alpha$ to $\alpha^{\text{new}}$.

Intervention (C), limiting the maximal number of contact per person, is modeled by prescribing a maximal node degree $M$ and then for each node $u$ with degree higher than $M$, randomly chosen connections of $u$  are cut  until at most $M$ connections remain, see more details in Section \ref{s:interventions}.

We choose the parameters $M, L, \alpha^{\text{new}}$ so that the average degree is the same after all  interventions.
See Figure \ref{fig:networks-intro} for a visualization of the intervention methods.

We summarize our most important, qualitative findings. For quantitative values see Section \ref{s:interventions}.

{\bf The height of the first peak drops}. In all interventions, the height of the first peak is dropping, by at least as much as the shrinkage in average node degree (intervention A) and even more with other interventions. The most effective intervention in this respect is the hard no-travel rule. See left on Figure \ref{fig:first-second-peak-g2}.

{\bf Elongated first peak.} The time and duration of the first peak is later/longer than without intervention. For keeping physical distance, the time-shift is the least apparent (a few days), while for intervention B it is the most profound, a factor of $10$, see bottom of Figure \ref{fig:girg-interventions-firstpeak}, as well as Table \ref{table:heights}.

\begin{figure}[t]
\begin{center}
\begin{subfigure}{\textwidth}
		\centering
		\includegraphics[width=0.8\textwidth]{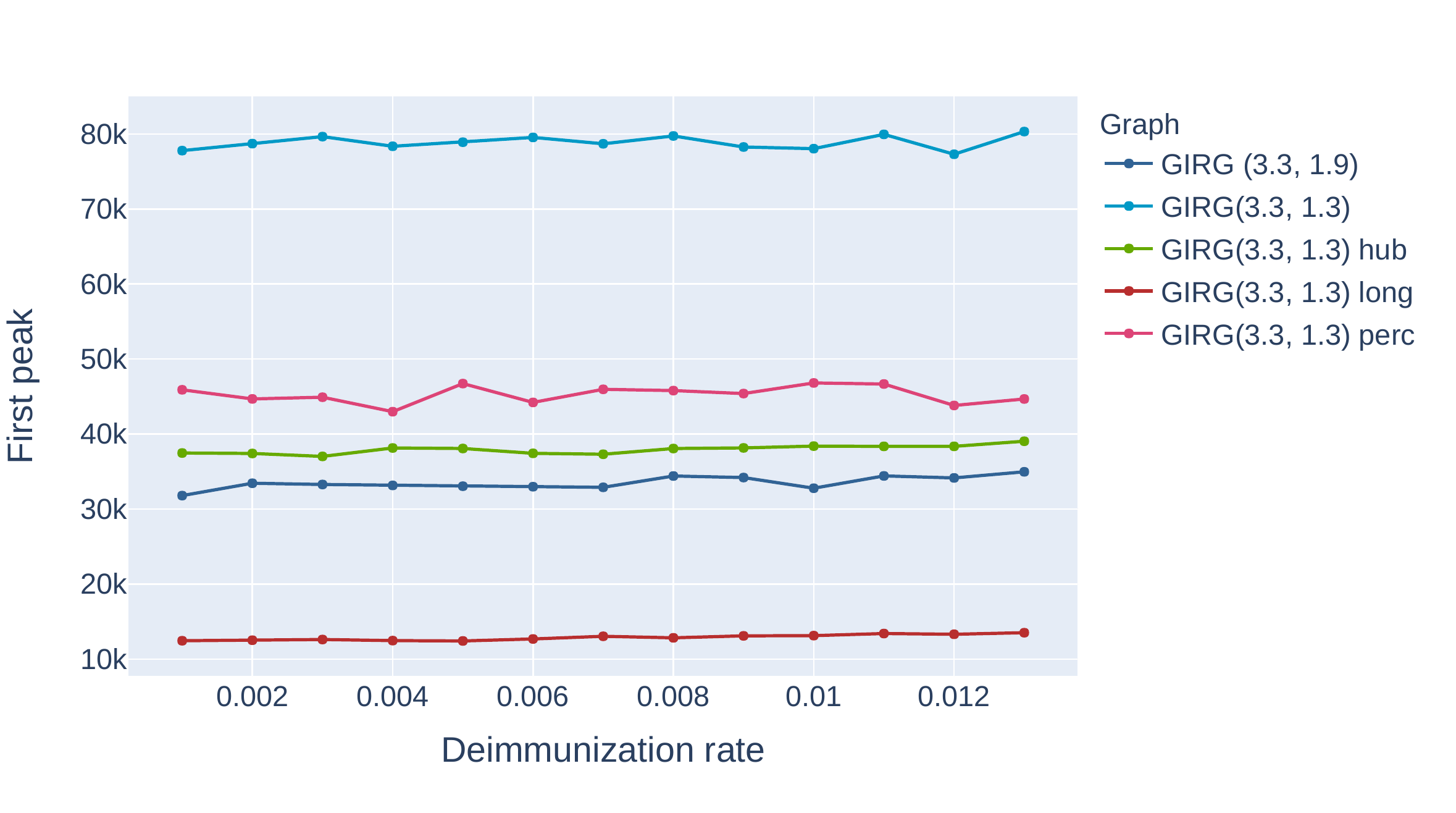}
		\caption{First peak under interventions}
	\end{subfigure}
	\begin{subfigure}{\textwidth}
		\centering
		\includegraphics[width=0.8\textwidth]{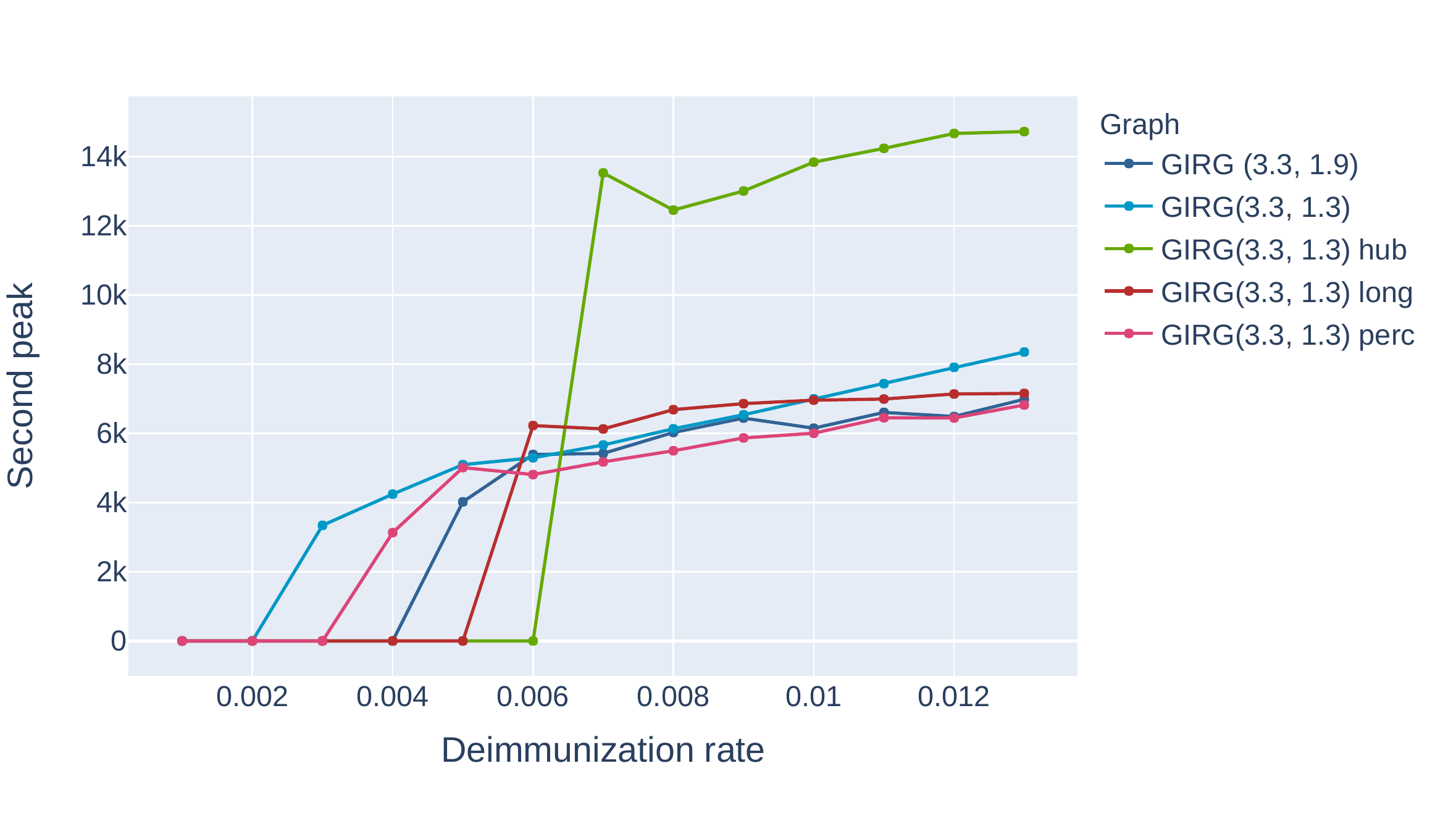}
		\caption{Second peak under interventions}
	\end{subfigure}
\end{center}
\caption{The effect of interventions on the first and second peak, as a function of $\eta$ (the inverse of the average immunity period). Abbreviations in the legend: `hub': limiting maximal node degree, `long': strong restriction on travel, `perc': social distancing. The first entry is increasing $\alpha$, a weak restriction on travel. GIRG$(3.3,1.3)$ is the original network. We see that the height of the first peak is insensitive to the immunity length, and that cutting long edges is most effective in reducing the first peak. For the second  peak, limiting node degree pushes the critical $\eta$ for appearance of the second peak from $0.002$ to $0.006$, however, above that the second peak is twice as high as for other interventions. See also Figures \ref{fig:first-second-peak-g1}, \ref{fig:first-second-peak-g2-v2}.}
\end{figure}\label{fig:first-second-peak-g2}

{\bf Long-survival probabilities drop.} Each intervention pushes the probability of the system entering phase (2M) lower for a fixed parameter setting $\beta, \gamma, \eta$. When increasing $\eta$ (thus decreasing the immunity length), at a critical $\eta_t$ the system abruptly goes from dominantly phase 2S from dominantly phase 2M, (a sharp threshold). The sharp threshold $\eta_t$ of long-survival is increased under all interventions, see Figure \ref{fig:survival-interventions}.  In this respect, interventions B and C perform best.

 {\bf Higher second peak.} Intervention B and C result in a higher second peak, or even make a second peak appear where originally it was not present. The worst intervention in this respect is intervention C (limiting the maximum degree), where the second peak can be as high as $1/3$ of the first peak. See Figure \ref{fig:first-second-peak-g2} and \ref{fig:first-second-peak-g1}. For infinite variance degree GIRG, on Fig. \ref{fig:first-second-peak-g1}) the strong travel restriction also causes a second peak, while social distancing and increasing $\alpha$ does not.

A possible explanation for a higher second peak is the following. Under intervention C (limiting node degree) and Intervention B (cutting long-edges), there are no more hubs in the network: nodes with very high degree. Once infected, hubs infect a large proportion of their neighbors roughly at the same time. Moreover, hubs are much closer to each other (in terms of graph distance) than the average distance in the graph, thus the infection can travel quickly between hubs. To summarize, hubs \emph{synchronize} the system. Once the hubs are removed, this synchronization is also removed from the system and oscillations similar to the grid appear. Limiting node degree without travel restrictions has the strongest effect on the removal of hubs. Limiting travel reduces the degree of hubs, but do not remove them completely, since nodes that were hubs before the intervention, have typically many local connections as well, see Figure \ref{fig:change-graph2}.

 {\bf More oscillations to reach equilibrium}. Interventions C and B, while they are most effective regarding the first peak,  introduce high oscillations in the system, and the time to reach equilibrium can be much longer than in the original model (from a few days to more than a thousand days). Again, these oscillations are explained by the lack of synchronization. See Figures \ref{fig:girg1-interventions} and  \ref{fig:girg2-interventions} for these later effects.

\medskip

To summarize, we conclude that the effect of interventions vary to a high extent and depend on the precise way the intervention changes the network topology. Again, ODE-approximations fail to capture these various effects.

\section{Network models}\label{s:models}
\subsection{Scenario 1: Lattice network models}
Arguably the simplest geometric network model is the nearest neighbor lattice. To avoid boundary effects, and make the network completely homogeneous, we shall study epidemics on the torus $\Z_n^2$ instead of a square that we define here:
\begin{definition}[Nearest neighbor tori]\label{def:grid} Arrange nodes on a square grid, and label them with two coordinates $u=(x,y)$, for $x,y\in\{1,2\dots, n\}$. To obtain $\Z_n^2$, connect $u=(x_1,y_1)$ to $v=(x_2,y_2)$ when either $x_1=x_2$ and $|y_1-y_2| = 1 \mod n$ or  $y_1=y_2$ and $|x_1-x_2| = 1 \mod n$. To obtain $\wit Z_n^2$, connect $u=(x_1,y_1)$ to $v=(x_2,y_2)$ when $|x_1-x_2|\le 1 \mod n$ and $|y_1-y_2| \le 1 \mod n$.
The networks have  $N=n^2$ nodes, see Figure \ref{fig:grid-graph}.
\end{definition}
Lattice models are homogeneous and the average distance within the network is large, it grows polynomially with the number of nodes. Writing $d_G(u,v)$ for the number of connections on the shortest path between nodes $u$ and $v$, (also called: graph distance), the average distance becomes
\be\label{eq:dist-lattice} \overline{\mathrm{Dist}}(N):= \frac{1}{{N \choose 2} } \sum_{u,v} d_G(u,v) = \Theta(\sqrt{N}). \ee
\subsection{Scenario 2: The configuration model and other a-geometric network models}
A-geometric random network models are often used as null-models to compare network data to purely random networks. Here we describe a commonly used model, the configuration model, and mention similar alternatives. The configuration model is a well-studied object in mathematical network science, dating back to Bollob\'as \cite{Boll1980} and Molloy and Reed \cite{MollReed95, MollReed98}. The main advantage of this simple model is that it can mimic the degree distribution of real-life networks.

\begin{definition}[Configuration model]\label{def:config} Fix $N\ge 1$ the number of nodes. Prescribe to each  node $u \in \{1,2,\dots, n\}$ its node-degree $\deg(u)\ge 0$, so  that the total degree $h_N:=\sum_{u\le N} \deg(u)$ is even. To form a graph, to each node $u$ we assign $\deg(u)$ number of half-edges and the half-edges are then paired uniformly at random to form edges. The resulting random multi-graph is the configuration model. The \emph{erased configuration model} can be obtained by erasing all self-loops and multiple edges between nodes.
\end{definition}
See Figure \ref{fig:config} for an illustration of the algorithm, and Figure \ref{fig:config2} for two realizations.
 Observe that we pair half-edges randomly,  hence, even though the prescribed degrees might be deterministic, the configuration model  is a random (multi)graph. It is a multi-graph since multiple connections between nodes and even connections to the same node may occur, however, these become proportionally insignificant when the size of the graph is sufficiently large. An easy way to obtain a simple graph from the configuration model is to erase second, third, etc edges between nodes, as well as self-connections. It is shown that the erasure does not affect the empirical distribution of the model in any significant way \cite{Hofbook}.  For a mathematical overview on the properties of the configuration model and erased configuration model, see \cite{Hofbook}.

In real-life networks, an extreme variability of node degree is often observed, see \cite{albert2002statistical, DorMen02, newman2011structure}. Extreme node degree variability results in the presence of a few individuals with extreme influence on spreading processes, the \emph{hubs or superspreaders} \cite{newman2011structure}. Mathematically, this extreme degree variability can be expressed using the empirical distribution of node degrees, that follows a  \emph{power-law}:
\begin{equation}\label{power-law}
\mathbf{Prob}(\mathrm{deg}(u)\ge x) \asymp \frac{1}{x^{\tau-1}}.
\end{equation}
with exponent $\tau >2$.
 In our simulations we study the case when the empirical degree distribution follows such a power law, i.e., the node-degrees satisfy \eqref{power-law}, for some parameter $\tau>2$. In this case, the node-degrees in the erased configuration model version also follow \eqref{power-law}, with the same exponent $\tau$.

 Other popular a-geometric models include the Chung-Lu and the similar Norros-Reittu model \cite{ChungLu02.1, ChungLu02.2, NorRei06}, and preferential attachment models (also known as the Barab\'asi-Albert model) \cite{BarAlb99}.
 In the Chung-Lu and the Norros-Reittu model, only the expected degrees of nodes are prescribed, rather than their exact degree. It is shown that these models, with similar parameter settings, behave qualitatively similarly to the configuration model, see   \cite{Hofbook} for references. The preferential attachment model is a growing network model, that is best suitable to study the evolution of networks on a longer time-scale. For a relatively short time-frame typical to epidemic spread, a time snapshot of the preferential model could be used. Mathematical results suggest that snapshots of this model behave, again, qualitative similarly to the configuration model with similar degree structure, although numerical values might differ \cite{berger2005spread, DorMen02, DerMor09, DomHofHoog10}. As a result, we choose the configuration model for baseline comparison.

An important feature of the configuration model is the \emph{small-world property}. Heuristically, this means that two arbitrary nodes in the network can be connected via very short paths,  using only a few connections.  Mathematically, the small world phenomenon means that the average graph distance  $\overline{\mathrm{Dist}}(N)=\sum_{u,v} d_G(u,v)/{N \choose 2}$ in the network is small, at most
logarithmic in the network size. For the configuration model, when $\tau>3$, the network is proven to be a small world \cite{HofHoog08, HofHoogZnam05}, while, when $\tau \in (2,3)$, the average graph distance is ultra small, meaning a double-logarithm of the network size:
\be\label{eq:smallw} \overline{\mathrm{Dist}}(N) =  \begin{cases}\Theta( \log N) &\text{ when } \tau>3 \\
\Theta(\log \log N) &\text{when } \tau \in (2,3). \end{cases}\ee
In the former case, when $\tau>3$, the shape of the epidemic curve is shown to converge in the large network limit, see \cite{BarbRein13, BhaHofKom14, SvaLucWin14}.

 While the configuration model easily accommodates power-law node degrees, and has the small-world property, it neither contains communities nor clustering \cite{DorMen02, SteHofLee19}. Empirically, clustering quantifies the effect commonly known as  ``a friend of a friend is also likely to be my friend''. Mathematically, clustering means the presence of \emph{triangles} in the network. Communities are parts of the network that have significantly more connections than the same number of randomly selected nodes.
One could argue that clustering and communities in human contact networks often arise due to spatial effects: people living nearby tend to know each other with higher probability.  To accommodate clustering and communities, as well as keep node degree variability high, we introduce our last model of investigation.

\subsection{Scenario 3: Geometric Inhomogeneous Random Graphs}
The last network class we run the S-I-T-S epidemic on is a mixture of pure geometric and purely random network models.
For this we use a general Geometric Inhomogeneous Random Graph (GIRG) model, that is a state-of-the-art model for real-world social and technological networks, embedded in geometric space.
		The presence of underlying geometry underpins the model: individuals are embedded in space, just like in real life, allowing for local community structures to be present in the contact network, leading to strong clustering \cite{BriKeuLen19, SteHofLee19}. The GIRG incorporates edges bridging spatial distance on all scales, ranging from short to long-range edges, as well as a high variability of node degree: in fact, the model is \emph{scale-free} in two respects: both in spatial distance that edges cover, and in node-degree variability \cite{BriKeuLen19}.

Contact- or activity networks of humans have been found to show similar behavior, including heavy-tailed degree distributions, strong clustering, and community structures \cite{albert2002statistical,boccaletti2006complex, muchpei13, newman2011structure,newman2018networks}, as well as heavy-tailed distance-distribution for edges, see references in \cite{IlNag13}.

Recently, several spatial random graph models were  developed to incorporate these features: hyperbolic random graphs \cite{BodFouMul15, papadopoulos2010greedy,GugPanPet12}, scale-free percolation \cite{DeiHofHoo13}, and GIRGs \cite{BriKeuLen15, BriKeuLen19,bringmann2016average}. These models can be unified into a general model containing all these three models as special cases. The qualitative behavior of the three models is the same. The following definition is general, and in its last sentence we specify it for GIRG. The underlying space can be the earth's surface, or $\R^2$, the two-dimensional Euclidean space. We denote by $x\wedge y$ the minimum of two numbers $x,y$.
\begin{definition}[Geometric Inhomogeneous Random Graph (GIRG)]\label{def:sclm} Fix $N\ge 1$ the number of nodes. Assign to each  node $u \in \{1,2,\dots, n\}$ a \emph{fitness} $w_u >0$, and a  \emph{spatial location} $\Phi(u)$.  Fix $\alpha>0$. For any pair of nodes $u,v$ with fixed $w_u, w_v, \Phi(u), \Phi(v)$, connect them by an edge with probability
\begin{equation}\label{longedge}
	\mathbf{Prob}( u \text{ is connected to } v) \asymp \Big(\frac{w_u  w_v}{\mathrm{dist}(\Phi(u),\Phi(v))^{2}}\Big)^\alpha \wedge1,
\end{equation}
where $\mathrm{dist}$ is a distance function\footnote{Mathematically, $\mathrm{dist}$ is a metric on the underlying space, (e.g.\ $\| \, \cdot \|_2$). Here, the formula is given for a two-dimensional model. In higher dimension, the exponent $2\alpha$ in the denominator should change to $\dim\cdot \alpha$, with $\dim$ the dimension of the model.}, i.e., it measures the distance between the locations $\Phi(u), \Phi(v)$. When $\Phi(u)$ is chosen uniformly in a box of volume $N$, we obtain the GIRG.
\end{definition}
Any or all of the vertex set, fitnesses, and spatial locations can be random (see Figure~\ref{fig:GIRG}).
GIRGs have a natural interpretation: the fitnesses express the ability of nodes to have many connections, $\Phi$ embeds them in space, and $\alpha$ is the \emph{long-range} parameter: the smaller $\alpha$ is, the more the model favors longer connections.
 The parameter space of GIRG is rich enough to model many desired features observed in real networks:
\begin{itemize}
\item[(a)] extreme variability of the number of neighbors (degrees),
\item[(b)] connections  present on all  length-scales,
\item[(c)] small and ultra-small distances,
\item[(d)] strong clustering,
\item[(e)] local communities.
\end{itemize}
In real-life networks, (a) corresponds to the presence of a few individuals with extreme influence on spreading processes, the \emph{hubs or superspreaders}.
Mathematically, (a) corresponds to the empirical distribution of node degrees following a  \emph{power-law}, as in \eqref{power-law} above,
 for some $\tau >2$.
Setting a power-law fitness distribution for $w_u$ in GIRG yields that the degrees satisfy \eqref{power-law}, while setting $w_u\equiv 1$ results in a (fairly) homogeneous network with some longer edges, that is known in the literature as \emph{(continuum) long-range percolation} \cite{Bisk04}.
The abundance of long-range connections is tuned by $\alpha$ in \eqref{longedge}: the smaller $\alpha$, the more likely are long-range connections (b).
The power-law exponent $\tau$ and the long-range parameter $\alpha$ tune the average graph distance $\overline{\mathrm{Dist}}(N)$, see  \cite{DeiHofHoo13, bringmann2016average, Bisk04, DeHaWu15}:
\be \label{eq:dist-girg}\overline{\mathrm{Dist}}(N)
= \begin{cases}
\Theta(\log \log N) & \text{when } \tau\in(2,3), \alpha>1\\
\Theta \big((\log N)^\zeta\big)  & \text{when } \tau>3, \alpha\in(1,2) \\
\Theta (\sqrt{N})  & \text{when } \tau>3, \alpha>2.
 \end{cases}\ee
Comparing this to the average distance in the configuration model in \eqref{eq:smallw} and to the lattice models in \eqref{eq:dist-lattice}, one sees that geometry and long-range connections play a role when $\tau>3$, and the model interpolates between the small-world configuration model and the lattice.

Empirically, clustering quantifies the effect commonly known as  ``a friend of a friend is also likely to be my friend''. Mathematically, clustering means the presence of \emph{triangles} in the network.
Communities and clustering are naturally present in SCLM, because the model favors connections between nodes that are close to each other in space (see Fig.\ \ref{fig:GIRG}).
Thus, GIRG incorporates all five desired features.

	For the spread of information or infections in such networks, it is known that large-degree nodes and many triangles have opposing effects.
	On the one hand, nodes of large degree (also called hubs, super-spreaders, or influencers) contribute to fast dissemination, and foster explosive propagation of information or infections~\cite{gruhl2004information,pastor2015epidemic,pastor2007evolution,dorogovtsev2008critical}. On the other hand, clustering and community structures provide natural barriers that slow down the process, while long-range edges accelerate the spread~\cite{karsai2011small,merler2015spatiotemporal,bajardi2011dynamical,janssen2017rumors,isham2011spread}.

\subsection{Scenario 4: S-I-T-S compartmental model}\label{s:ODE}
 For baseline comparison to continuous epidemic models, we use a continuous S-I-T-S model, where there are only three possible states, \emph{susceptible (S), infected (I) or temporarily immune (T)}, in a total population of $N$ individuals. Susceptible individuals may become infected via a contact to infectious individuals, while infectious individuals heal and thus become temporarily immune. The assumption is that individuals make contact with a random other individual at rate $\beta$. Immune individuals loose their immunity at a certain rate and become susceptible again.
\begin{figure}[t]\begin{center}
		\includegraphics[width=0.85\columnwidth]{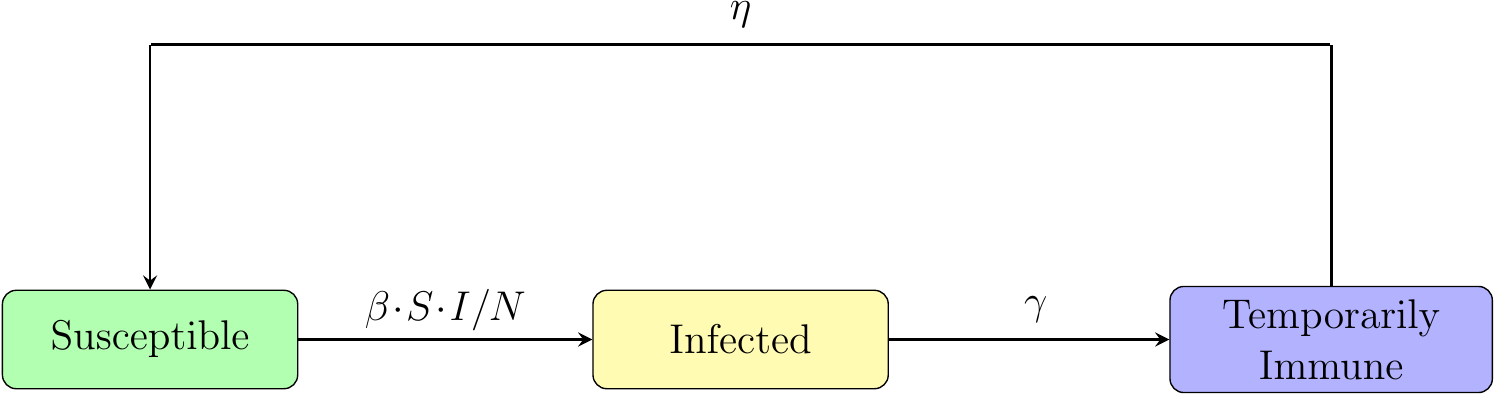}
		\caption{Schematic diagram of  states and their transition rates in the continuous compartmental model. Infected individuals make contact at rate $\beta$ to randomly chosen individuals, as a result, the number of susceptibles increases at rate $\beta\! \cdot\! S\!\cdot\! I/N$. This is a mean-field approximation of the graph model in Section \ref{s:epi-girg}, see Figure \ref{fig:comp1}.}
\label{fig:comp2}.
\end{center}\end{figure}
The deterministic ordinary differential equation (ODE) model is governed by the three main parameters:
 \begin{itemize}
\item $\beta$: rate of infecting a susceptible individual when being infected,
\item $\gamma$: rate of healing, and becoming temporarily immune,
\item $\eta$: rate of  losing temporal immunity and becoming susceptible again.
\end{itemize}
See Figure \ref{fig:comp2} for a schematic diagram.
We assume the population size is $N$, denote by $S(t), I(t), T(t)$ the number of susceptible, infected and temporarily immune individuals at time $t$, then the corresponding ordinary differential equation becomes:
\be\ba
\frac{\dd S}{\dd t} &= -\frac{\beta I S}{N} + \eta T \\
\frac{\dd I}{\dd t} &= +\frac{\beta I S}{N} - \gamma I\\
\frac{\dd T}{\dd t} &= + \gamma I - \eta T
\ea\nonumber\ee
For the infection to survive,  $\beta>\gamma$ is necessary, otherwise $\lim_{t\to \infty} I(t)=0$.  When $\beta>\gamma$ holds, this ODE has a stationary state, which is explicitly computable:
\be\label{eq:equilibrium} S_\infty= N\frac{\gamma}{\beta} , \quad  I_\infty= N \frac{\eta}{\eta+\gamma}\Big(1-\frac{\gamma}{\beta}\Big), \quad T_\infty= N\frac{\gamma}{\eta+\gamma}\Big(1-\frac{\gamma}{\beta}\Big).\nonumber \ee
Observe that the equilibrium proportion of infected is roughly linear in $\eta$, as long as $\eta\ll \gamma$.
One can also compute the basic reproduction number, $R_0=\beta/\gamma$, and observe that $T_\infty= I_\infty\gamma/\eta$.
Qualitatively, the ODE exhibits either (sub)critical behavior, implying extinction, or supercritical behavior, as described in Section \ref{s:phases-sits} and depicted in Figures \ref{fig:ode1}--\ref{fig:ode3}.
 On these figures, for  the supercritical case one can observe that
 \begin{itemize}
 \item[a)] the height of the first peak does not depend on the length of the immune period; only on $\beta,\gamma$.
 \item[b)] for fixed $\beta,\gamma$, increasing the length of the immune period results in deepening the valley between the first and the second peak. The minimum between the first and second peak is always strictly positive, however, it approaches $0$ as $1/\eta$ tends to infinity. For somewhat realistic values ($\beta=0.38, \gamma=0.14$, the minimum gets near $0$ around $1/\eta\approx210$).
 Since this is a continuous system, the second peak is always present, explaining the absence of phase 2S in this compartmental model. \end{itemize}

\section{Modeling intervention methods on GIRGs}\label{s:interventions}
\subsection*{A) Social distancing measures: percolation}
In the current Covid-19 outbreak, many governments issued a collection of social distancing measures: keeping $1.5$ meter distance from each other, wearing face masks,  hand-gloves, washing hands frequently, and so on. Such measures result in reducing the number of contact moments
between individuals and thus resulting the chance for the virus to spread. When the social distancing measures are not combined with travel restrictions, this results in the underlying contact network to lose its connections randomly.

Hence, we model the effect of social distancing by \emph{randomly removing} a certain proportion of edges in the network. This method is called \emph{percolation} in the literature \cite{Grim99}. The same method is used to model social distancing in e.g. \cite{CovidKumarAdhiTempRandGeomIndia2020}, where percolation is applied to the configuration model.  We  denote the resulting network by $G^{\text{perc}}$.

\subsection*{B) Travel restrictions}
In the current Covid-19 outbreak, many governments  issued everybody to stay at home and only travel when necessary. Outside country travel is not allowed for some countries, e.g. Spain, Italy, Germany, Hungary. This results in the underlying contact network to lose many of its long connections.

We model  the removal of long-connections in the network in two different ways: a (seemingly) milder and a stronger intervention measure.

 \medskip
{\bf B(w): Weak travel restriction.}
One way to reduce the number of long-range connection in the network is by increasing the parameter $\alpha$ in GIRG. This intervention does not cause a `hard cutoff' for the length of connections: some long-range connections might remain, but the intervention is strong enough to push the distances in the network from poly-logarithmic to polynomial, see \eqref{eq:dist-girg}.

Increasing  $\alpha$, when $\tau>3$, decreases the probabilities of long edges: indeed, in \eqref{longedge}, for two nodes $u,v$ with locations $\Phi(u), \Phi(v)$, the ratio
\be R_{u,v}:=\frac{w_u w_v}{\mathrm{dist}(\Phi(u), \Phi(v))^2}\nonumber\ee
is raised to the power $\alpha$, so as long $R_{u,v}<1$, increasing $\alpha$ reduces the chance of a long connection being present.

When $\tau>3$, the node fitnesses are, typically, not high enough\footnote{Node fitnesses are less than $\sqrt{n}$ for all nodes, with high probability} so that, typically, $R_{u,v}<1$, hence the intervention of increasing $\alpha$ is very effective.
The average distance increases from poly-logarithmic to polynomial in the network size.  On the other hand, when $\tau\in(2,3)$, node pairs with ratio $R_{u,v}>1$ happen frequently enough (even on global scale, nodes with fitness $> \sqrt{n}$ are present), hence, increasing $\alpha$ turns out to be ineffective in reducing the average distance in the network, since many long-range edges remain. This is in good correspondence with \eqref{eq:dist-girg}, where the average distance is doubly-logarithmic regardless of the value of $\alpha$ when $\tau\in(2,3)$.
We denote the resulting network by $G^\alpha$.

\medskip
{\bf B(s): Strong travel restriction.}
Another way to reduce the number of long-range connection in the network is by removing long-edges in the network.
Mathematically, we set a threshold length $L$, and for each edge $e=(u,v)$ present in the original network we draw an independent Exponential variable $X_{(u,v)}$ with mean $L$;  and remove the edge $e$ when
\be \mathrm{dist}(\Phi(u), \Phi(v))> X_{(u,v)} \nonumber\ee
This intervention does cause a `hard cutoff' for the length of connections at the threshold $L$. The intervention is also strong enough to push the distances in the network from poly-logarithmic to polynomial.
We denote the resulting network by $G^{\text{long}}$.

\subsection*{C) Limiting the maximal number of contacts}
Another intervention measure that one can imagine is to limit the maximal number of contact a person can have. On the contact level, this rule barely affects individuals with low number of contacts, but it aims to decrease the number of contact of `superspreaders' in the network: nodes that have a relatively high degree.

We model this effect by choosing a node-degree threshold $M$, and for each node $u$ that has, in the original network, more than  $M$ connections, we remove $\deg(u)-M$ many randomly chosen connections from node $u$. The removal of these connections, in turn, also reduces the number of connections of other vertices.
Thus, the resulting network has maximal node degree at most $M$. Since we removed connections randomly, some long connections still remain in the network if they were originally present.
We denote the resulting network by $G^{\text{hub}}$.

 \section{Simulation results}\label{s:results}

  In this section we present our results for the network based S-I-T-S epidemic model. We first comment on the general description of the types of phases that may occur that are mentioned in Section \ref{s:phases-sits}, then we compare our findings for each Scenario (the purely geometric lattice models, the mean field network configuration model, and their mixture geometric inhomogeneous random graphs) with respect to the various intervention strategies.

  Throughout this section, these underlying networks for the different scenarios (and interventions) are sampled once per parameter setting and kept constant over the simulation runs. On every graph, for every described parameter setting of the S-I-T-S model, the results are based on 100 simulations of the epidemic spreading starting from a single node. This node is sampled uniformly at random at the beginning of every simulation, and differs per run.

 Recall from Section \ref{s:phases-sits-graph} that the epidemic on networks can have two supercritical phases, one with a single peak, and one with many peaks, where the system survives for longer. The transition between these phases is stochastic, but undergoing a sharp threshold.

 {\bf Long-time survival probabilities.}  The probability of long survival (phase 2M) can be seen in Figure \ref{fig:survival} for eight different models as a function of $\eta$, all with the \emph{same average degree} $\E[\deg (u)]=8$: two configuration models with $\tau=2.5$ and $\tau= 3.3$, respectively, three geometric inhomogeneous random graphs (GIRG) with $(\tau, \alpha) = (2.5, 2.3), (3.3,1.3)$ and $(3.3, 2.3)$ respectively,
 and the modified lattice $\widetilde \Z^2$ (called `grid' on the figure). All networks have the same number of $N=160000$ nodes.
   We see that the configuration model with the smallest $\tau$ has the smallest threshold $\eta_t$, followed closely by GIRG with $(\tau, \alpha)=(2.5, 1.3)$. The grid $\wit Z_N^2$ and long range percolation (LRP) with $\alpha=2.3$ have the highest $\eta_t$ ($\approx0.085$ and $\approx0.01$), these are networks where the average distance is linear.
  Theory \cite{Hugosharp, FridKal96} suggest that the threshold $\eta_c(N)$ converges in the large network limit.

 {\bf The effect of geometry: long-range edges matter}
 Comparing the  S-I-T-S epidemic curves on the configuration model and on GIRGs, with matching parameters (degree distribution and average degree) can be seen in Figure \ref{fig:cm-girg-match}.
 When we run the epidemic with the same parameters $\beta, \gamma, \eta$ on  configuration models and on GIRGs, with the same degree power-law exponent $\tau=2.5$ and $\tau=3.3$, and same mean degree, we find that the epidemic curves are in good match, as long as the long range parameter $\alpha$ of GIRG is in the interval  $(1,2)$ when $\tau>3$. For $\tau\in(2,3)$, hubs dominate the network (see left picture on Figure \ref{fig:GIRG}), and the role of the parameter $\alpha$ is insignificant. This is in accordance with theoretical results \eqref{eq:dist-girg}, where distances are doubly-logarithmic regardless of the value $\alpha$ as long as $\tau\in(2,3)$.

 In Figure \ref{fig:girg-cmtau33-1}, we see that there is only a good match between the a-geometric configuration model and GIRG when $\tau<3$ or when $\tau>3$ and $\alpha \in (1,2)$.
 When $\tau>3$, the effect of the long range parameter $\alpha$ plays a crucial role. When $\alpha>2$, even when keeping the average degree and the degree distribution the same, the curve flattens by roughly $40\%$. The explanation for this is that in this parameter range, even though hubs are present in the network, they mostly connect to nearby nodes, and thus average distance is significantly larger  (see \eqref{eq:dist-girg}), thus the infection needs more time to spread, see right on Figure \ref{fig:GIRG}.

 We explain more about this effect under Intervention B(w), the weak travel restrictions in Section \ref{s:intervention} below.
  For the first two scenarios, the height of the first peak is  $80\cdot 10^3$ while in the second scenario, $\alpha=2.3$, the first peak only has magnitude $50\cdot10^3$ nodes, an almost $40\%$ decrease.

  The paper \cite{CovidMedoLongRangeSIR2020} studies the effect of the exponent $\alpha$ in more detail on an S-I-R epidemic (corresponding to the first peak here) on a related model, when \eqref{longedge} is replaced by
  \begin{equation}
  	\mathbf{Prob}( u \text{ is connected to } v) \asymp \Big(\frac{\max(w_u, w_v)}{\mathrm{dist}(\Phi(u),\Phi(v))^{2}}\Big)^\alpha \wedge1.
   \nonumber
  \end{equation}
  This is called the max-kernel in the literature, while \eqref{longedge} corresponds to the product-kernel.
  We mention, however, that the topology of networks generated by the max-kernel is significantly different from that of the product kernel,  and the model in \cite{CovidMedoLongRangeSIR2020} falls under a  different universality class, see \cite[Theorem 1.1]{GraHeyMonMor19}.

 \subsection{Intervention strategies.}\label{s:intervention}
 In this section we study the effect of four different intervention strategies: social distancing, a weak and a strong travel restriction,  and  limiting the maximal number of contacts per node. We apply each intervention to two initial graphs on $N=160000$ nodes: $G_1$, an GIRG with $(\tau,\alpha)=(2.5, 1.3)$ and average degree $9.7$, and $G_2$, an GIRG with  $(\tau,\alpha)=(3.3, 1.3)$ and average degree $8.7$.
 To be able to compare the effect of the different intervention strategies,  we choose the parameters of each intervention so the intervention removes about $40\%$ of all connections, resulting in four sub-networks of $G_1$ and $G_2$, respectively.

 After obtaining the sub networks, we run S-I-T-S on each of them with parameters $\mu=0.225, \beta=0.2$ and varying the values of $\eta$ from $0.001$ to $0.013$ at step-size $0.001$, corresponding to average immunity period $1/\eta$ decreased from $1000$ days  down to $77$ days.

 We visualize and study below the effect of each intervention on the S-I-T-S epidemic in five aspects:
 \begin{itemize}
 \item[i)] the change in the underlying contact network,
 \item[ii)] a visualization of the spread of the initial infection,
 \item[iii)] the probability of long-survival as a function of the immunity parameter $\eta$,
 \item[iv)] the height and location of the first and second peak,
 \item[v)] the full epidemic curve.
 \end{itemize}

 \subsection*{Without intervention}
 For a baseline comparison, we carry out the above listed four points on the initial networks:
 \begin{itemize}
 \item[i)] On the top left picture of Figures \ref{fig:change-graph1}  and \ref{fig:change-graph2}, we see similar networks than $G_1$ and $G_2$: GIRGs with $(\tau,\alpha)=(2.5,1.3)$ and $(3.3,1.3)$, respectively. The size is smaller, $N=1000$ nodes. Both graphs have approximately the same average degree $4.8$.
 \item[ii)] On the top left picture of Figures \ref{fig:G1-vis} and Figures \ref{fig:G2-vis}, the visualization of the first $19$ days of the epidemic are shown: the epidemic spreads very quickly on $G_1$, and also quite fast on $G_2$: (almost) the whole network is infected in both cases. This is explained by the next point, the time of the first peak is about $8$ and $14$ days, respectively.

 \item[iii)] The probability of long-survival (phase 2M) as a function of the immunity parameter $\eta$ can be seen in Figure \ref{fig:survival-interventions}: the top curves on the top/bottom picture show the curves for $G_1$ and $G_2$, respectively. Probabilities are calculated based on a hundred runs for each model and value of $\eta$.
 Observe that the survival curves drop for all intervention types, and stay below $0.7$ even after the sharp increase.

 \item[iv)] Table \ref{table:heights} shows the height and location of the first and second peak for $G_1, G_2$ (column O, original graph), while Figures \ref{fig:first-second-peak-g1}, \ref{fig:first-second-peak-g2-v2} plots these heights as a function of $\eta$.

 We see that the location and height of the first peak is insensitive to the change in $\eta$, it is $\approx 84*10^3$ on day $8$ for $G_1$, while $\approx 79*10^3$ on day $14$ on $G_2$.

 There is no second peak on $G_1$, the `star' shaped dots on the curve indicate the height of the equilibrium. Observe that this grows linearly with $\eta$, in accordance with the ODE-solution in \eqref{eq:equilibrium}.

  On $G_2$ there is a second peak, and its height is increasing with $\eta$, and its location is decreasing with $\eta$. However, there is no clear relation between the location and $1/\eta$, the average immunity length, see Table \ref{table:heights}.
  \item[v)] On Figure \ref{fig:twogirg-noint}, the full epidemic curve ($\eta=0.009$) can be seen,
 on top, the first $60$ days while on the bottom picture the first $500$ days. We indeed see that for $G_1$, when $\tau<3$, the system reaches equilibrium without a well-defined second peak.

 \end{itemize}

 \subsection*{A) Social distancing measures: percolation}
 Before we start with details, a short summary: social distancing seems to be the least effective intervention in terms of reducing peaks, but at least it does not influence the presence or the height of the second peak.
 \begin{itemize}
 \item[i)] On the top right picture of Figures \ref{fig:change-graph1} and \ref{fig:change-graph2}, we see the change in the network after percolation: even though the edge density and the average degree drops, long-range edges still remain present, and the graph looks qualitatively the same as the ones without intervention (top left). The long edges will carry the majority of the infections.
 \item[ii)] On the top right picture of Figures \ref{fig:G1-vis} and Figures \ref{fig:G2-vis}, the visualization of the first $19$ days of the epidemic are shown under social distancing: the epidemic spreads somewhat slower, but it still manages to spread to fill the space and infect a large part of the network. Observe that both top-right pictures are qualitatively similar to the top left pictures, where the same spread is shown without intervention.
 \item[iii)]
 On Figures \ref{fig:girg-interventions-firstpeak}, \ref{fig:girg1-interventions}, \ref{fig:girg2-interventions}  and \ref{fig:girg-interventions-eta13} we see the comparison of the epidemic curves: the first peak appears later on the percolated graphs, and their height is roughly $40\%$ less than the height of the original peak. On $G_1^{\text{perc}}$, there is no second peak, (just like on the original $G_1$), while the second peak on $G_2$ only drops by $10\%$, though it appears later. The equilibrium number of infected also drops, roughly by $33\%$.

 \item[iii)] On Figure \ref{fig:survival-interventions}, the long-survival probabilities of epidemics are shown as a function of $\eta$. Social distancing (called ``perc'' in the legend) has its sharp threshold $\eta_t=0.004$ for $G_1$ and $0.007$ for $G_2$, in this respect it is one of the less effective measures even with respect to survival probability.

 \item[iv)]Table \ref{table:heights} shows the height and location of the first and second peak under social distancing  (column A), while Figures \ref{fig:first-second-peak-g1}, \ref{fig:first-second-peak-g2-v2} plots these heights as a function of $\eta$.

 We see again that the location and height of the first peak is insensitive to the change in $\eta$, it is $\approx 52*10^3$ on day $11\pm1$ for $G_1$, while $\approx 45*10^3$ on day $22\pm1$ on $G_2$. Thus the first peak appears later on the percolated graphs, and their height is roughly $40\%$ less than the height of the original peak.

 There is no second peak under social distancing on $G_1$ (the large data are due to fluctuations in the equilibrium system), while on $G_2$ its height is increasing with $\eta$, and roughly $75--80\%$ of the original graph's second peak.

 \item[v)]
 On Figures \ref{fig:girg-interventions-firstpeak}, \ref{fig:girg1-interventions}, \ref{fig:girg2-interventions}  and \ref{fig:girg-interventions-eta13} we see the comparison of the epidemic curves, just like on the original graph, there is no second peak on  $G_1^{\text{perc}}$ either, while the second peak on $G_2$ only drops by $10\%$, though it appears later. The equilibrium number of infected also drops, roughly by $33\%$.

 \end{itemize}

 \subsection*{B) Travel restrictions}
 Before we start with details, a short summary: strong travel restriction is the most effective interventions in terms of reducing the first peak, but it elongates the epidemic enormously and causes later oscillations and often a higher second peak. The effectiveness of increasing $\alpha$ depends on the node-degree power-law exponent: when $\tau>3$, its effect is comparable to the strong travel restriction, but when $\tau\in(2,3)$, it is not effective or rather, it does not model travel restrictions well.

 \medskip

 {\bf B(w): Weak travel restriction.}

 \begin{itemize}
 \item[i)] On the middle left picture of Figures \ref{fig:change-graph1} and \ref{fig:change-graph2}, we see the change in the network after increasing $\alpha$ to a value that results in the average degree to drop to $2.6$: for $G_1$ (when $\tau=2.5$, Figure \ref{fig:change-graph1}), this intervention only effect node-pairs with $R_{u,v}>1$ and most long-range edges remain present, while many short and medium-scale edges are deleted. For $\tau<3$, increasing $\alpha$ is not a good model for the no-travel rule. However, for $\tau>3$ ($\tau=3.3$ on Figure \ref{fig:change-graph2}), most node-pairs have $R_{u,v}<1$ and thus increasing $\alpha$ does have a similar effect than cutting long-range edges (compare to middle right picture).   Observe that both pictures have the same average degree $2.6$, yet the middle left picture on Figure \ref{fig:change-graph2}) looks more sparse: this is because edges are much more localized in space.
 \item[ii)]  On the middle left pictures of Figures \ref{fig:G1-vis} and Figures \ref{fig:G2-vis}, the visualization of the first $19$ days of the epidemic are shown under the weak no-travel rule: for $G_1$ (Figures \ref{fig:G1-vis}), the remaining edges are mostly long, and thus the picture looks qualitatively similar to the top left, without intervention. However, for $G_2$  (Figure \ref{fig:G2-vis}), the intervention is very effective, and the epidemic spreads much slower. Within the first 19 days, it only reaches a small fraction of the network localized in various patches.

 \item[iii] On Figure \ref{fig:survival-interventions}, the long-survival probabilities of epidemics are shown as a function of $\eta$.
 The weak no-travel rule, increasing $\alpha$ (last row in the legend) has its sharp threshold $\eta_t=0.003$ for $G_1$ and $0.007$ for $G_2$, in this respect it is also among the less effective measures for both values of $\tau$.

 \item[iv)]Table \ref{table:heights} shows the height and location of the first and second peak under the weak no-travel rule (column Bw), while Figures \ref{fig:first-second-peak-g1}, \ref{fig:first-second-peak-g2-v2} plots these heights as a function of $\eta$.

 We see again that the location and height of the first peak is insensitive to the change in $\eta$, it is $\approx 50*10^3$ on day $10\pm1$ for $G_1$, while $\approx 33*10^3$ on day $29\pm2$ on $G_2$. We conclude that increasing $\alpha$ only model travel restrictions well when $\tau>3$. In this case, its effective, the only more effective measure is the strong travel restriction. On $G_2$, the height of the first peak is $44\%$  of the original peak, while on $G_1$, its $60\%$.

 We conclude that this intervention is very effective in terms of reducing the first peak on $G_2$.

 On $G_1^\alpha$, there is no second peak, (just like on the original $G_1$), while the second peak on $G_2^\alpha$ is about $80\%$ of the height of the second peak on $G_2$, though it appears later.

 \item[v)]
 On Figures \ref{fig:girg-interventions-firstpeak}, \ref{fig:girg1-interventions}, \ref{fig:girg2-interventions}  and \ref{fig:girg-interventions-eta13}, we see the comparison of the epidemic curves, On $G_1^\alpha$, there is no second peak, just like on the original $G_1$.

 \end{itemize}
 \medskip
 We mention that increasing $\alpha$ already has an effect on the epidemic curve when the average degree is kept constant, see Figure \ref{fig:girg-cmtau33-1} that shows the effect of increasing $\alpha$ on the epidemic curve: for the configuration model ($\tau=3.3$) and for GIRG with $\tau=3.3$ and $\alpha=1.3$,  herd immunity is reached first when $80000$ nodes are infected, while in the second scenario, $\alpha=2.3$, the first peak only has magnitude $50000$ nodes, an almost $40\%$ decrease.

 For short average immunity length, (large $\eta$), later peaks can be seen at the bottom picture of Figure \ref{fig:girg-cmtau33-1}, where we observe that the intervention is effective in the sense that the stationary proportion of infected nodes is lower (roughly, $15\%$ decrease: $4.1k$ vs $4.8k$ nodes). However, the amplitude of the oscillations do increase when $\alpha=2.3$ (vs $\alpha=1.3$) and thus the second peak is higher by $\approx15\%$ ($7.3k$ vs $6.3k$). This is in accordance with intuition: flattening the curve results in an elongated first peak of the epidemic and a higher second peak.
 \medskip

 {\bf B(s): Strong travel restriction}
 \begin{itemize}
 \item[i)] On the middle right picture of Figures \ref{fig:change-graph1} and \ref{fig:change-graph2}, we see the change in the network after applying the no-travel rule with a parameter $L$ so that the average degree drops by $40\%$: for $G_1$  (when $\tau=2.5$, Figure \ref{fig:change-graph1}), already  $L=7$ resulted in a $40\%$ drop, while for $G_2$, $L=5$ was needed. (when $\tau=2.5$, Figure \ref{fig:change-graph2}).
 Again, observe that both pictures have the same average degree $2.6$, yet the middle left picture on Figure \ref{fig:change-graph2}) looks more sparse: this is because edges are much more localized in space.

 \item[ii)] On the middle right pictures of Figures \ref{fig:G1-vis} and Figures \ref{fig:G2-vis}, the visualization of the first $19$ days of the epidemic are shown under the hard no-travel rule: for both networks,  the effect of cutting long-range edges is drastic: during the course of the first $19$ days, the epidemic barely manages to leave a small area near its source (for $G_2$) and only infects a smaller part of the network (for $G_1$).
 \item[iii)] On Figure \ref{fig:survival-interventions}, the long-survival probabilities of epidemics are shown as a function of $\eta$. The hard no-travel rule, (denoted by ``long'' in the legend) has its sharp threshold $\eta_t=0.005$ for $G_1$ and $0.01$ for $G_2$, in this respect it is one of the most effective measures in containing the epidemic: even with average immunity length  $111$ days, we only observe a single peak for half of the runs.

 \item[iv)]
 Table \ref{table:heights} shows the height and location of the first and second peak under the weak no-travel rule (column B),
 while Figures \ref{fig:first-second-peak-g1}, \ref{fig:first-second-peak-g2-v2} plots these heights as a function of $\eta$.
 We see again that the location and height of the first peak is insensitive to the change in $\eta$, it is $\approx 39*10^3$ on day $25\pm1$ for $G_1^{\text{long}}$, while $\approx 13*10^3$ on day $80\pm2$ on $G_2^{\text{long}}$, and so the height of the first peak is  only $45\%$  of the original peak, while on $G_2^{\text{long}}$, its only $15\%$.

 However, the first peak appears much later, by a factor $3$ later on $G_1$, and a factor $\approx 5$ later on $G_2$.

 We conclude that this intervention is the most effective in terms of reducing the first peak on $G_2$, but it elongates the epidemic.

 This intervention does cause a second peak to appear also $G_1$.
 The height of second peak  increases only slightly with $\eta$, and thus, for low values of $\eta$, the second peak is \emph{higher} on $G_2^{\text{long}}$ than on $G_2$, while for high values of $\eta$ it is lower. On $G_1^\alpha$, the height of the second peak is above the equilibrium proportion on $G_1$ for lower values of $\eta$, but not for higher values.

  So, the intervention causes the second peak to be rather insensitive to the value of $\eta$. It would be interesting to investigate this further.

 \item[v)] On Figures \ref{fig:girg-interventions-firstpeak}, \ref{fig:girg1-interventions}, \ref{fig:girg2-interventions}  and \ref{fig:girg-interventions-eta13}, we see the comparison of the epidemic curves:
 this intervention is the most effective on both $G_1$ and $G_2$, the peak is much lower and appears later.

 Thus, the price for a (much) lower peak in case of the strong travel restriction is a very elongated epidemic curve.

 \end{itemize}

 \subsection*{C) Limiting the maximal number of contacts}
 Before we start with details, a short summary: even though  this intervention has comparable effects to social distancing in reducing the first peak, it is causing very high second peak. We gave a plausible explanation for this: removing hubs removes the synchronization from the system, causing high oscillations and a long time to reach equilibrium.

 \begin{itemize}
 \item[i)] On the bottom picture of Figures \ref{fig:change-graph1} and \ref{fig:change-graph2}, we see the change in the network after applying the rule of limiting the maximal node-degree. The maximal degree  $M$  is chosen so that the average degree drops by $40\%$: on the bottom  picture of Figures \ref{fig:change-graph1}, $M=5$, while  $M=4$ on the bottom  picture of Figures \ref{fig:change-graph1}.
 Observe that limiting the maximal degree, but removing edges randomly to achieve that, results in a network where there are a still many long-range edges, but \emph{no hubs} remain. Due to the presence of the remaining long edges, the network looks more densely connected compared to the hard and weak no-travel rules.

 \item[ii)] On the bottom pictures of Figures \ref{fig:G1-vis} and Figures \ref{fig:G2-vis}, the visualization of the first $19$ days of the epidemic are shown when the maximal node-degree is limited: for both networks,  the effect  is apparent, but more drastic on $G_2$, when $\tau>3$ During the course of the first $19$ days, the epidemic grows in a scattered way, mostly using the remaining long edges. The effect is somewhere between the weak and the hard no-travel rule in terms of its capability to restrain the spread.

 \item[iii)] On Figure \ref{fig:survival-interventions}, the long-survival probabilities of epidemics are shown as a function of $\eta$. Limiting the maximal degree, meaning no hubs,  (denoted by ``hub'' in the legend) has its sharp threshold $\eta_t=0.008$ for $G_1$ and $0.01$ for $G_2$, so in terms of long survival it is  the most effective measure in containing the epidemic (together with the strong travel restriction for $G_2$).

 \item[iv)]
 Table \ref{table:heights} shows the height and location of the first and second peak under the weak no-travel rule (column B), while Figures \ref{fig:first-second-peak-g1}, \ref{fig:first-second-peak-g2} plots these heights as a function of $\eta$.
 We see again that the location and height of the first peak is insensitive to the change in $\eta$, it is $\approx 49*10^3$ on day $25\pm1$ for $G_1^{\text{hub}}$, while $\approx 37*10^3$ on day $41\pm2$ on $G_2^{\text{hub}}$. Interestingly, this intervention has roughly the same effect as social distancing in terms of reducing the first peak, especially on $G_1$, the difference is a few percentages only:
 On $G_1^{\text{hub}}$, the height of the first peak is  roughly $60\%$  of the original peak, while on $G_2^{\text{hub}}$, its only $50\%$.

 This intervention, unlike social distancing, causes a second peak to appear on $G_1^{\text{hub}}$, that again increases very slowly with $\eta$ and is roughly $1/5$ of the height of the first peak on $G_1^{\text{hub}}$.
 The height of second peak on $G_2^{\text{hub}}$ is surprisingly high: roughly a factor $2$ higher than that on $G_2$. Its height, compared to the first peak on $G_2^{\text{hub}}$, is also significant: as big as $1/3$ of the first peak.

 So, the intervention causes the second peak to be rather insensitive to the value of $\eta$.
  It would be interesting to investigate this asynchronization, the surprisingly high second peak, and its insensitivity to $\eta$ further.

 \item[v)] On Figures \ref{fig:girg-interventions-firstpeak}, \ref{fig:girg1-interventions}, \ref{fig:girg2-interventions}  and \ref{fig:girg-interventions-eta13}, we see the comparison of the epidemic curves: also further peak are quite high, and the system takes a very long time to equilibriate.

 Thus, the \emph{price for a lower first peak is a very high second peak}, followed by quite high later peaks and a long time for the system to equilibriate.
 \end{itemize}

 \subsubsection*{Monotonicity of survival probabilities}
 On Figures \ref{fig:survival} and \ref{fig:survival-interventions}, some curves do not appear to be monotone increasing, but rather, after the sharp increase they seemingly develop a small peak of survival probabilities. We believe that this is only due to the increased fluctuations around the critical value $\eta_t$. For one model, namely, $G_2$ under weak no-travel rule (which is an instance of a GIRG with $(\tau, \alpha)=(3.3,1.9)$), we have tested this on a larger scale simulation. We have run $500$ runs for each value of $\eta$, and plotted the survival probabilities as a function of $\eta$ on Figure \ref{fig:survival-500}.

 On this larger scale, the curve appears to be monotone, and we believe that this is indeed the case.

 \subsection{S-I-T-S on lattice network models.}\label{s:lattice}
 Epidemic models on lattice networks have been under investigation for decades, see e.g.\ the book \cite{Ren91}, or the survey \cite{Riley07}.
 There are close relations to percolation cluster growth models  see e.g. \cite{CarGRa85, Gras83}, and stochastic lattice gas models \cite{SouTom10}. In the more applied sciences, similar epidemic models have been developed under the name agent based modeling, see e.g. \cite{PerDra09} and references therein, or the survey \cite{Riley07}.
 The  paper \cite{SouTom10} conducts a simulation study of a similar model\footnote{The parameters there can be translated as follows: $a=\eta$, $b/4=\beta$, $c=\gamma$. The only difference between the S-I-R-S model studied in \cite{SouTom10} and the S-I-T-S model we study is in the infection probability of a node $u$ when it  has more than one infected neighbor: there, the probability of infection is additive, here, we draw the infections from each infected neighbor independently with probability $\beta$.} to the S-I-T-S model studied here, but focusing on the critical regime (and critical exponents) rather than the supercritical regime investigated here.

 Arguable, due to their homogeneous nature, lattice models do not represent realistic epidemics well.
 Nevertheless, they provide an interesting comparison, especially, when we apply homogenization methods to GIRG, such as the strong no travel rule or limiting the maximal node degree.
 The S-I-T-S on lattice networks has again subcritical, critical and supercritical phases.
 Here we briefly summarize our findings for supercritical S-I-T-S on lattice networks.
 Within the supercritical  phase, we distinguish the two sub-phases 2S and 2M as before, see Figure \ref{fig:grid-epidemic}. Between these phases, there is a sharp transition between the two phases.

 However, the shape of the peak is different from the mean-field configuration model and that of GIRGs: the number of infected grows \emph{linearly}, not exponentially, followed by a linear decline and extinction. The linear growth corresponds to the linear circumference of graph distance balls in two dimensions. The epidemic proceeds like a wave. The number of nodes ever infected grows quadratically at the beginning, corresponding to quadratic volume growth in the model.

\section{Conclusion and outlook}
We have studied the first and second peak in an epidemic spreading model with temporary immunity on three different underlying networks. These results are compared to the classical compartmental model that is based on ordinary differential equations (ODE). In the networks a supercritical epidemic either survives, or it dies out after a single peak. The latter case is not possible in ODE-based models. Moreover, as the variability of the node degrees in the underlying network increases, the amplitude of the oscillations after the first peak become smaller for supercritical epidemics.

The geometric inhomogeneous random graph is the network that matches reality best, since the nodes are embedded in a geometric space, and its node-degree distribution has a heavy tail. These networks allow for intuitive modeling of several intervention methods: social distancing, travel restrictions, and meeting a limited number of people. Classical compartmental models do not offer this intuitive modeling. We compare the effect of these interventions by `permanently' modifying the underlying contact network, while maintaining the same average node-degree across interventions. We find that the strong travel restrictions are most effective in elongating and diminishing the first peak. Travel restrictions and meeting a limited number of people result in a higher second peak, where the latter restriction yields the highest second peak.

To make the model more realistic, several adjustments can be made: other compartments could be added, such as an exposed or quarantined state. Moreover, one could study a scenario where interventions are only applied once certain thresholds in the number of infected nodes are exceeded. Lastly, we mention that our framework could also be used to investigate the effect of geometry for other intervention methods, such as a mobile app.

	\clearpage
	\newpage
  \begin{appendices}
  \clearpage
  	\newpage
  \section{The underlying network models.}
  \begin{figure}[ht]\begin{center}
  		\includegraphics[width=0.45\columnwidth]{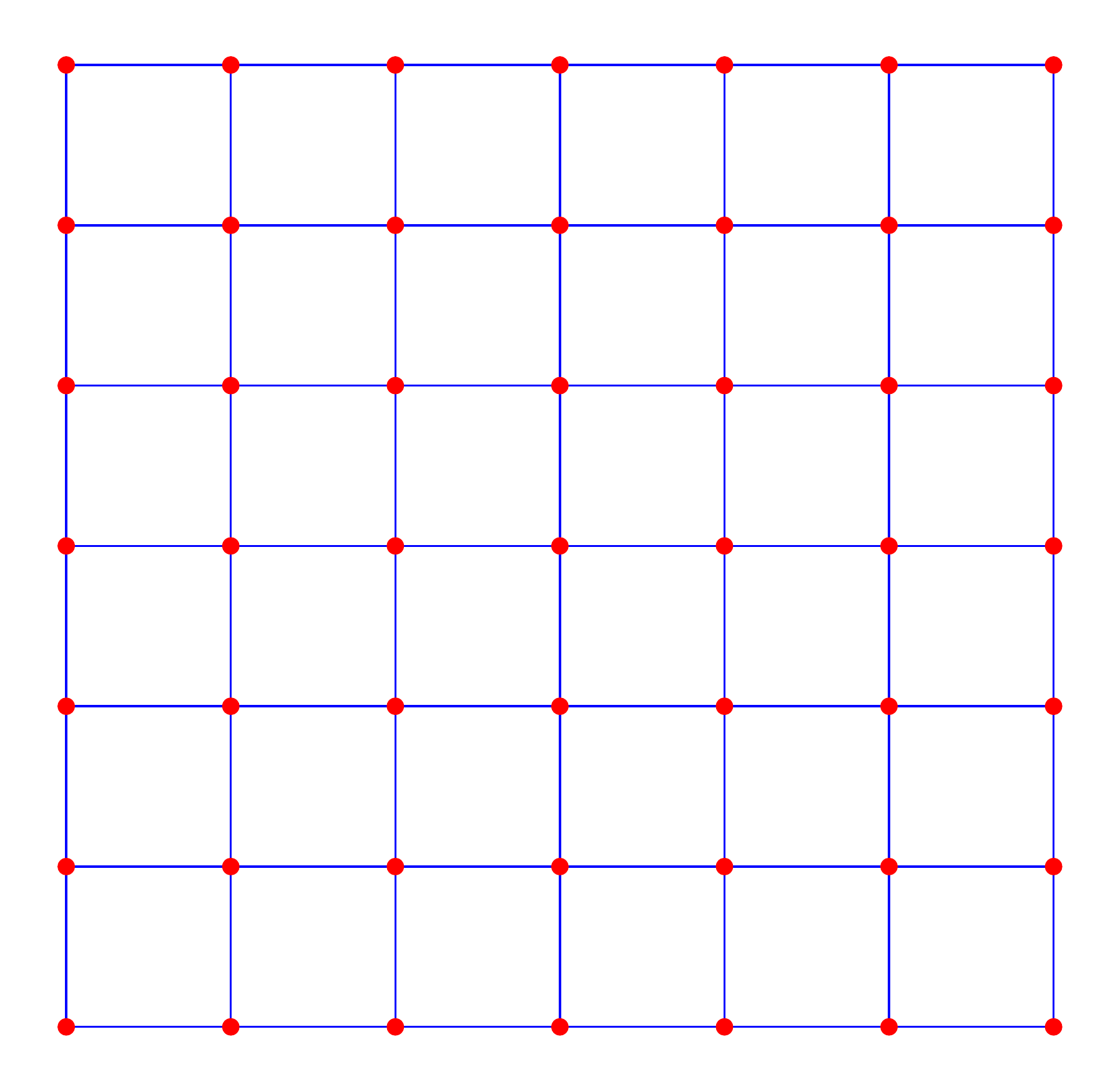}
  		\includegraphics[width=0.45\columnwidth]{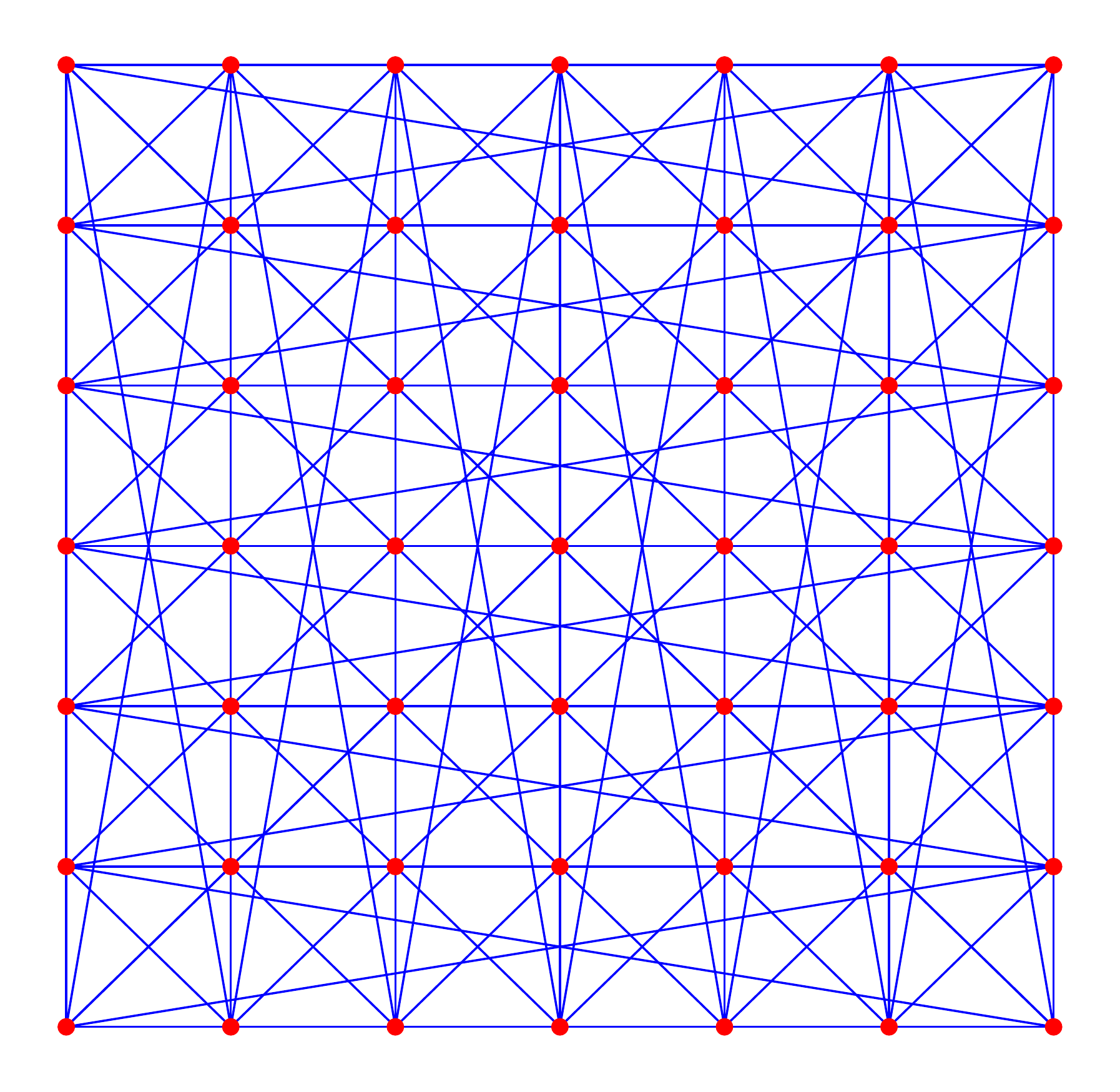}
  		\caption{\emph{Left:} The two dimensional torus $\Z_7^2$ on $N=49$ nodes. Each node has four neighbors. \emph{Right:} The modified two dimensional torus $\wit \Z_7^2$ on $N=49$ nodes. Each node has $8$ neighbors. The nodes in the bottom row are also connected to the nodes in the top row, and the nodes in the very left column are also connected to the nodes in the very right column.}
  \label{fig:grid-graph}
  \end{center}\end{figure}
  \begin{figure}[b]\begin{center}
  		\includegraphics[width=0.3\columnwidth]{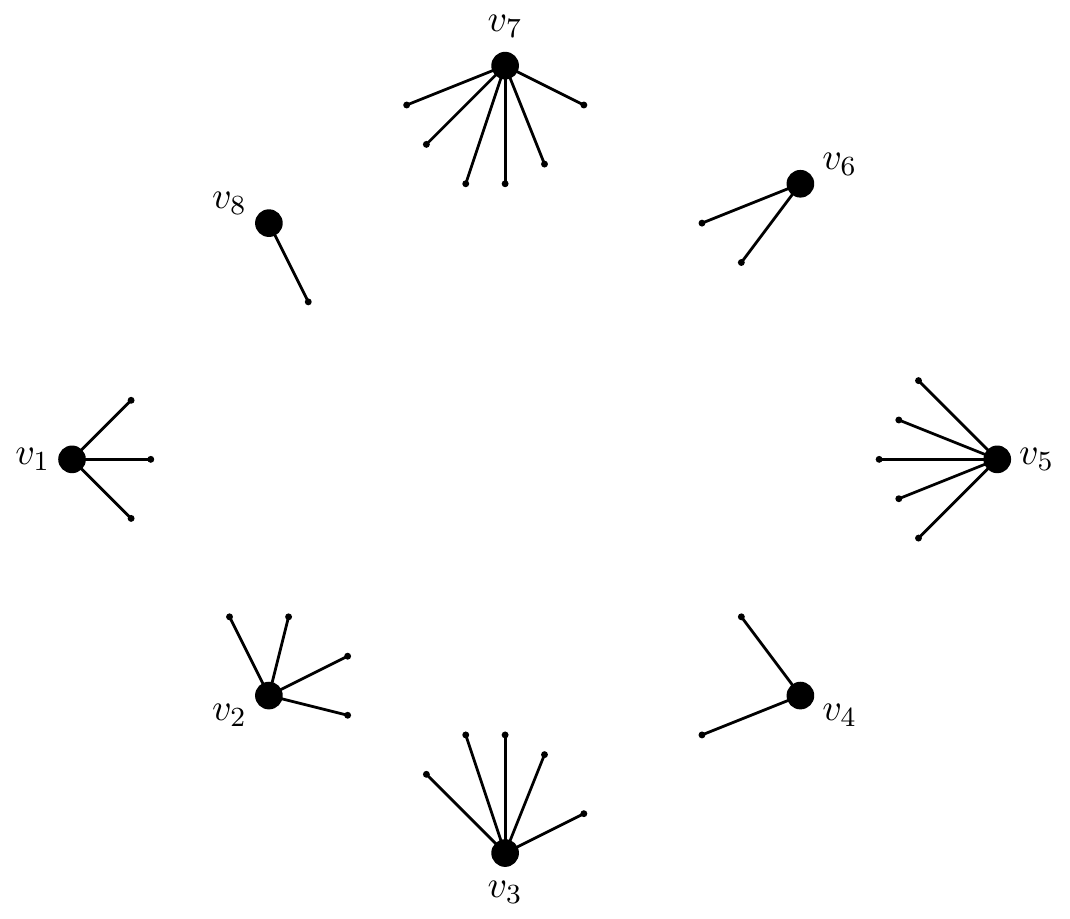}
  		\includegraphics[width=0.3\columnwidth]{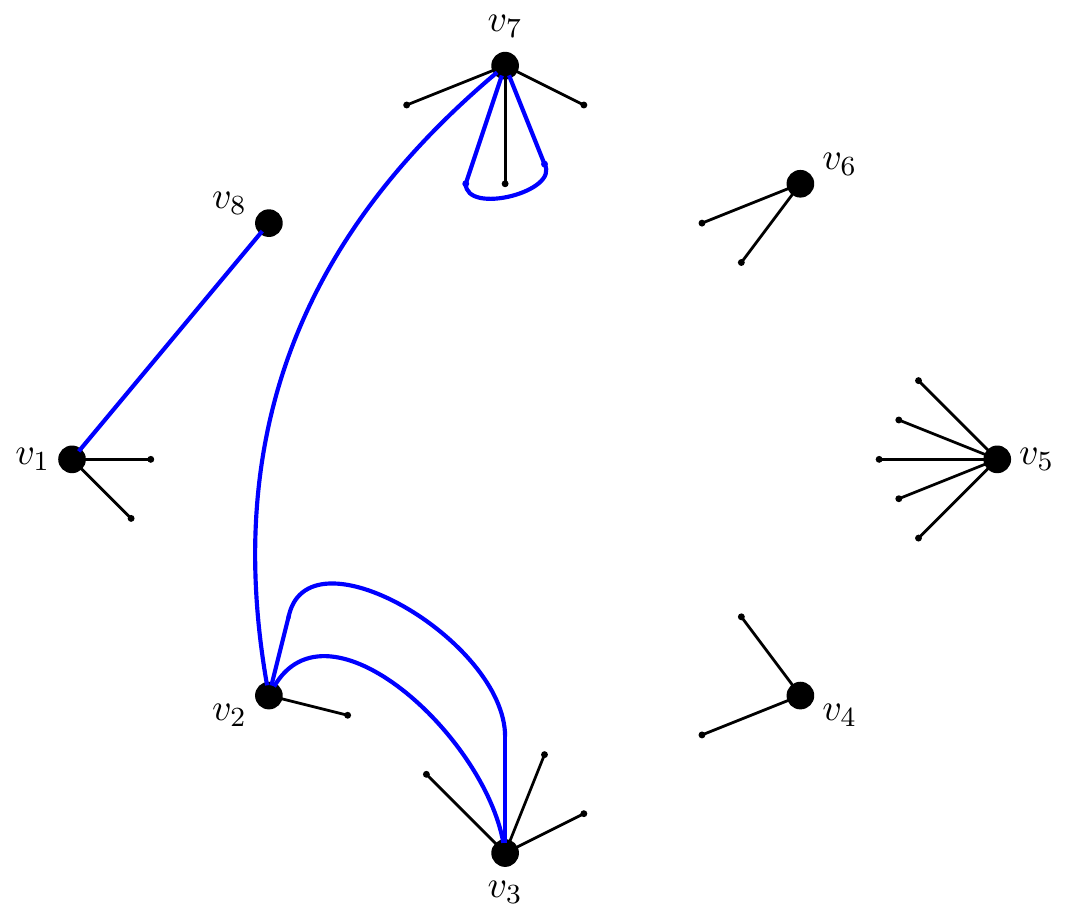}
  		\includegraphics[width=0.3\columnwidth]{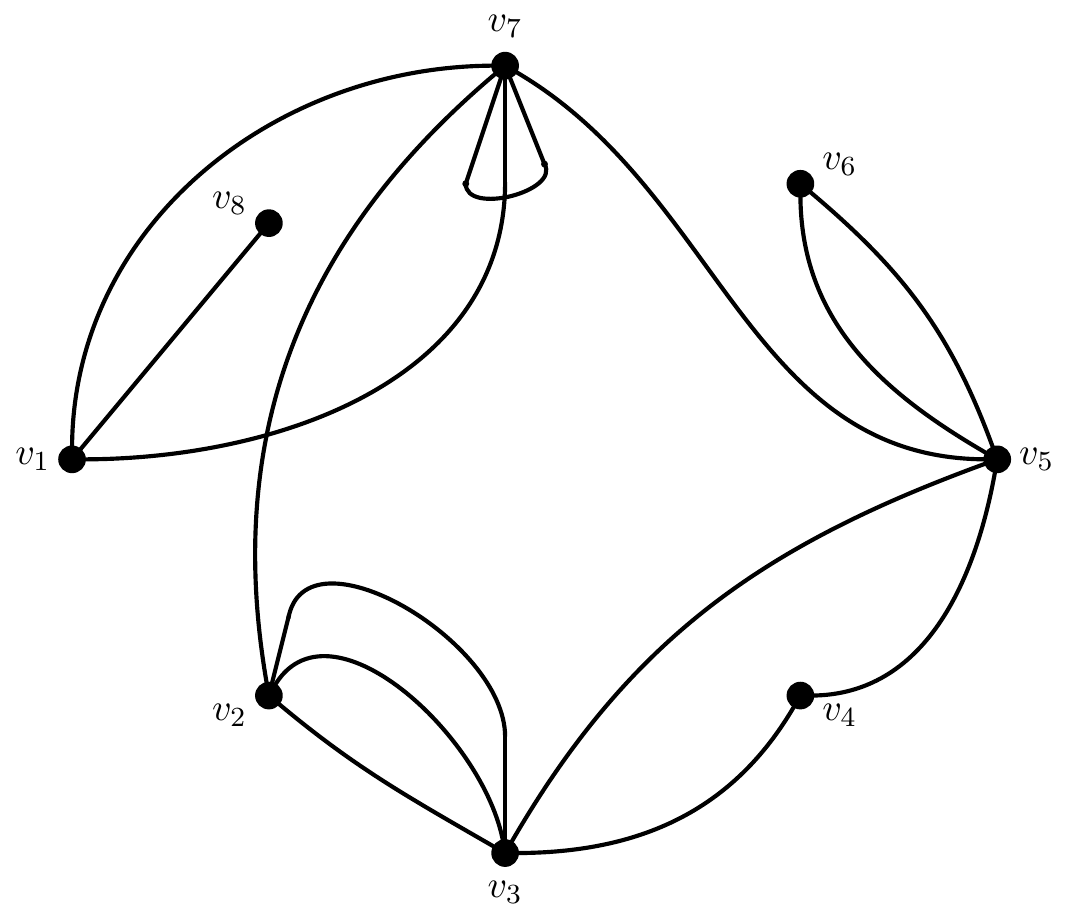}
  		\caption{{\small The algorithm producing the configuration model. \emph{Left: }First, for each node in the network we prescribe its degree and draw it as half-edges. \emph{Middle:} Then, we match half-edges randomly to form edges (connections). This may be done in a sequential way, by always choosing a uniform pair from the remaining half-edges. Here, an intermediate stage is shown when there are five edges formed. \emph{Right:} All half-edges are matched. The output is a random graph called the configuration model.}}
  \label{fig:config}
  \end{center}\end{figure}
  \begin{figure}\begin{center}
  		\includegraphics[width=0.45\columnwidth]{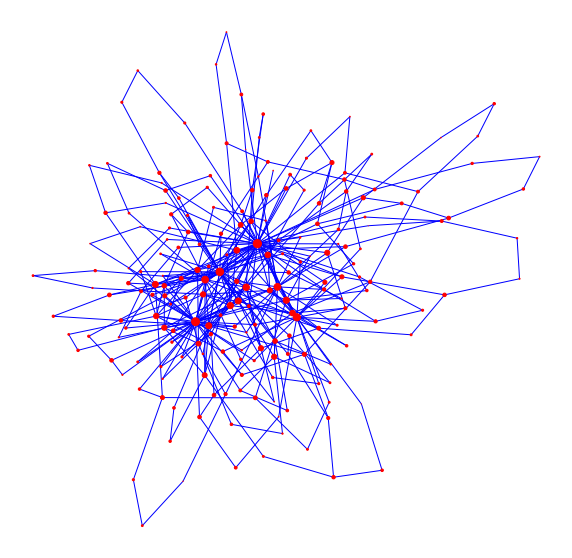}
  		\includegraphics[width=0.45\columnwidth]{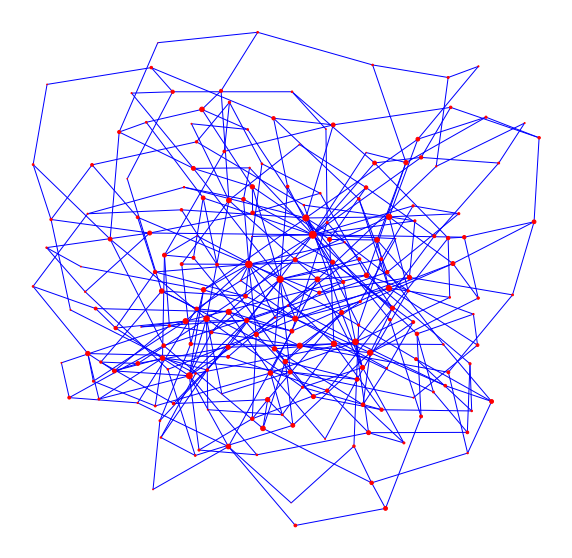}
  		\caption{{\small Two examples of the configuration model. The $N = 200$  nodes have average degree $4$, with power-law degree distribution, with exponent $\tau =2.9$ (\emph{left}) and $\tau=3.3$ (\emph{right}).} }
  \label{fig:config2}
  \end{center}\end{figure}

  \begin{figure}[t]\begin{center}
        \includegraphics[width=0.45\columnwidth]{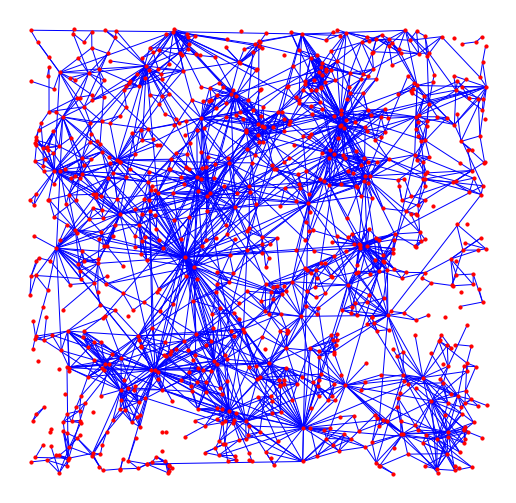}
  		\includegraphics[width=0.45\columnwidth]{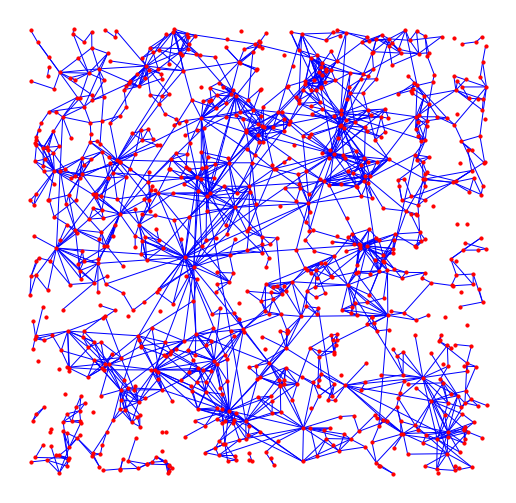}
  		\caption{{\small Two examples of Geometric Inhomogeneous Random Graphs (GIRGs). The $N = 1000$  nodes are placed randomly into a square of area $N$. Each node draws a random fitness from a power law distribution with exponent $\tau = 2.95$ (\emph{left}) and $\tau=3.3$ (\emph{right}).
  		We used the same location for nodes and the same underlying uniform variables to simulate fitnesses in both cases: for a uniform variable $U_v\in[0,1]$, we set the fitness of node $v$ to $W^{(2.95)}_v:=U_v^{-1/1.95}$ on the left, while  $W^{(3.3)}_v:=U_v^{-1/2.3}$ on the right.
  		Each pair of nodes with positions $x_1,x_2$ and weights $w_1,w_2$, respectively, is connected with probability $p^{(\tau)} = 0.5 (1\wedge 0.2 (w_1^{(\tau)} w_2^{(\tau)} |x_1-x_2|^{-d})^\alpha)$, where $\alpha = 2.5$. Connections are again generated in a coupled way, using the same set of uniform variables for the two pictures, thresholded at $p^{(2.95)}$ and $p^{(3.3)}$, respectively.}}

  		\label{fig:GIRG}
  \end{center}\end{figure}

  \clearpage
  \newpage

  \section{The effect of geometry and interventions}
  \begin{figure}[ht]\begin{center}
  		\includegraphics[width=\textwidth]{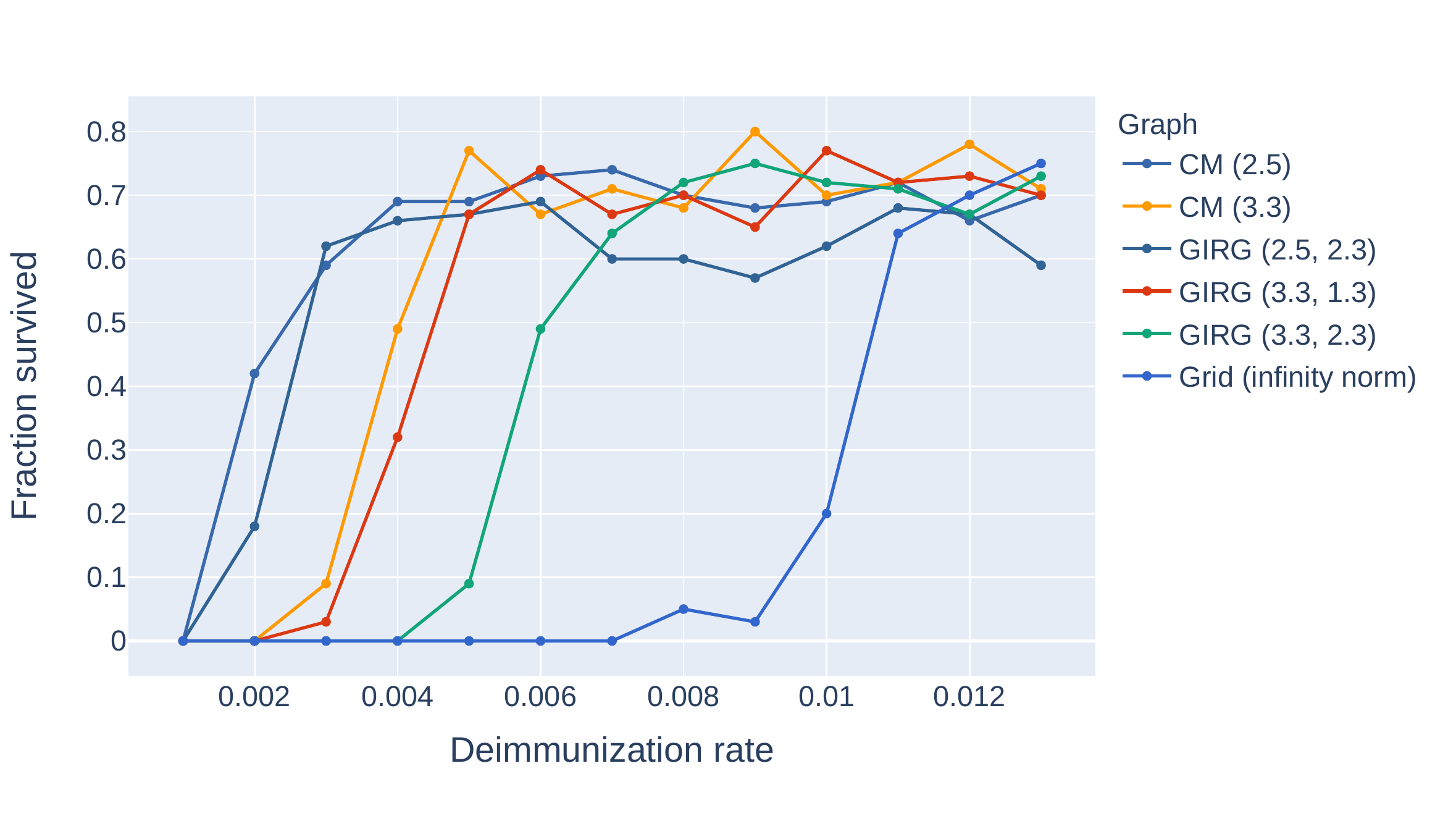}
  		\caption{A comparison of survival probability of S-I-T-S on eight different networks with the same average degree, as a function of the $\eta$, the rate of losing immunisation. Survival probabilities were computed using a $100$ runs for all parameter values ($\eta$) and models.  The infection and healing rates $\beta=0.225,  \gamma=0.2$ for all cases.  CM: the configuration model with $\tau=2.5, 3.3$,  GIRG: GIRG power-law fitnesses with $(\tau, \alpha) = (2.5, 2.3), (3.3,1.3), (3.3, 2.3)$, and Grid: the modified lattice $\widetilde \Z^2$. All networks have the same number of $N=160000$ nodes, and the same average degree $\E[\deg (u)]=8$. For all networks we see a sharp transition at a critical value of $\eta$ where the system moves from phase 2P (single peak followed by extinction) to 2M (survival with multiple peaks) with overwhelming probability.
  				} \label{fig:survival}.
  \end{center}\end{figure}
  \clearpage
  \newpage

  \begin{figure}[b]\begin{center}
  		\includegraphics[width=1\textwidth]{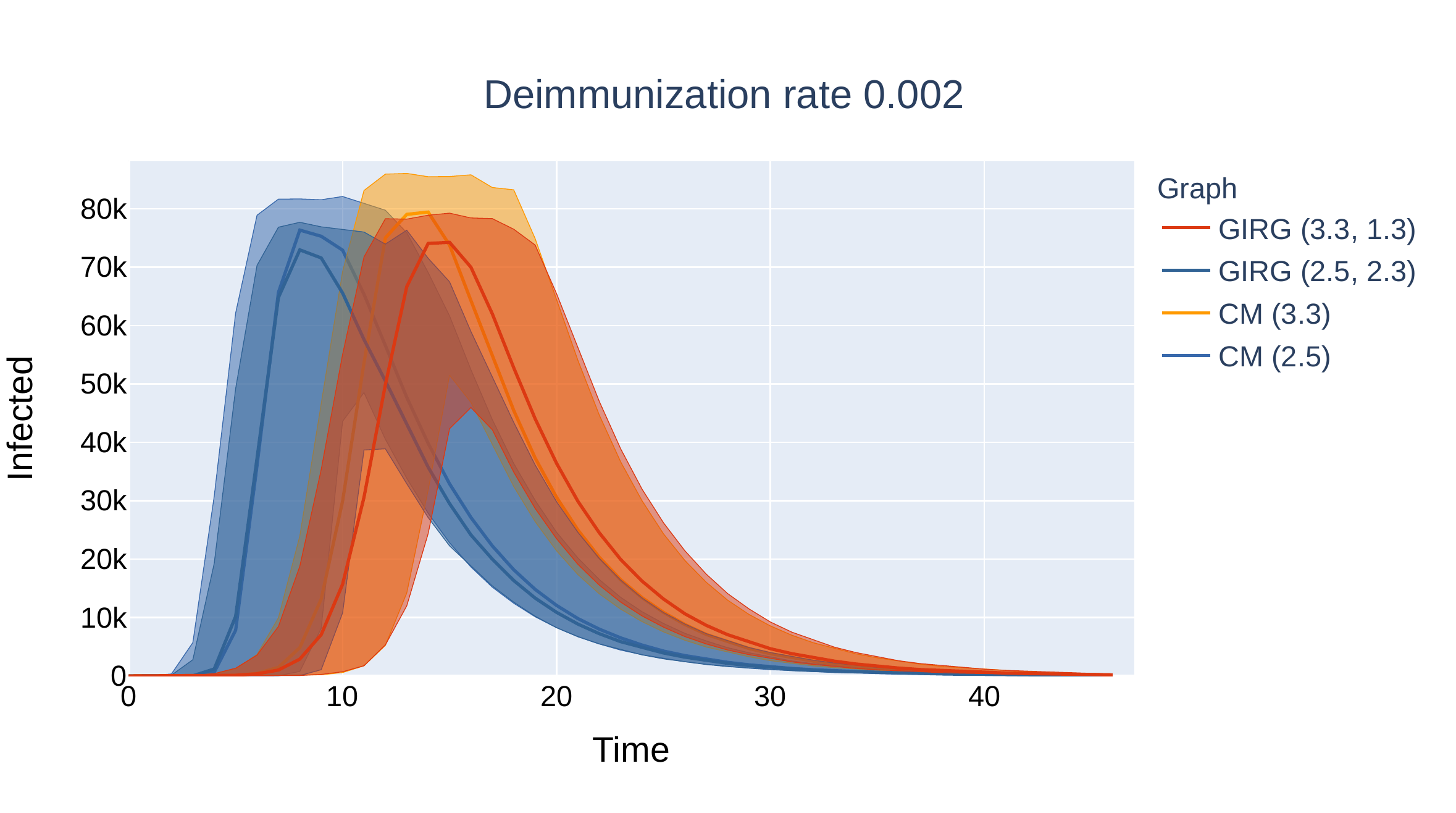}
  		\includegraphics[width=1\textwidth]{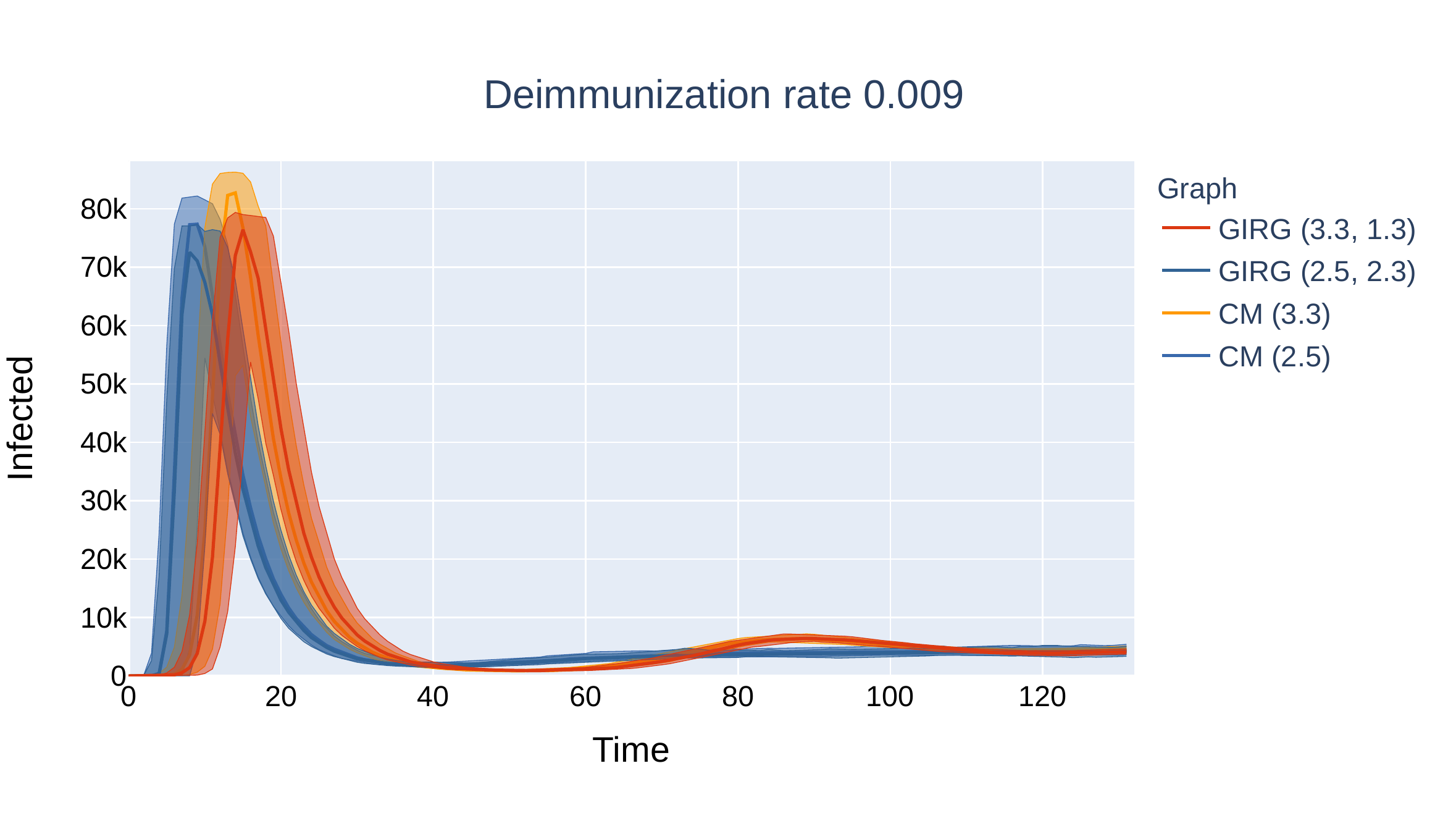}
  		\caption{No intervention. Mean field network vs spatial scale-free networks with many long-range edges.
  	A comparison of the number of infected on the configuration model versus GIRG with matching parameters. The continuous line shows the median of infected individuals, the shaded area is covers $95\%$ of all runs for which the number of infected nodes was positive. Each model has average degree $8$.  Observe that the epidemic curve of configuration model with $\tau=2.5$ (CM(2.5), orange) matches that of GIRG with $(\tau,\alpha)=(2.5, 2.3)$ (GIRG(2.5,2.3), purple), while the epidemic curve of the configuration model with $\tau=3.3$  (CM(3.3), yellow) matches that of that of GIRG with $(\tau,\alpha)=(3.3, 1.3)$ (GIRG(3.3,1.3), blue)  \emph{Top}: $\eta=0.002$ ($500$ days immunity), a single peak can be observed.  \emph{Bottom}: $\eta=0.009$ ($111$ days immunity), many peaks can be observed.
  				} \label{fig:cm-girg-match}.
  \end{center}\end{figure}
  \clearpage
  \newpage
  \begin{figure}[h]\begin{center}
  	\begin{subfigure}{\textwidth}
  		\centering
  		\includegraphics[width=1\textwidth]{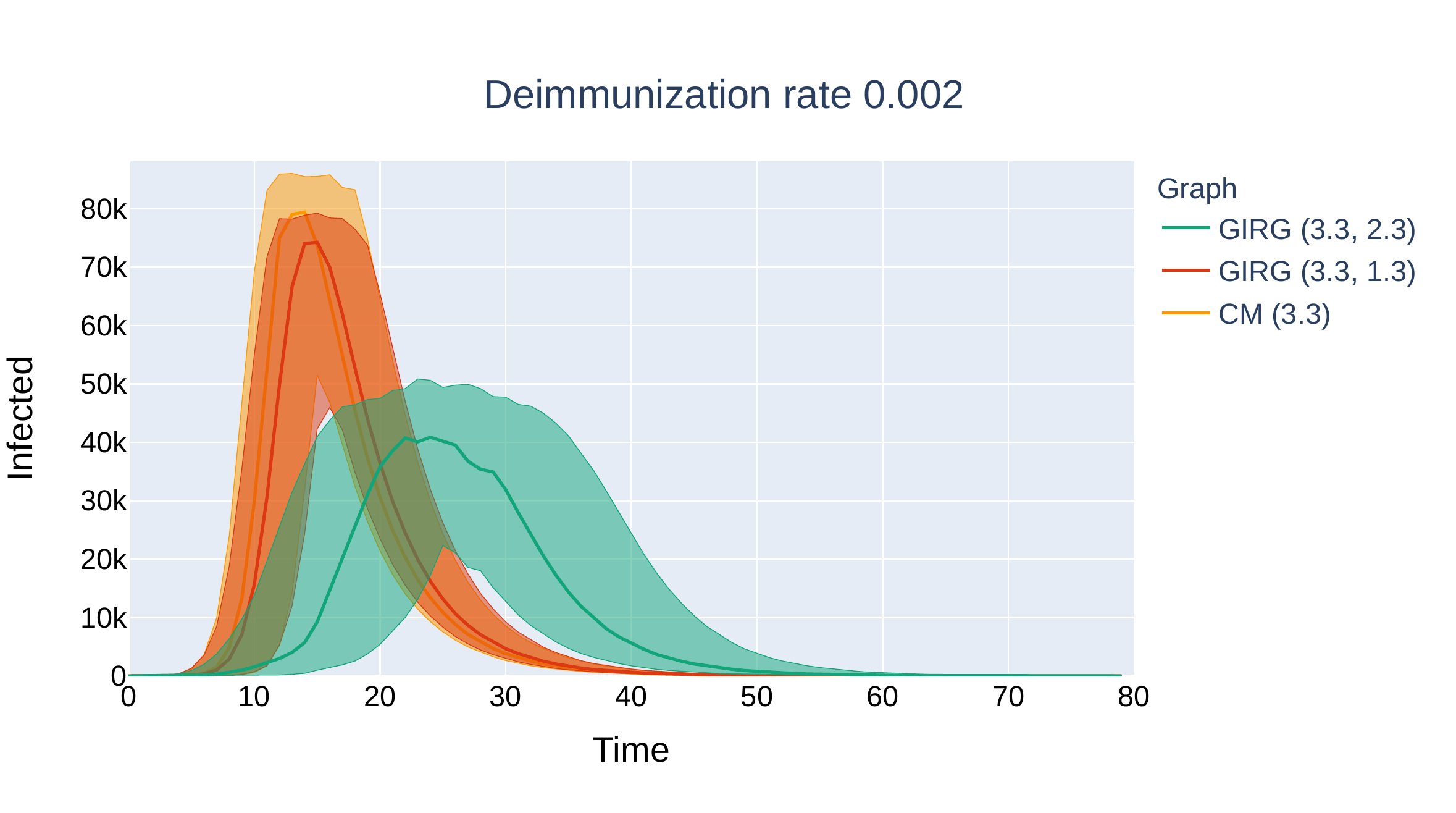}
  		\caption{}
  		\label{fig:girg-cmtau33-1top}
  	\end{subfigure}
  	\begin{subfigure}{\textwidth}
  		\centering
  		\includegraphics[width=1\textwidth]{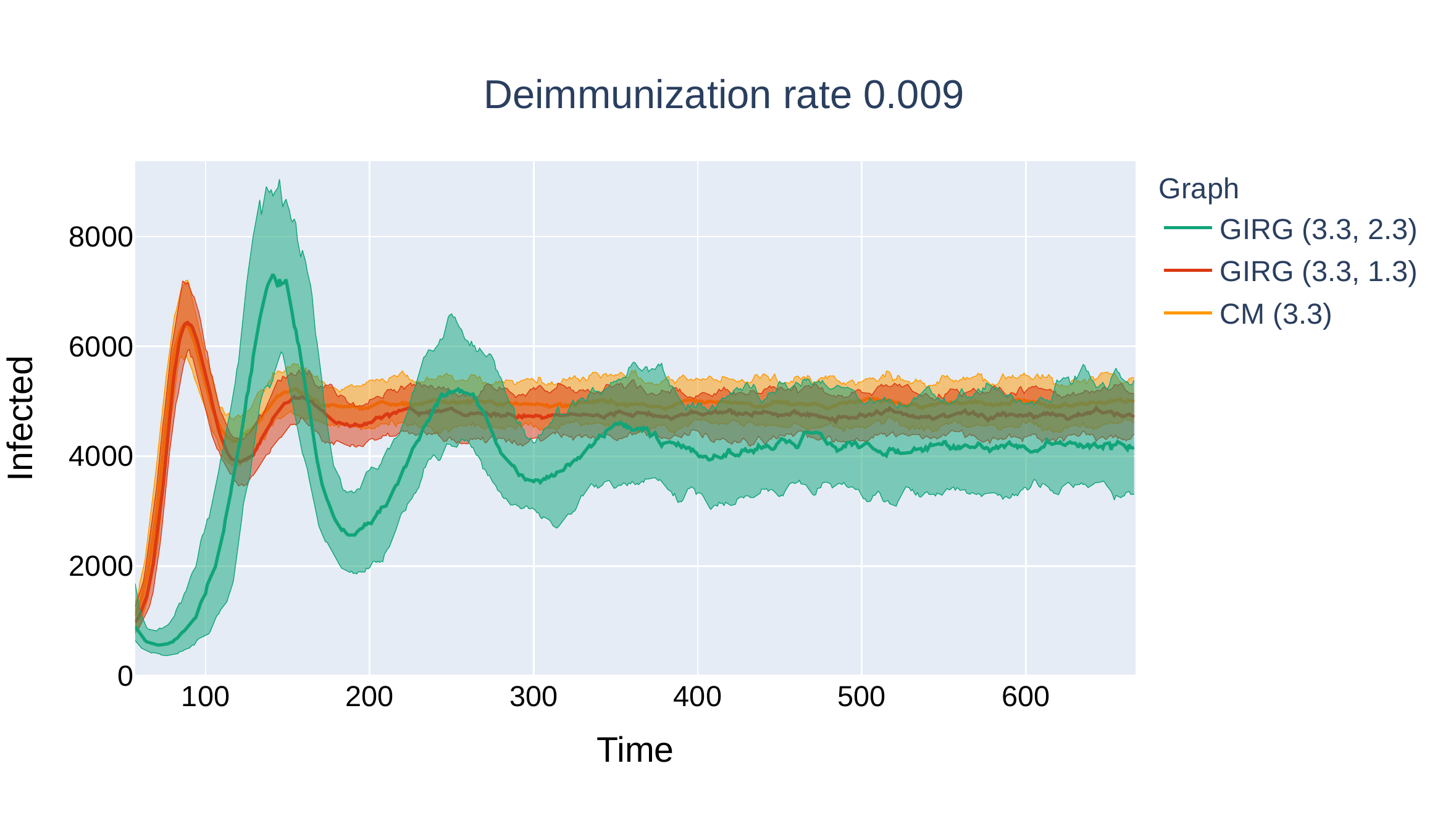}
  		\caption{}
  		\label{fig:girg-cmtau33-1bottom}
  	\end{subfigure}
  		\caption{No intervention: The effect of geometry.  Each model has average degree $8$, and power-law exponent $\tau=3.3$. We see that there is only a match between the mean field network configuration model (CM) when  $\alpha<2$, hence, there are many long-range connections in the network.
  		The epidemic curve of configuration model with $\tau=3.3$ (CM(3.3), orange) matches that of GIRG with $(\tau,\alpha)=(3.3, 1.3)$ (GIRG(3.3,1.3), dark blue). When $\alpha$ is increased to $2.3$ (GIRG(3.3,2.3), light blue). This effects the first peak of epidemic curve to flatten, its magnitude is shrunk by almost $40\%$, even though the average degree is tuned to remain  the same.
  		\emph{Top:} $\eta=0.002$ (500 days immunity) On all three models, a single peak can be observed, and all runs die out within $70$ days.  		\emph{Bottom}: $\eta=0.009$, the first peak has the same height as on to the top picture and is not shown.  Without intervention (GIRG(3.3,1.3)), the second and further peaks of the epidemic behave similarly to the mean-field network (CM(3.3)). With intervention, (GIRG(3.3,2.3)), both the oscillation period as well as its amplitude are higher, but the average stationary proportion of infected nodes is lower.}
  		 \label{fig:girg-cmtau33-1}.
  		\end{center}\end{figure}

  \clearpage
  \newpage
  	\begin{figure}\begin{center}
  			\begin{subfigure}{0.49\textwidth}
  				\centering
  				\includegraphics[width=1\textwidth]{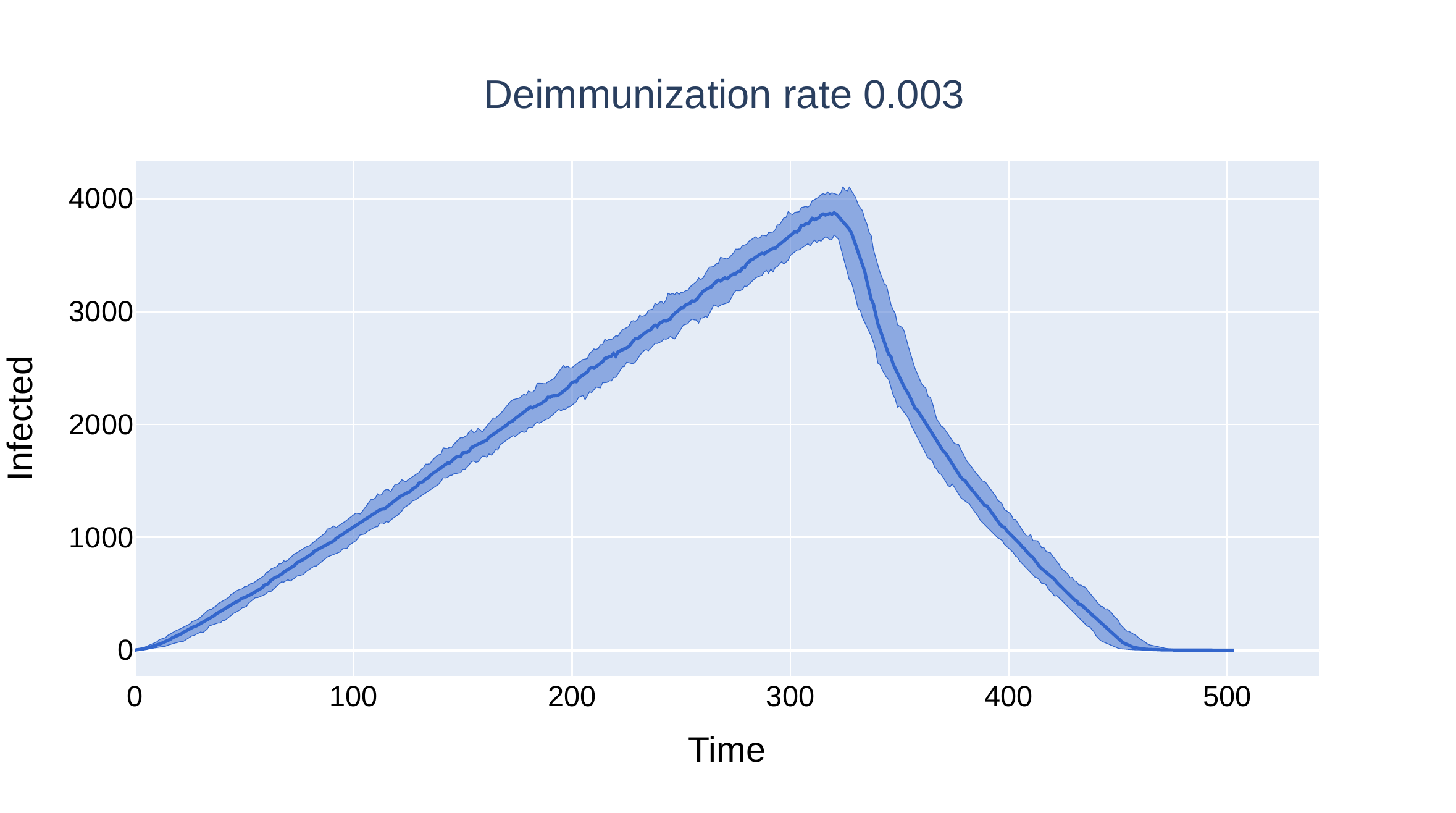}
  				\caption{Extinction after single peak (2S), $\eta=0.003$}
  			\end{subfigure}
  			\begin{subfigure}{0.49\textwidth}
  				\centering
  				\includegraphics[width=1\textwidth]{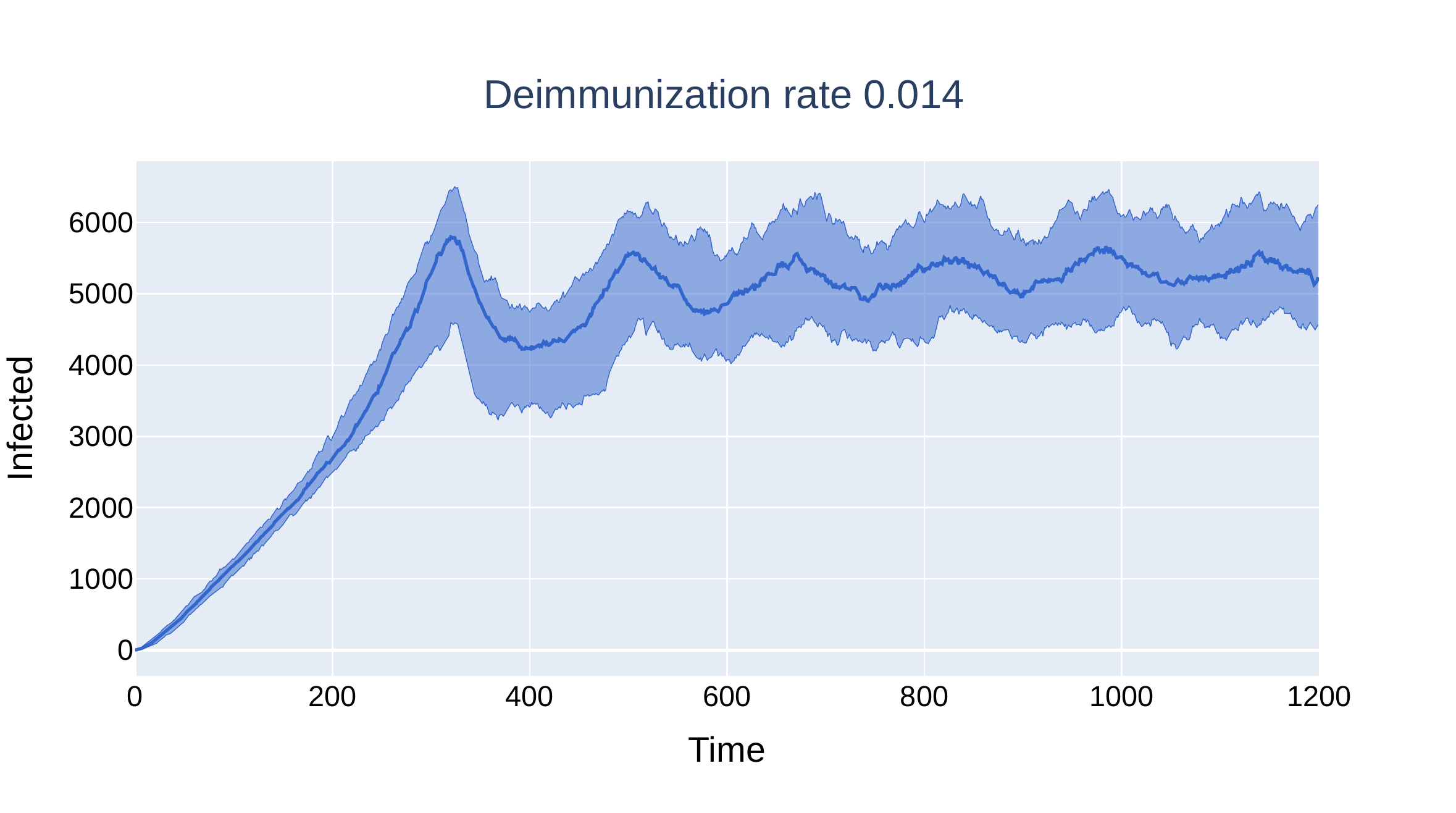}
  				\caption{Survival, many peaks (2M), $\eta=0.014$}
  			\end{subfigure}
  			\caption{A comparison of the number of infected on the torus $\wit Z^2_n$ with $n=400$, in total $N=160000$ nodes, $\gamma=0.2$, $\beta=0.225$ for two values of $\eta$, see in subfigure captions. The continuous curve shows the median, the shaded area $95\%$ of all runs of all runs for which the number of infected nodes was positive.} \label{fig:grid-epidemic}
  		\end{center}
  	\end{figure}

  		\begin{figure}\begin{center}
  			\begin{subfigure}{0.49\textwidth}
  				\centering
  					\includegraphics[width=\textwidth]{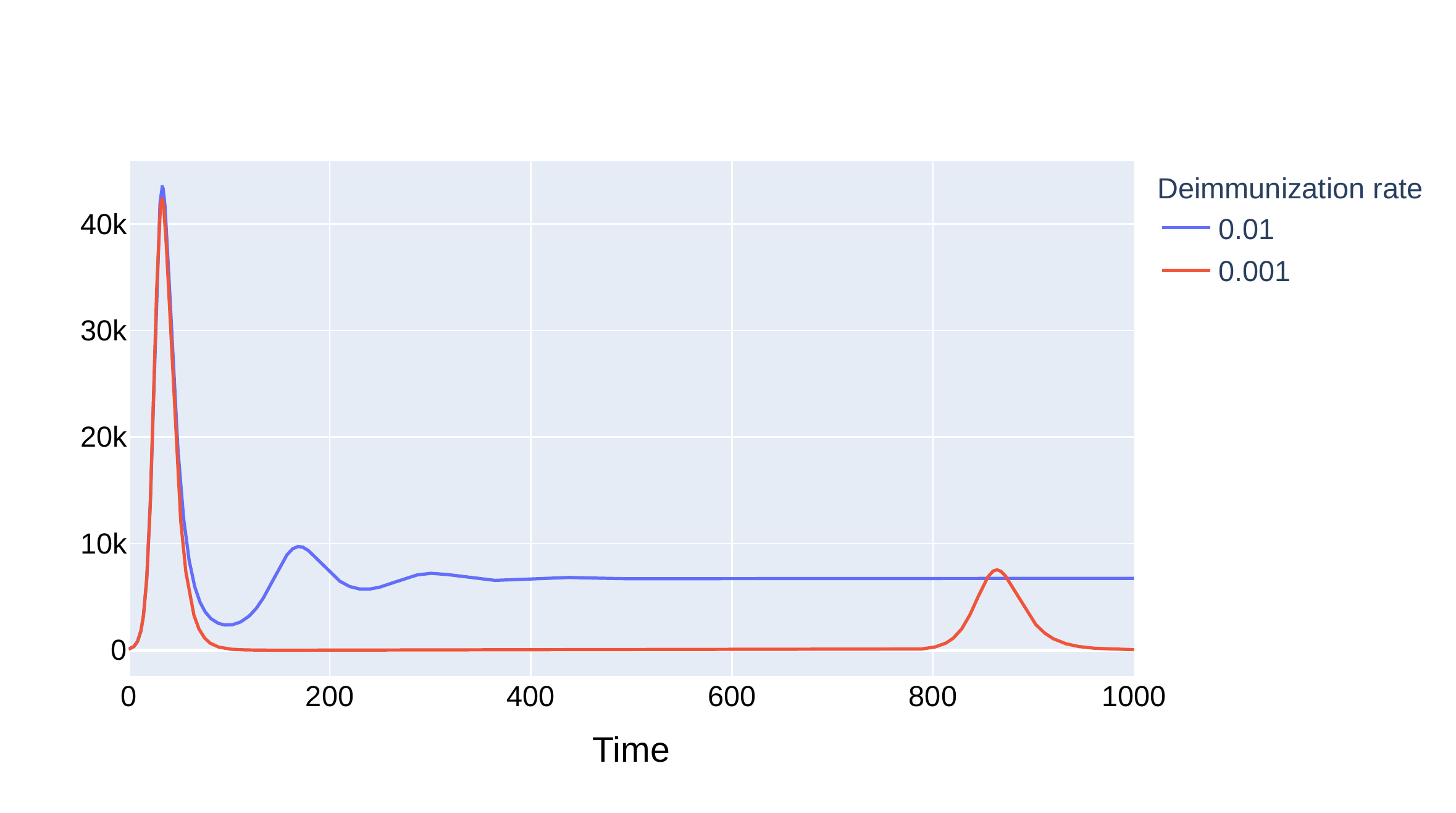}
  				  \caption{Supercritical behavior for $\beta=0.38$.}
  			\label{fig:ode1}
  		\end{subfigure}
  		\begin{subfigure}{0.49\textwidth}
  			\centering
  			\includegraphics[width=\textwidth]{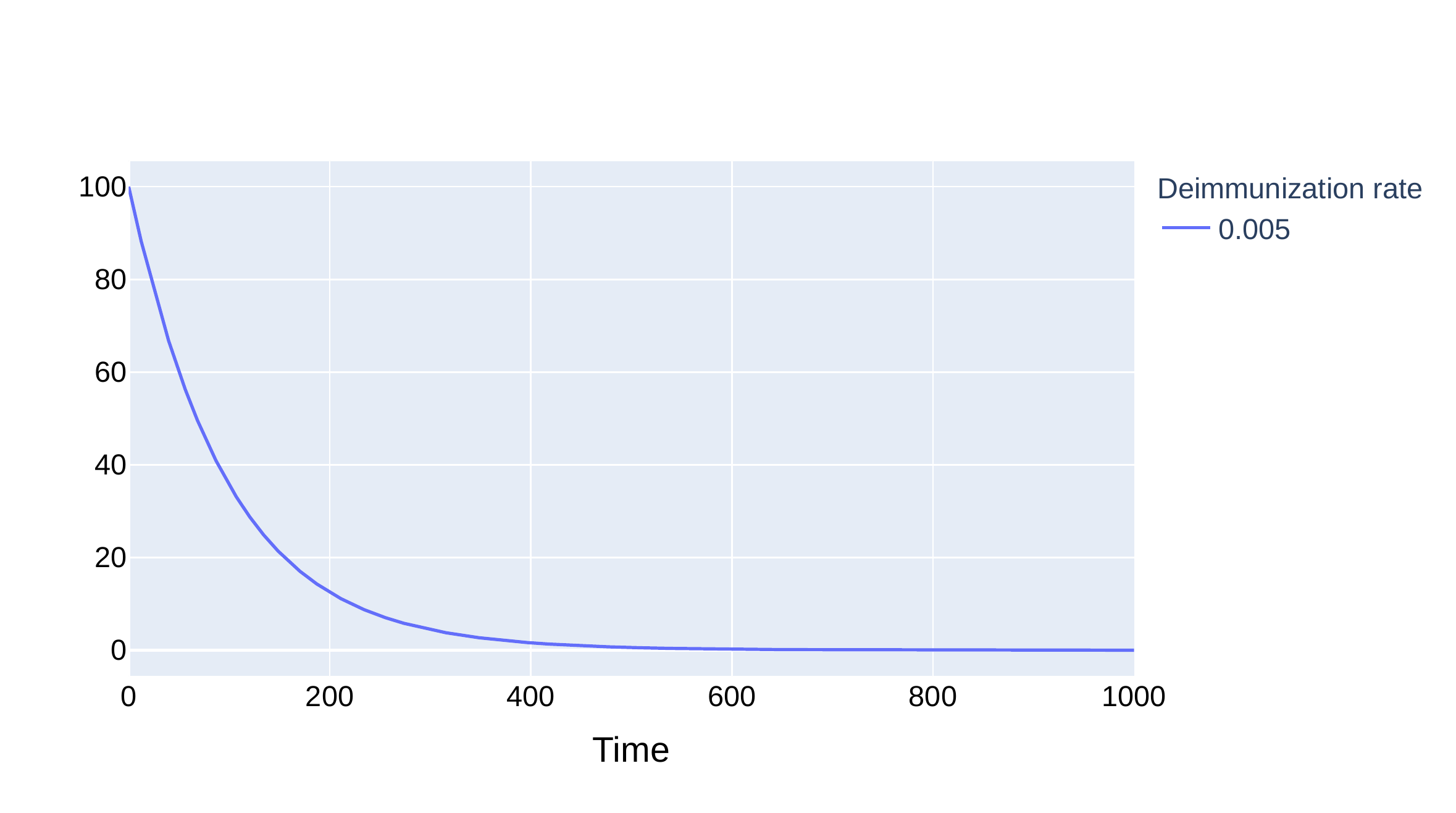}
  			\caption{Subcritical behavior for $\beta=0.13, \eta=0.005$. }
  	\label{fig:ode3}.
  	\end{subfigure}
  		\end{center}
  		\caption{Solutions of the ODE when $N=160000, \gamma=0.14$ with 100 initially infected nodes.
  		The infection almost seemingly disappears from the system before the second peak appears when $\eta=0.001$. The infection does not turn into a pandemic when $\beta<\gamma$.}
  		\end{figure}

  		\end{appendices}
  \clearpage
  \newpage
  \clearpage
  \begin{figure}\begin{center}
  	\begin{subfigure}{0.49\textwidth}
  		\centering
  		\includegraphics[width=0.9\textwidth]{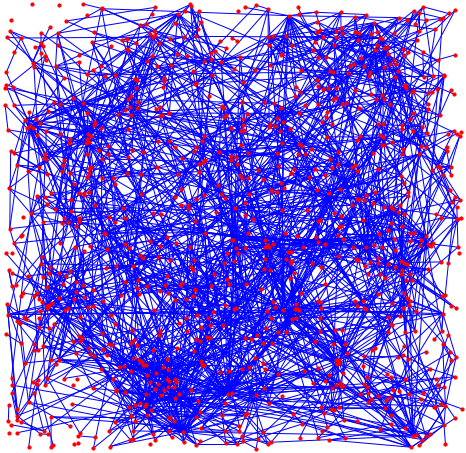}
  		\caption{No intervention.}
  	\end{subfigure}
  	\begin{subfigure}{0.49\textwidth}
  		\centering
  		\includegraphics[width=0.9\textwidth]{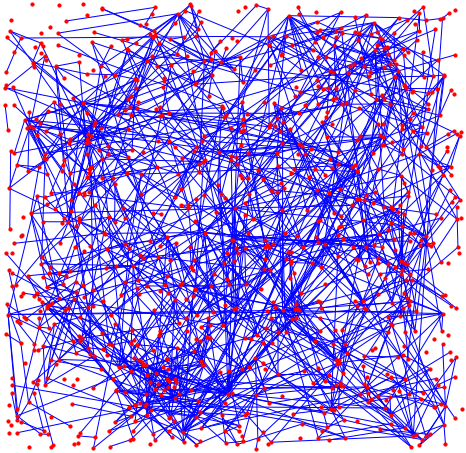}
  		\caption{Social distancing.}
  	\end{subfigure}
  	\begin{subfigure}{0.49\textwidth}
  		\centering
  		\includegraphics[width=0.9\textwidth]{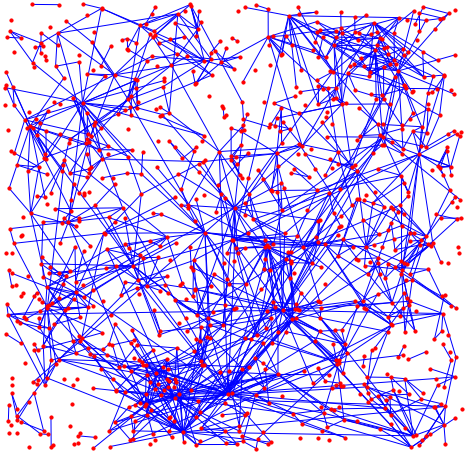}
  		\caption{Weak no-travel rule.}
  	\end{subfigure}
  	\begin{subfigure}{0.49\textwidth}
  		\centering
  		\includegraphics[width=0.9\textwidth]{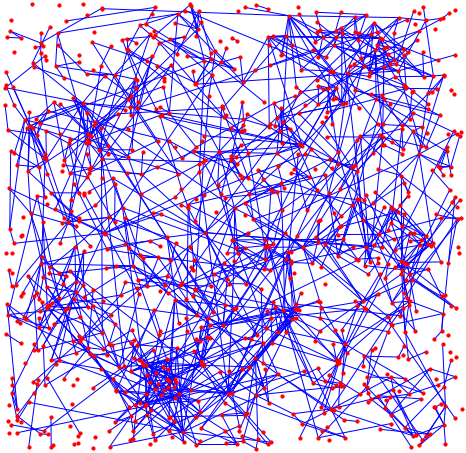}
  		\caption{Truncating long edges.}
  	\end{subfigure}
  	\begin{subfigure}{0.49\textwidth}
  		\centering
  		\includegraphics[width=0.9\textwidth]{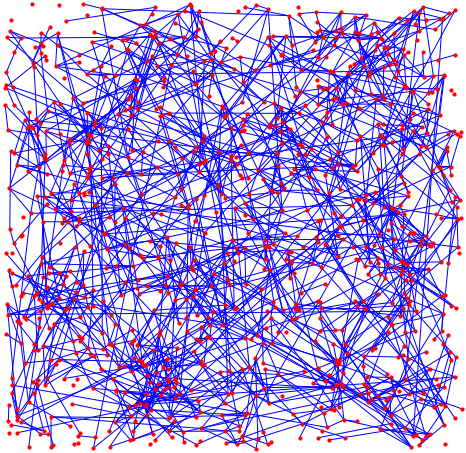}
  		\caption{Limiting maximal node degree.}
  	\end{subfigure}
  		\caption{Visualization of the change in the underlying contact network under interventions.
  		The network on the top left is similar to $G_1$: a GIRG with $(\tau, \alpha)=(2.5,1.3)$, average degree $4.8$ on $N=1000$ nodes. All interventions result in an average degree $\approx 2.6$. }
   \label{fig:change-graph1}
  		\end{center}\end{figure}

  \begin{figure}[h]
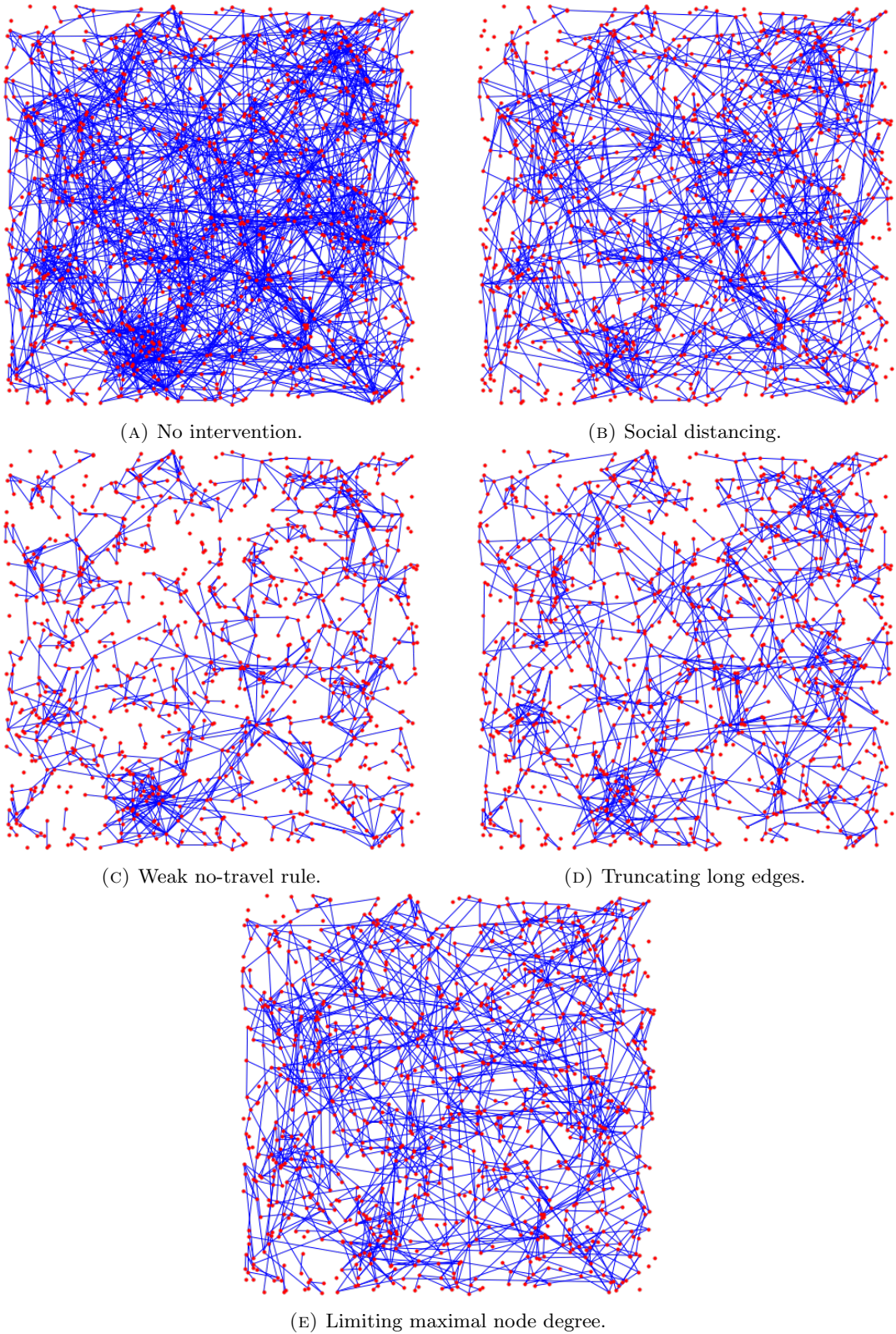
\begin{center}
  	\begin{subfigure}{0.49\textwidth}
  		\centering
  		\includegraphics[width=0.9\textwidth]{girg2-noint-cropped.png}
  		\caption{No intervention.}
  	\end{subfigure}
  	\begin{subfigure}{0.49\textwidth}
  		\centering
  		\includegraphics[width=0.9\textwidth]{girg2-perc06-cropped.png}
  		\caption{Social distancing.}
  	\end{subfigure}
  	\begin{subfigure}{0.49\textwidth}
  		\centering
  		\includegraphics[width=0.9\textwidth]{girg2-alpha28-cropped.png}
  		\caption{Weak no-travel rule.}
  	\end{subfigure}
  	\begin{subfigure}{0.49\textwidth}
  		\centering
  		\includegraphics[width=0.9\textwidth]{girg2-long10-cropped.png}
  		\caption{Truncating long edges.}
  	\end{subfigure}
  	\begin{subfigure}{0.49\textwidth}
  		\centering
  		\includegraphics[width=0.9\textwidth]{girg2-hub5-cropped.png}
  		\caption{Limiting maximal node degree.}
  	\end{subfigure}
  		\caption{Visualization of the change in the underlying contact network under interventions.
  		The network on the top left is similar to $G_2$: a GIRG with $(\tau, \alpha)=(3.3,1.3)$, average degree $4.8$ on $N=1000$ nodes. All interventions result in an average degree $\approx 2.6$, just like on Figure \ref{fig:change-graph1}, even though the pictures here look much more sparse.}
   \label{fig:change-graph2}
  		\end{center}\end{figure}

  \begin{figure}[h]\begin{center}
  		\begin{subfigure}{0.49\textwidth}
  			\centering
  			\includegraphics[width=0.9\textwidth]{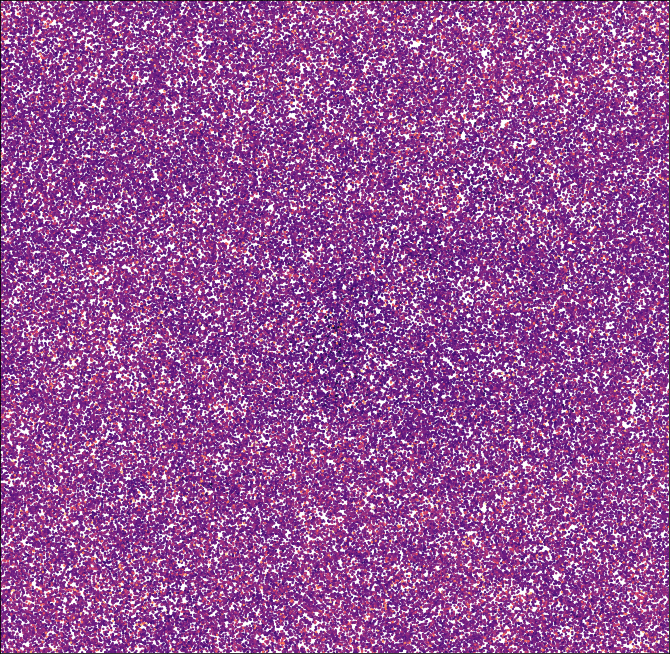}
  			\caption{No intervention.}
  		\end{subfigure}
  		\begin{subfigure}{0.49\textwidth}
  			\centering
  			\includegraphics[width=0.9\textwidth]{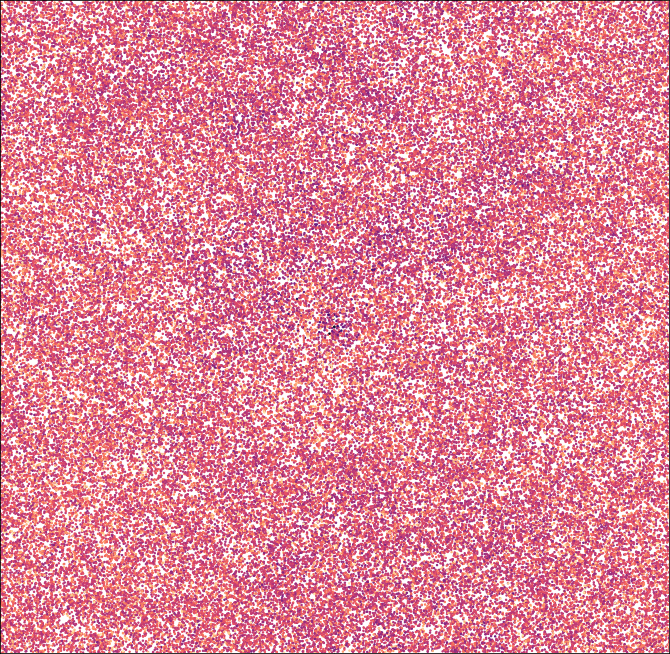}
  			\caption{Social distancing.}
  		\end{subfigure}
  		\begin{subfigure}{0.49\textwidth}
  			\centering
  			\includegraphics[width=0.9\textwidth]{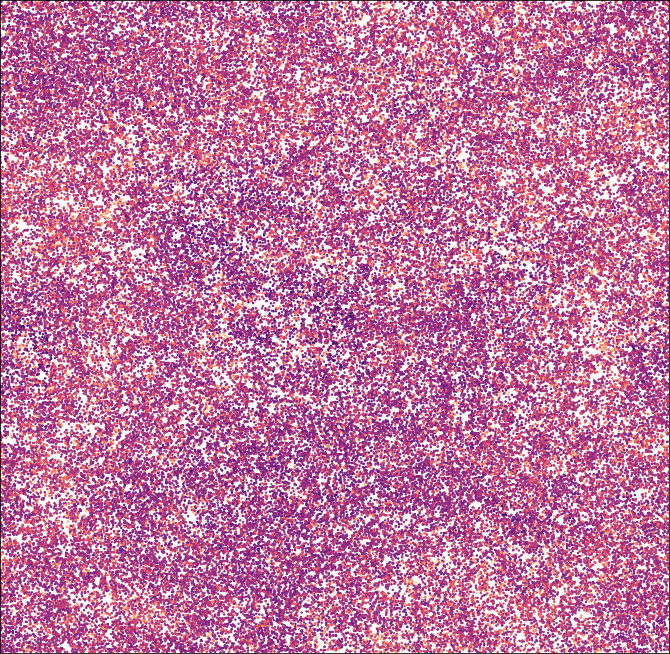}
  			\caption{Weak no-travel rule.}
  		\end{subfigure}
  		\begin{subfigure}{0.49\textwidth}
  			\centering
  			\includegraphics[width=0.9\textwidth]{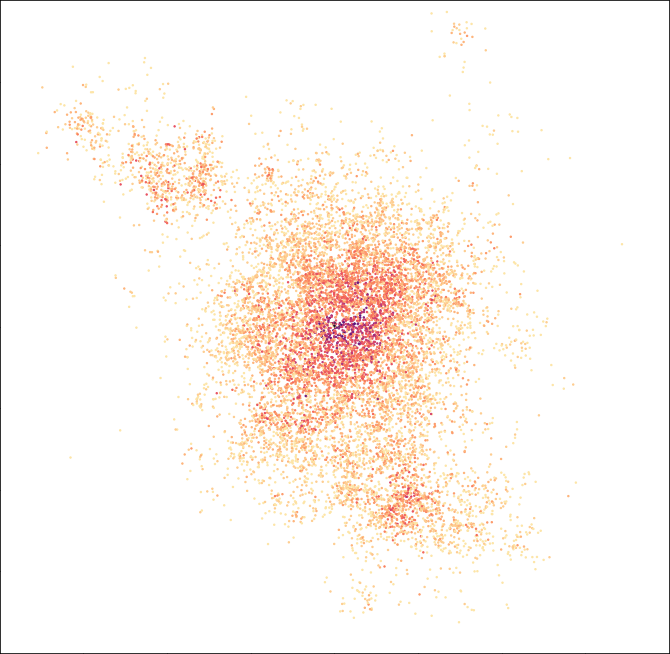}
  			\caption{Truncating long edges.}
  		\end{subfigure}
  		\begin{subfigure}{0.49\textwidth}
  			\centering
  			\includegraphics[width=0.9\textwidth]{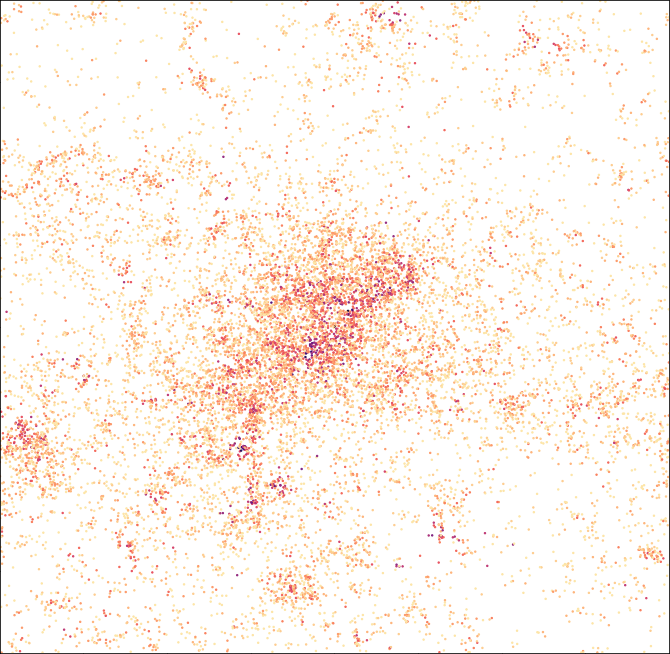}
  			\caption{Limiting maximal node degree.}
  		\end{subfigure}
  		\caption{Visualization of the initial $19$ days on $G_1$ under interventions.
  		The network on the top left is $G_1$, a GIRG with $(\tau, \alpha)=(2.5,1.3)$ and average degree $9.6$ on $N=160000$ nodes. All interventions result in an average degree $5.7$.  The darker the color, the earlier the infection. We chose typical runs to illustrate the effect of interventions.}
   \label{fig:G1-vis}
  		\end{center}\end{figure}

  \begin{figure}[h]\begin{center}
  	\begin{subfigure}{0.49\textwidth}
  		\centering
  		\includegraphics[width=0.9\textwidth]{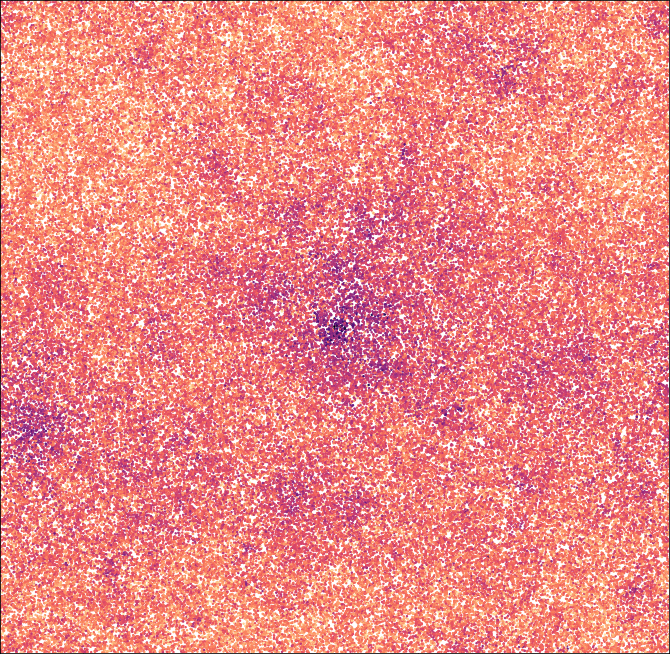}
  		\caption{No intervention.}
  	\end{subfigure}
  	\begin{subfigure}{0.49\textwidth}
  		\centering
  		\includegraphics[width=0.9\textwidth]{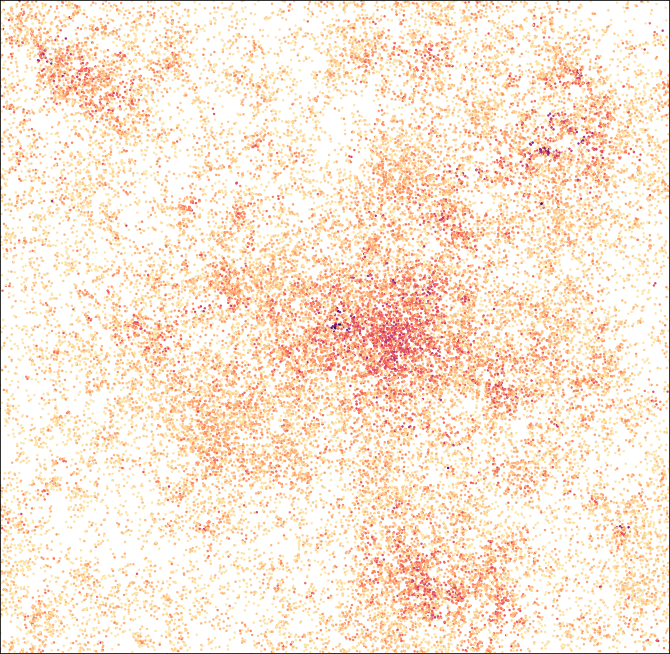}
  		\caption{Social distancing.}
  	\end{subfigure}
  	\begin{subfigure}{0.49\textwidth}
  		\centering
  		\includegraphics[width=0.9\textwidth]{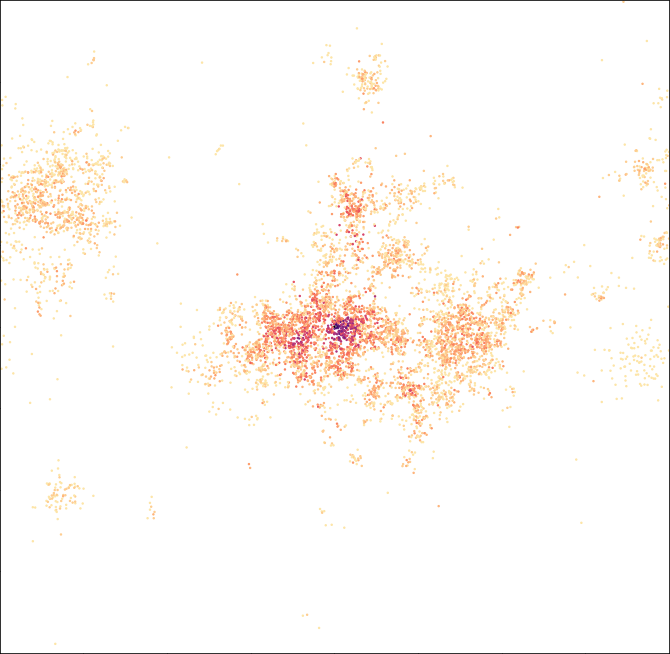}
  		\caption{Weak no-travel rule.}
  	\end{subfigure}
  	\begin{subfigure}{0.49\textwidth}
  		\centering
  		\includegraphics[width=0.9\textwidth]{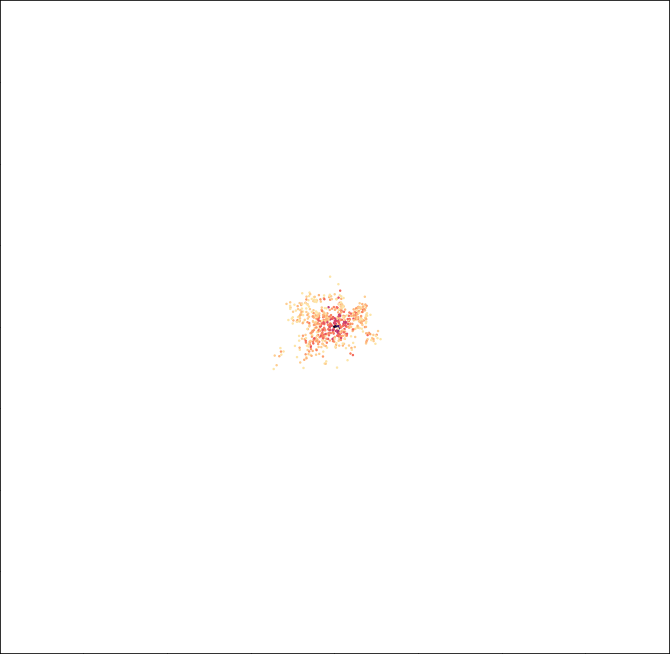}
  		\caption{Truncating long edges.}
  	\end{subfigure}
  	\begin{subfigure}{0.49\textwidth}
  		\centering
  		\includegraphics[width=0.9\textwidth]{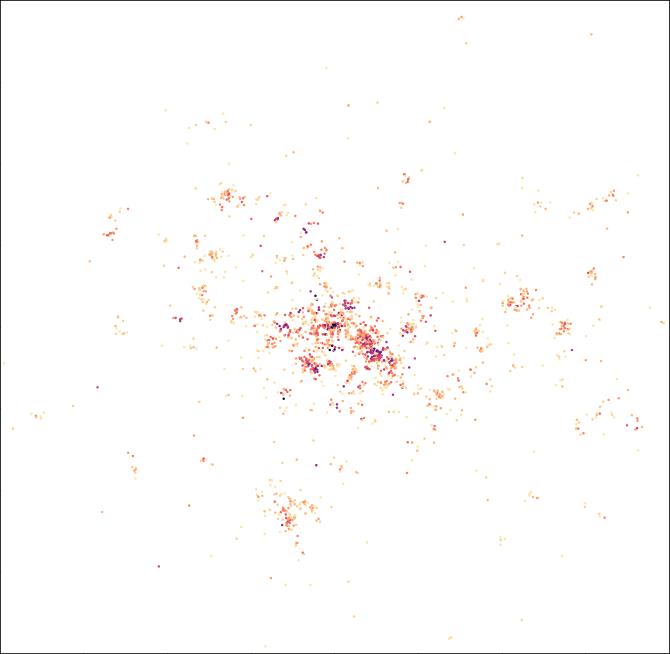}
  		\caption{Limiting maximal node degree.}
  	\end{subfigure}
  		\caption{Visualization of the initial $19$ days on $G_2$ under several intervention measures.
  		The network on the left is $G_2$, a GIRG with $(\tau, \alpha)=(3.3,1.3)$ and average degree $8.7$ on $N=160000$ nodes. All interventions result in an average degree $4.9$.  The darker the color, the earlier the infection. Color scaling is the same as on Figure \ref{fig:G1-vis}.}
   \label{fig:G2-vis}
  		\end{center}\end{figure}

  		\begin{figure}[h]\begin{center}
  		\begin{subfigure}{1\textwidth}
  		\includegraphics[width=1\textwidth]{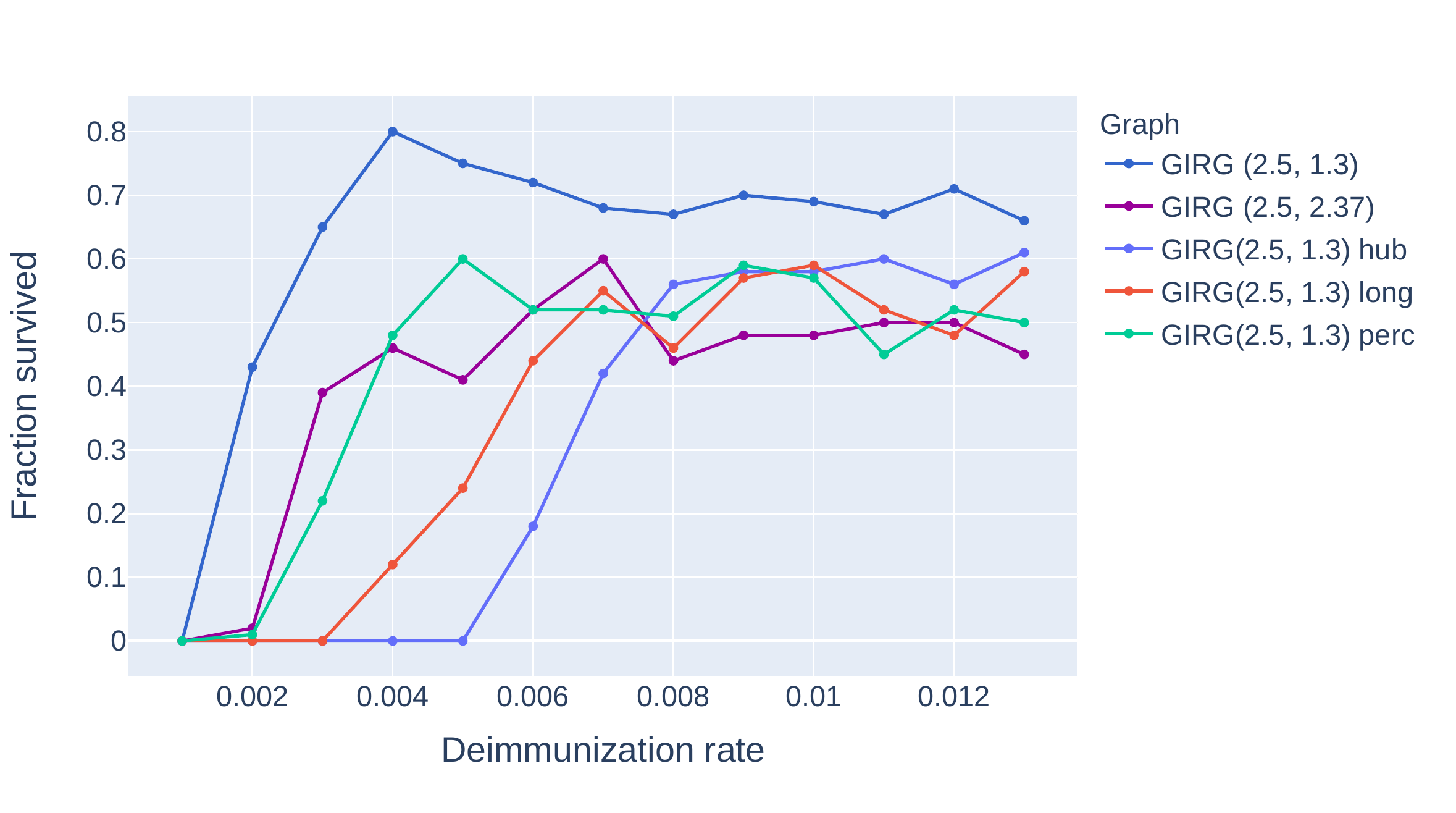}
  		\caption{Survival probabilities under interventions on $G_1$, a GIRG with $(\tau, \alpha)=(2.5,1.3)$.}
  		\end{subfigure}
  		\begin{subfigure}{1\textwidth}
  		\includegraphics[width=1\textwidth]{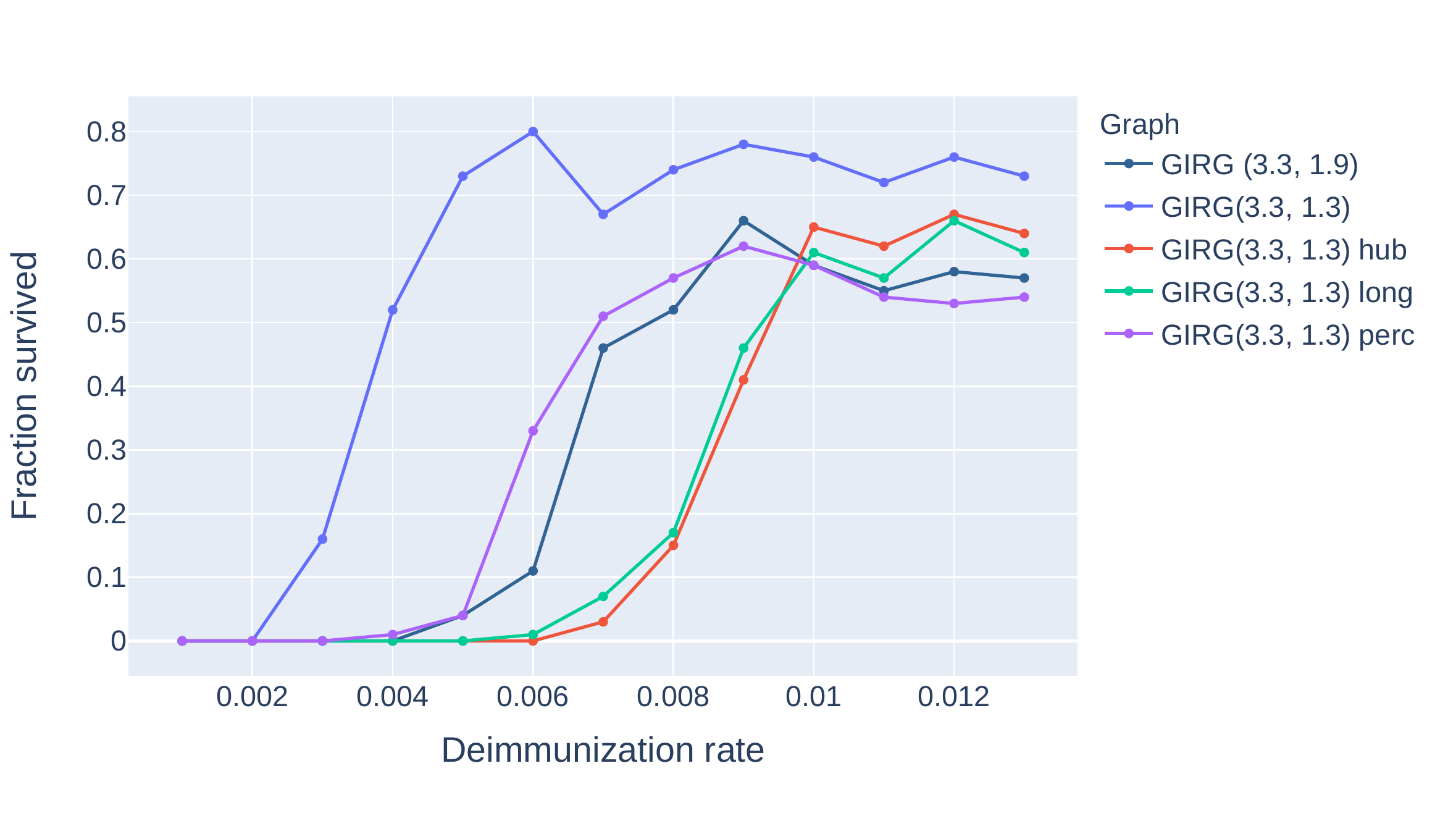}
  		\caption{Survival probabilities under interventions on $G_2$, a GIRG with $(\tau, \alpha)=(3.3,1.3)$.}
  		\end{subfigure}
  		\caption{The effect of interventions: the long-survival probabilities (phase 2M) are decreasing significantly under each intervention strategy. The four interventions are: (A) social distancing, denoted as "perc", (B) limiting maximal degree, denoted as "hub", (Ch) a hard no-travel rule, denoted as "long"  on the legend, and (Cw) a weak no-travel rule: increasing $\alpha$.  We see that in all interventions, the long-survival probabilities are below the original curve. The threshold $\eta$ is lowest for percolation and highest for limiting the maximal degree, while truncating long-edges and increasing $\alpha$ in the network both show an interesting non-monotonous survival probability curve. \emph{Top:} The initial network is $G_1$, a GIRG with $(\tau,\alpha)=(2.5, 1.3)$, and $\Ev[\deg(u)]=9.6$. \emph{Bottom:} The initial network is $G_2$, a GIRG with $(\tau,\alpha)=(3.3, 1.3)$, and $\Ev[\deg(u)]=8.7$. } \label{fig:survival-interventions}.
  \end{center}\end{figure}

  		\begin{figure}[b]\begin{center}
  		\includegraphics[width=1\textwidth]{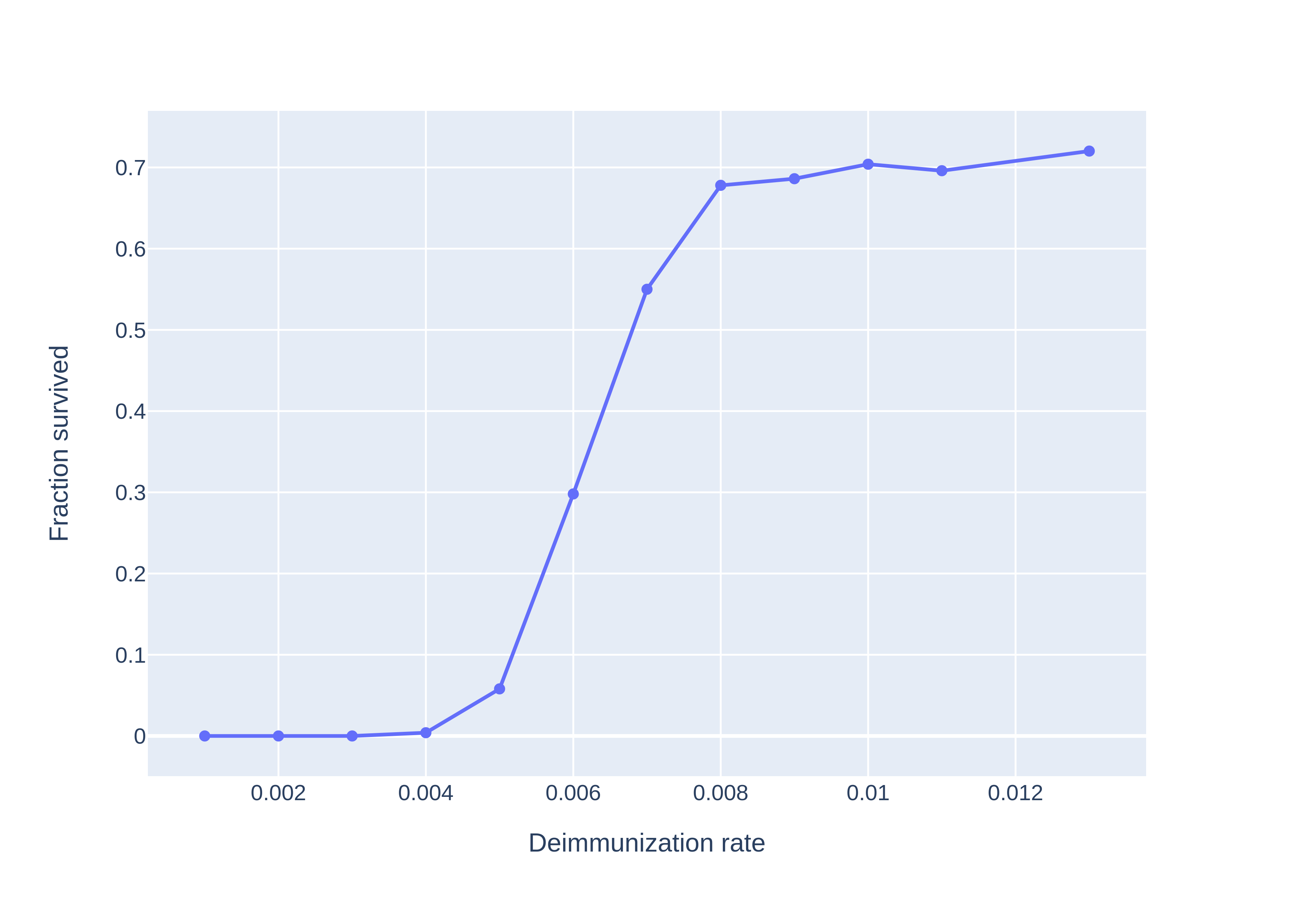}
  		\caption{Monotonicity of  survival probabilities as a function of $\eta$. For the graph
  		$G_2^\alpha$, a GIRG$(3.3, 1.9)$ (weak no-travel rule), we have tested the monotonicity of the survival probability curve, by increasing the number of runs to $500$ for each value of $\eta$. The curve appears to be monotonous, with a sharp threshold near  $\eta_t\approx0.007$.
  		} \label{fig:survival-500}.
  \end{center}\end{figure}

  \begin{figure}
  \begin{center}
  \begin{subfigure}{\textwidth}
  		\centering
  		\includegraphics[width=\textwidth]{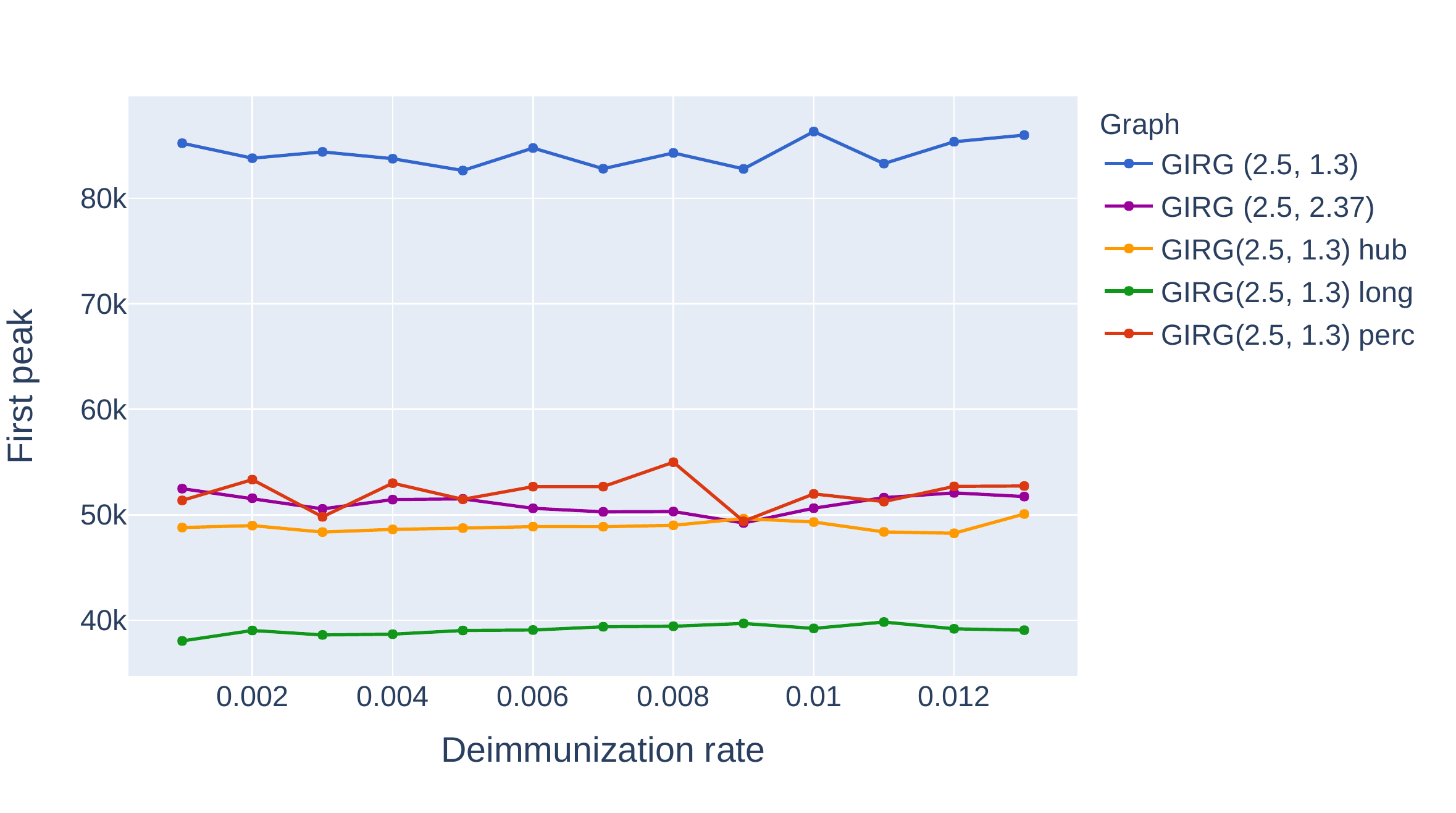}
  		\caption{First peak: interventions}
  	\end{subfigure}
  	\begin{subfigure}{\textwidth}
  		\centering
  		\includegraphics[width=\textwidth]{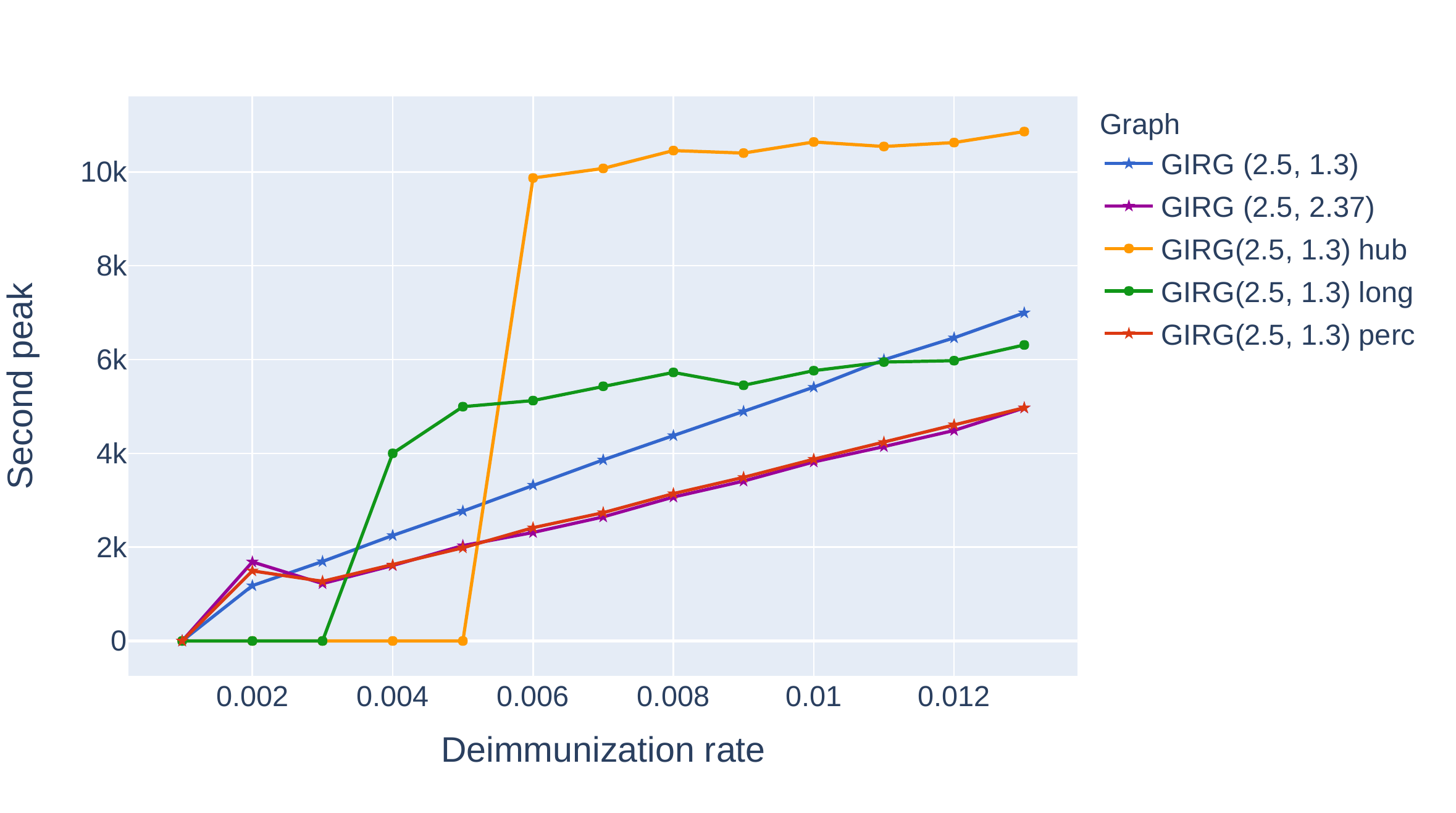}
  		\caption{Second peak: interventions}
  	\end{subfigure}
  \end{center}
  \caption{The effect of interventions on the first and second peak on $G_1$, as a function of $\eta$ (the inverse of the average immunity period). Abbreviations in the legend: `hub': limiting maximal node degree, `long': cutting long edges, `perc': social distancing. The first entry is increasing $\alpha$. GIRG$(2.3,1.3)$ is the original network. We see that the height of the first peak is insensitive to the immunity length, and that cutting long edges is most effective in reducing the first peak. For the second  peak, limiting node degree pushes the critical $\eta$ for appearance of the second peak from basically $0$ to $0.005$, however, above that the second peak is twice as high as for other interventions. Cutting long-edges also has a similar effect, although to a lesser extent. See also Figure  \ref{fig:first-second-peak-g2}. Star indicates that there is no second peak, and the height we see is the height of the equilibrium, see the top figure on Figure \ref{fig:girg1-interventions}.}\label{fig:first-second-peak-g1}
  \end{figure}

  \begin{figure}
  \begin{center}
  \begin{subfigure}{\textwidth}
  		\centering
  		\includegraphics[width=\textwidth]{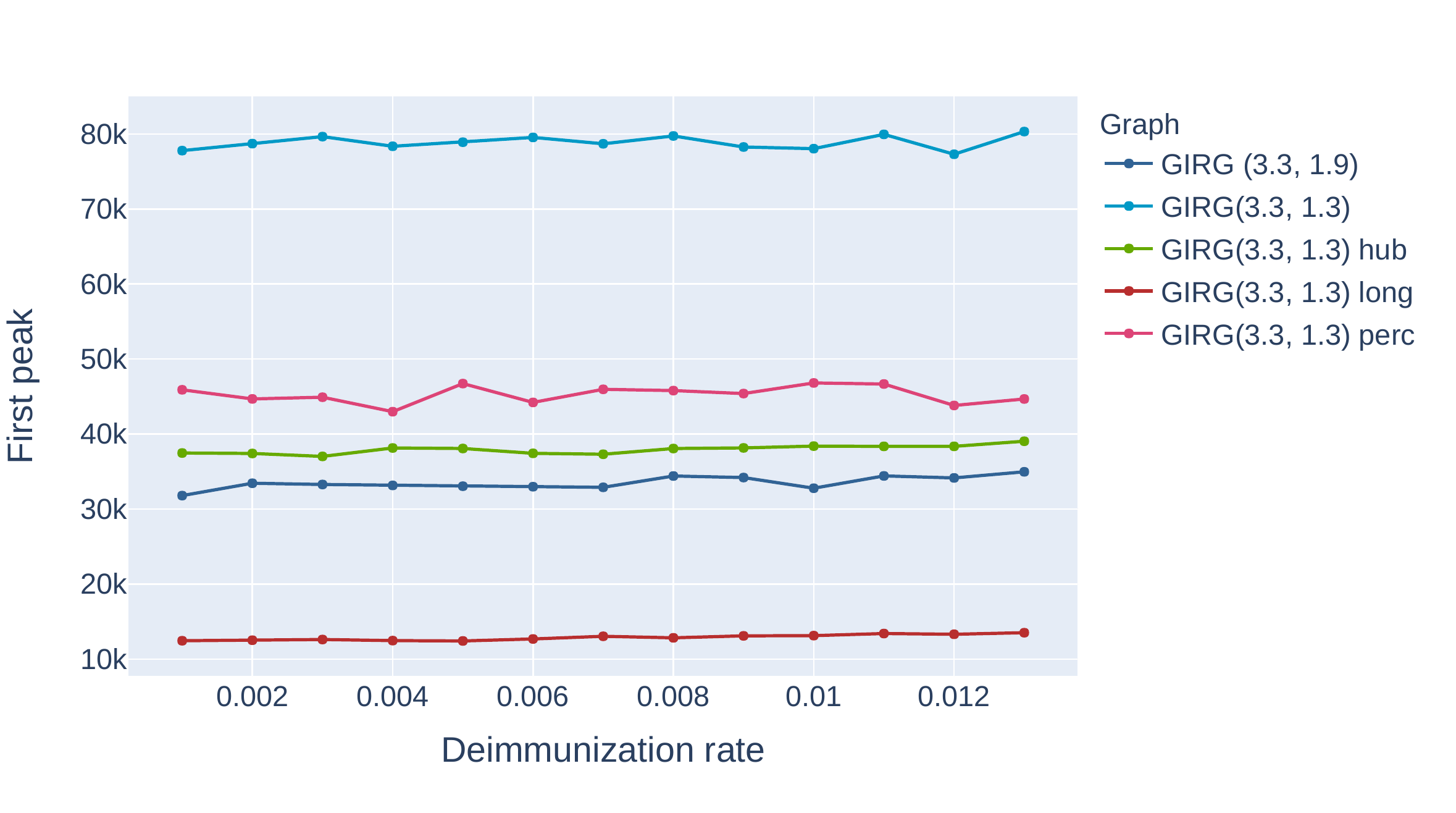}
  		\caption{First peak under interventions}
  	\end{subfigure}
  	\begin{subfigure}{\textwidth}
  		\centering
  		\includegraphics[width=\textwidth]{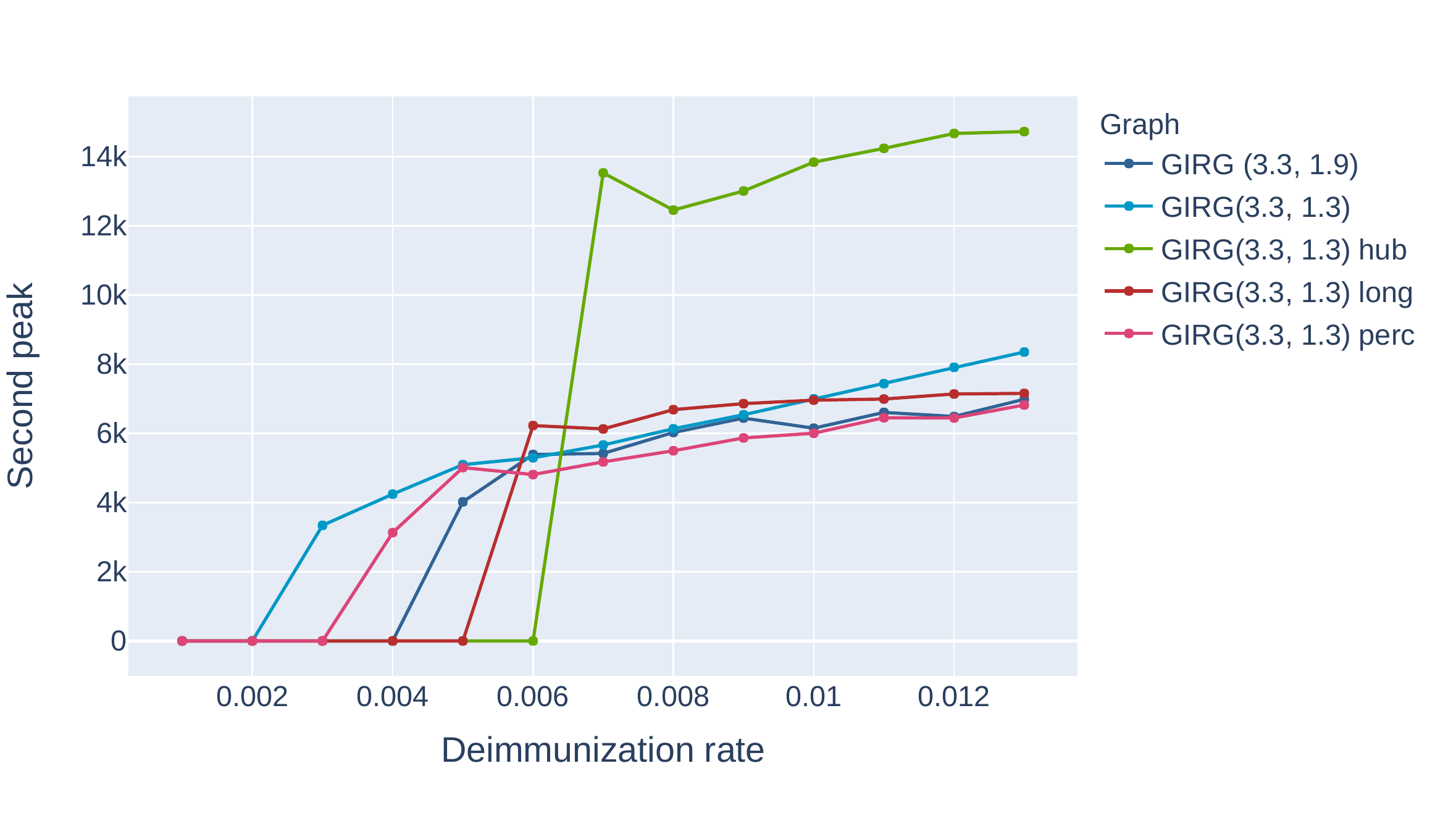}
  		\caption{Second peak under interventions}
  	\end{subfigure}
  \end{center}
  \caption{The effect of interventions on the first and second peak on $G_2$, as a function of $\eta$ (the inverse of the average immunity period). Abbreviations in the legend: `hub': limiting maximal node degree, `long': cutting long edges, `perc': social distancing. The first entry is increasing $\alpha$. GIRG$(3.3,1.3)$ is the original network. We see that the height of the first peak is insensitive to the immunity length, and that cutting long edges is most effective in reducing the first peak. For the second  peak, limiting node degree pushes the critical $\eta$ for appearance of the second peak from $0.002$ to $0.006$, however, above that the second peak is twice as high as for other interventions. See also Figure  \ref{fig:first-second-peak-g1}.}\label{fig:first-second-peak-g2-v2}
  \end{figure}

  \clearpage
  \newpage

  \begin{figure}[t]\begin{center}
  		\includegraphics[width=0.9\textwidth]{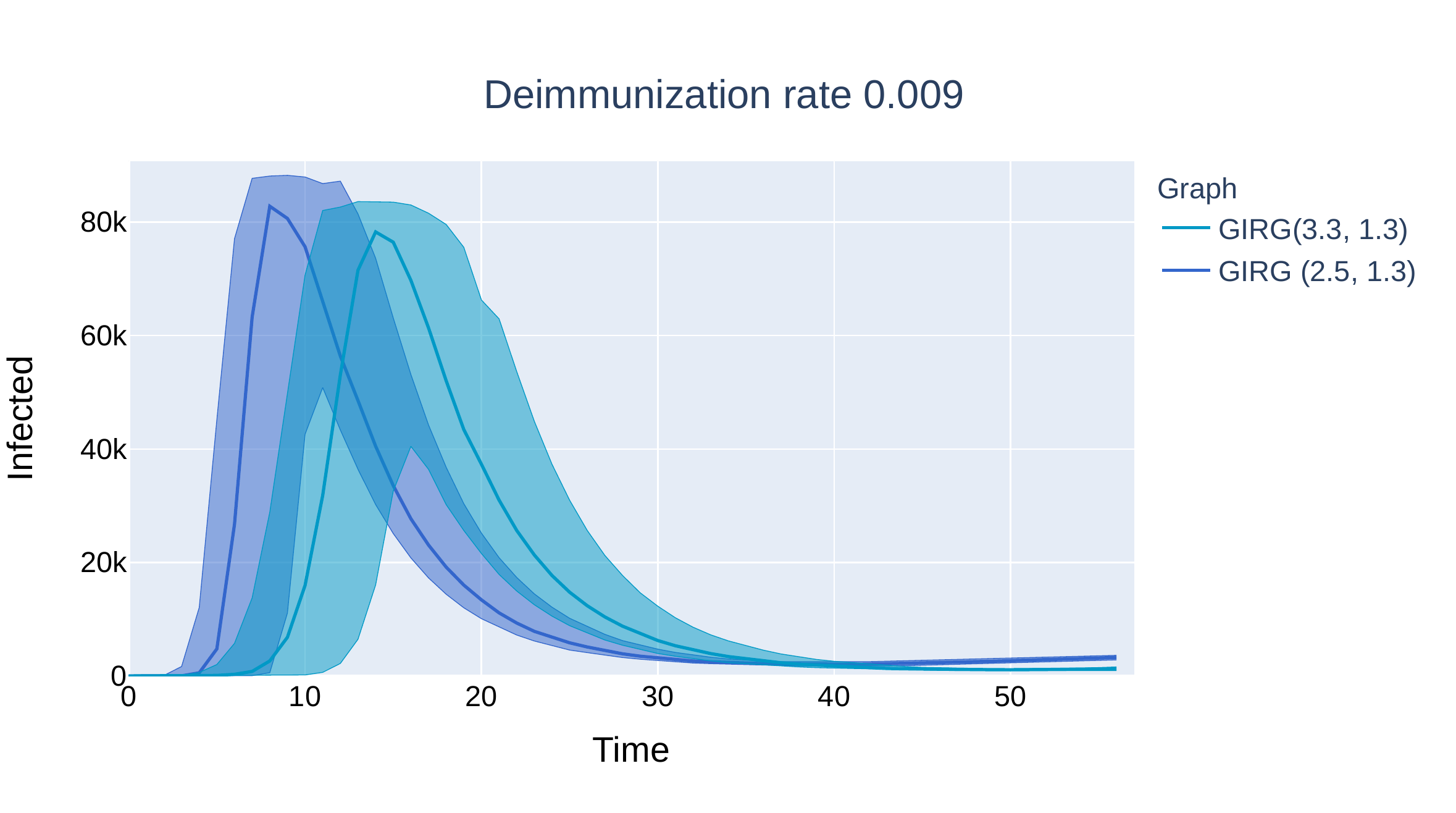}
  		\includegraphics[width=0.9\textwidth]{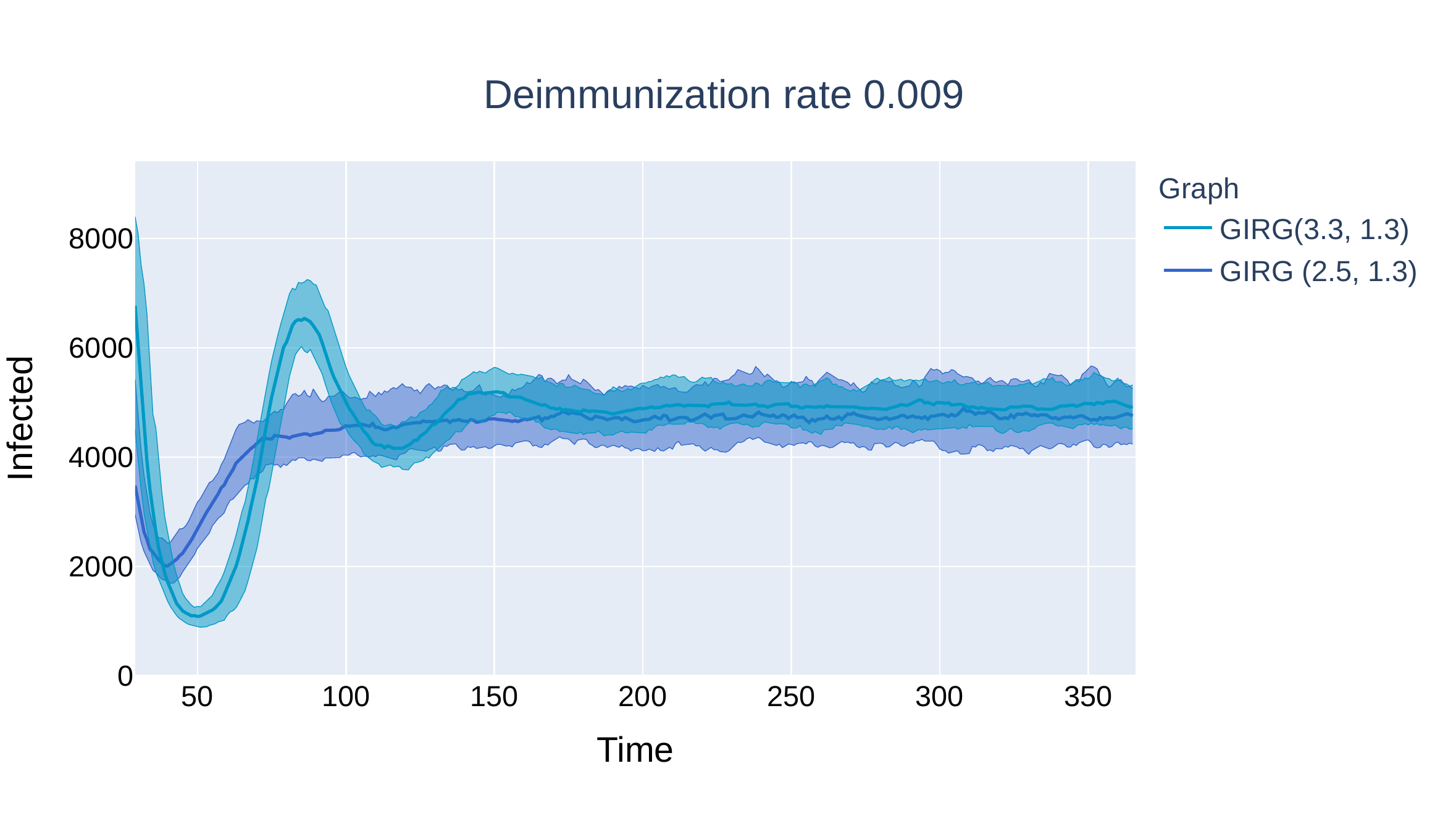}
  		\caption{No intervention: the epidemic curves for the networks $G_1, G_2$ on which we apply the intervention methods. $G_1$ is a GIRG with $(\tau,\alpha)=(2.5, 1.3)$ and average degree $9.6$, while $G_2$ is a GIRG with $(\tau,\alpha)=(2.5, 1.3)$ and average degree $8.7$. \emph{Top:}
  	The first peak of the epidemic. \emph{Bottom:} The second peak and stabilization of the number of infected. The average height  and location of the first peak is $\overline H_1(G_1)=82785$ on day $8$ for $G_1$, while it is  $\overline H_1(G_2)=78263$ on day $14$ for  GIRG with $(\tau,\alpha)=(3.3, 1.3)$. On $G_2$, we see a second peak at day $79$ of average height $\overline H_2(G_2)=6538$, and the equilibrium number of infected $\overline E(G_2)\approx 5000\pm 100$.	The equilibrium number of infected $\overline E(G_1)\approx 4700\pm100$ is reached without a second peak on $G_1$.
  				} \label{fig:twogirg-noint}.
  \end{center}\end{figure}

  \begin{figure}[t]\begin{center}
  		\includegraphics[width=0.9\textwidth]{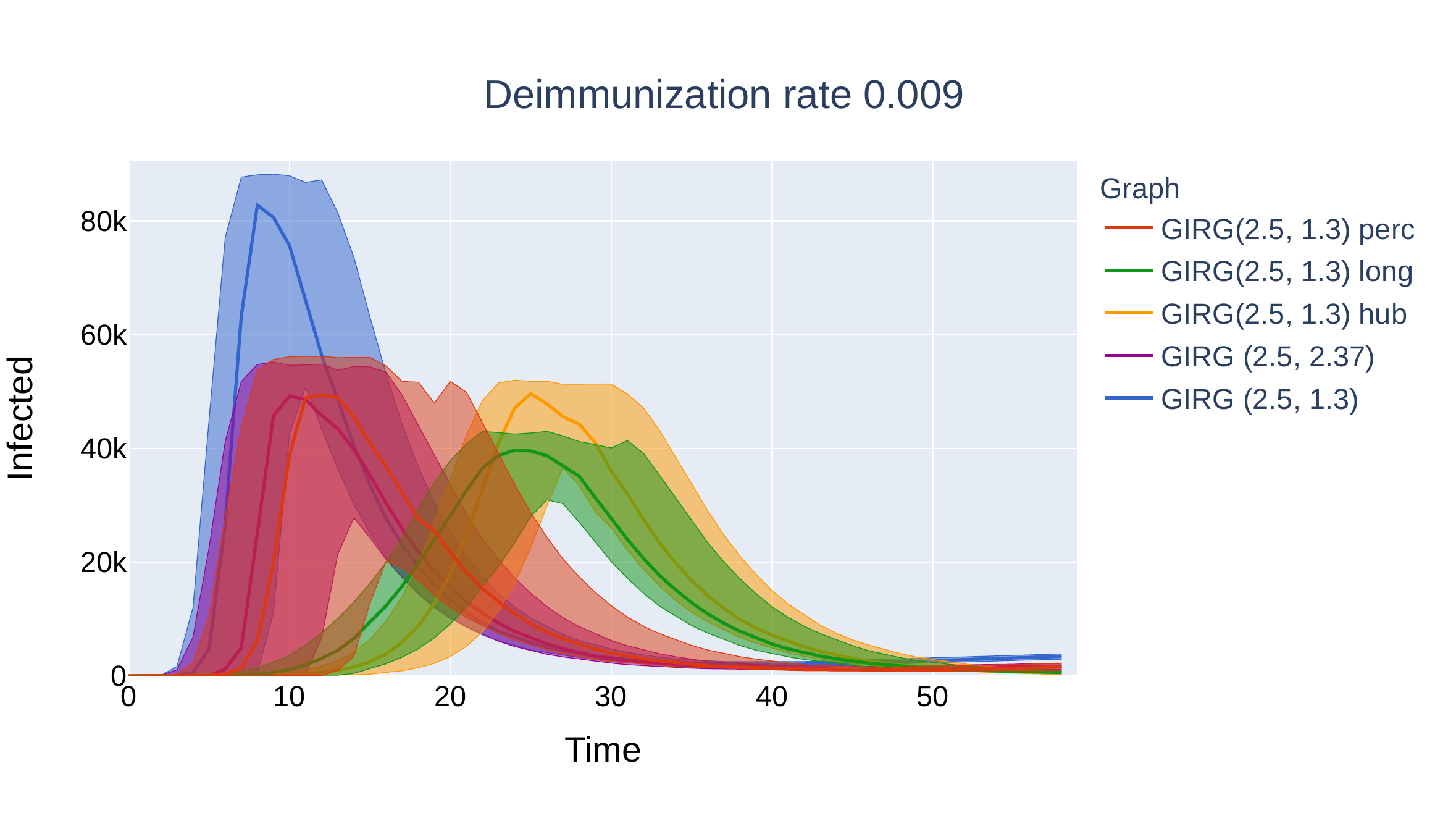}
  		\includegraphics[width=0.9\textwidth]{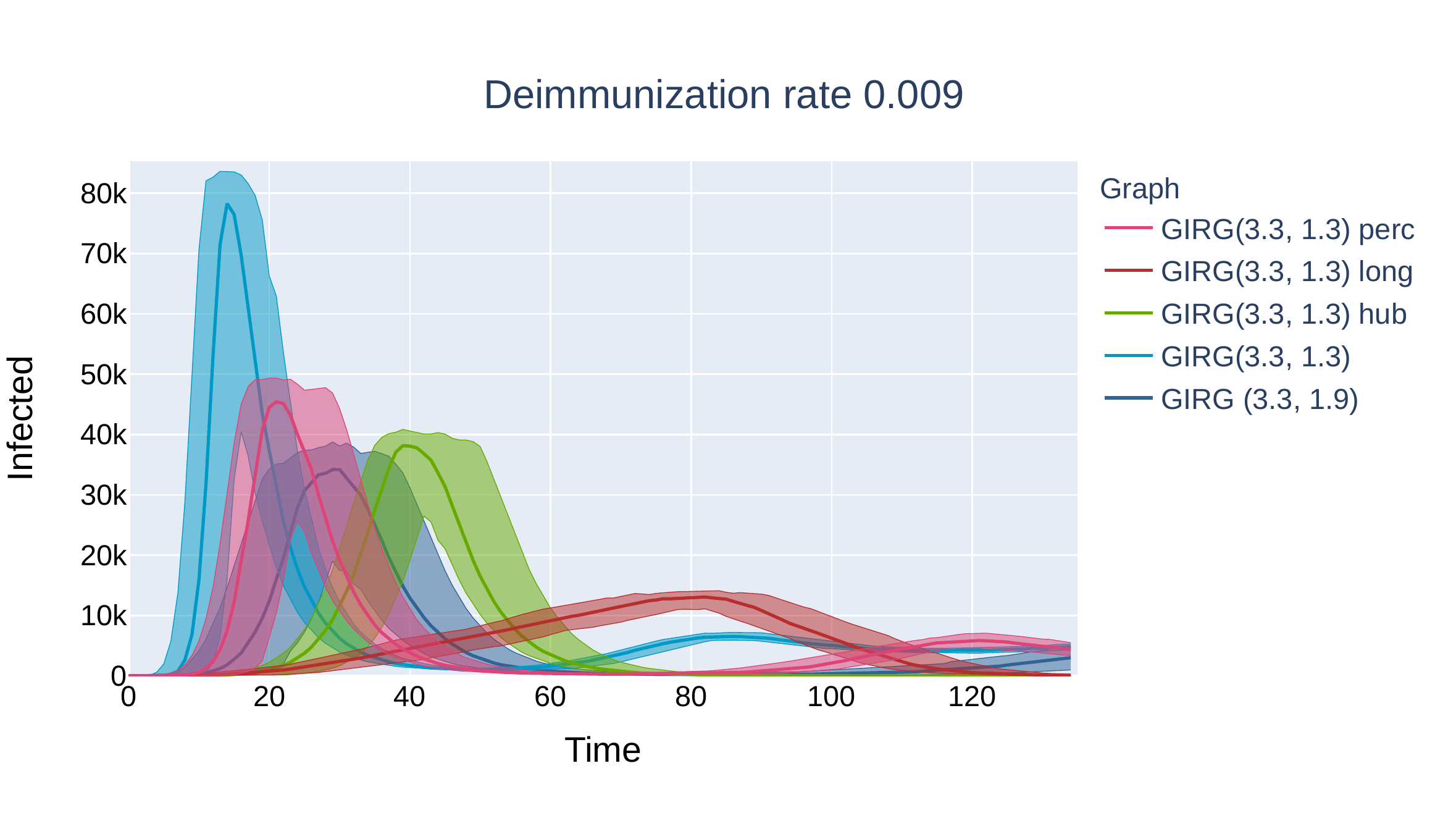}
  		\caption{The first peak under interventions. \emph{Top:} The effect of interventions on $G_1$: Social distancing (``GIRG$(2.5, 1.3)$ perc''), the hard no travel rule (``GIRG$(2.5, 1.3)$ long''), limiting the maximal number of contact (``GIRG$(2.5, 1.3)$ hub''), and the weak no-travel rule (``GIRG$(2.5, 2.37)$''). The first peak disappears before day $60$ in all cases. The curves for the weak no-travel rule and social distancing are almost identical. \emph{Bottom:} The effect of interventions on $G_1$: Social distancing (``GIRG$(3.3, 1.3)$ perc''), the hard no travel rule (``GIRG$(3.3, 1.3)$ long''), limiting the maximal number of contact (``GIRG$(3.3, 1.3)$ hub''), and the weak no-travel rule (``GIRG$(3.3, 1.9)$''). The hard no travel rule is most effective in flattening the curve: both in its height as well as the day of the peak.	 } \label{fig:girg-interventions-firstpeak}.
  \end{center}\end{figure}

  \begin{figure}[t]\begin{center}
  		\includegraphics[width=0.9\textwidth]{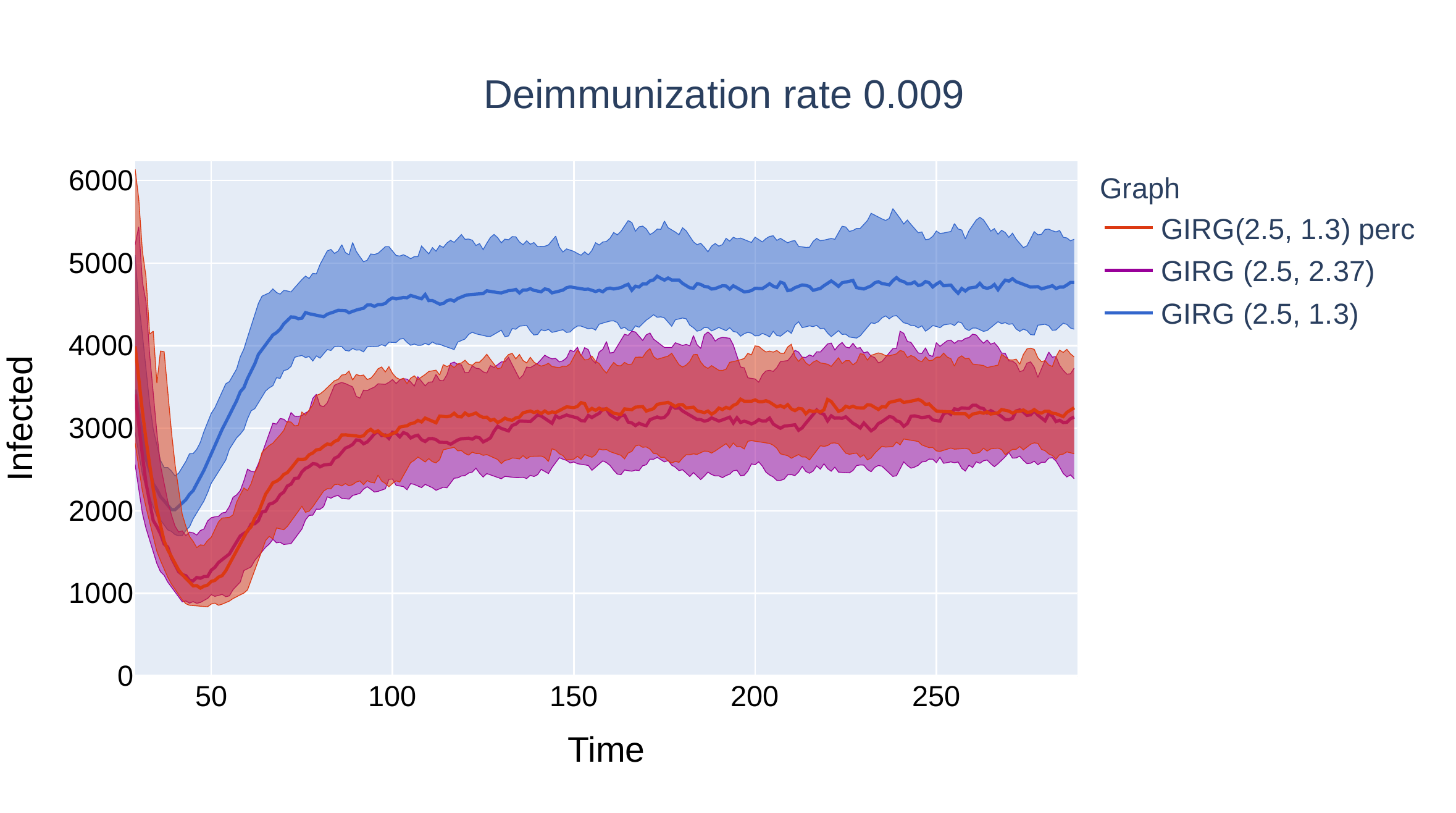}
  		\includegraphics[width=0.9\textwidth]{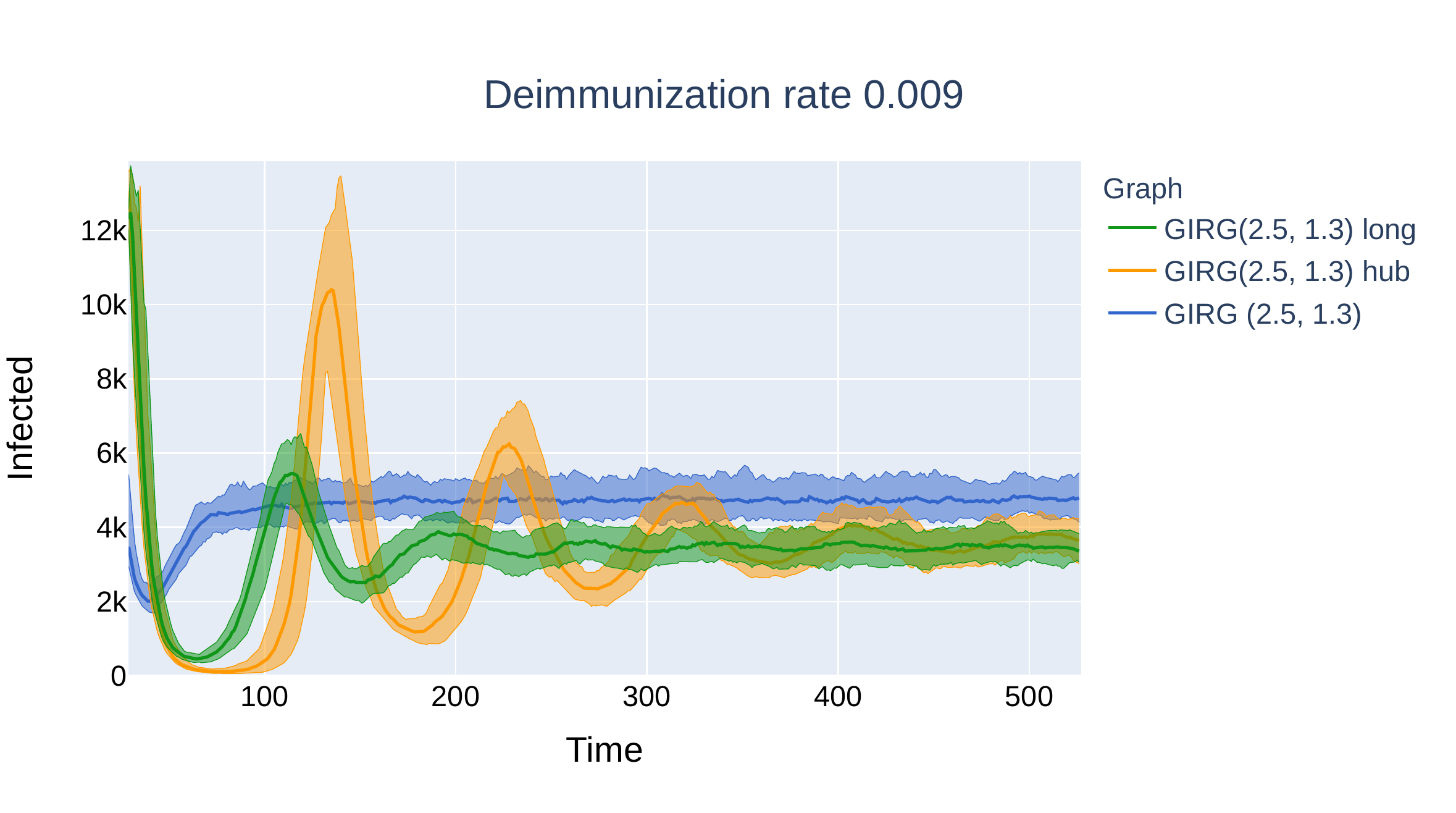}
  		\caption{Later effects of interventions on $G_1$: \emph{Top}: Social distancing (``GIRG$(2.5, 1.3)$ perc''), as well as the weak no-travel rule (``GIRG$(2.5, 2.37)$''). No second peak can be observed, just like on the original network. \emph{Bottom:} Limiting the maximal number of contact (``GIRG$(2.5, 1.3)$ hub''), and the hard no travel rule (``GIRG$(2.5, 1.3)$ long'"): second and further peaks are periodically present, with decreasing magnitude and period roughly the average immunity length. $G_1$ is a GIRG with $(\tau,\alpha)=(2.5, 1.3)$ and average degree $9.6$ on $N=160000$ nodes. All networks after intervention have average degree $\approx 4.9$.
  	 } \label{fig:girg1-interventions}.
  \end{center}\end{figure}

  \begin{figure}[t]\begin{center}
  		\includegraphics[width=0.9\textwidth]{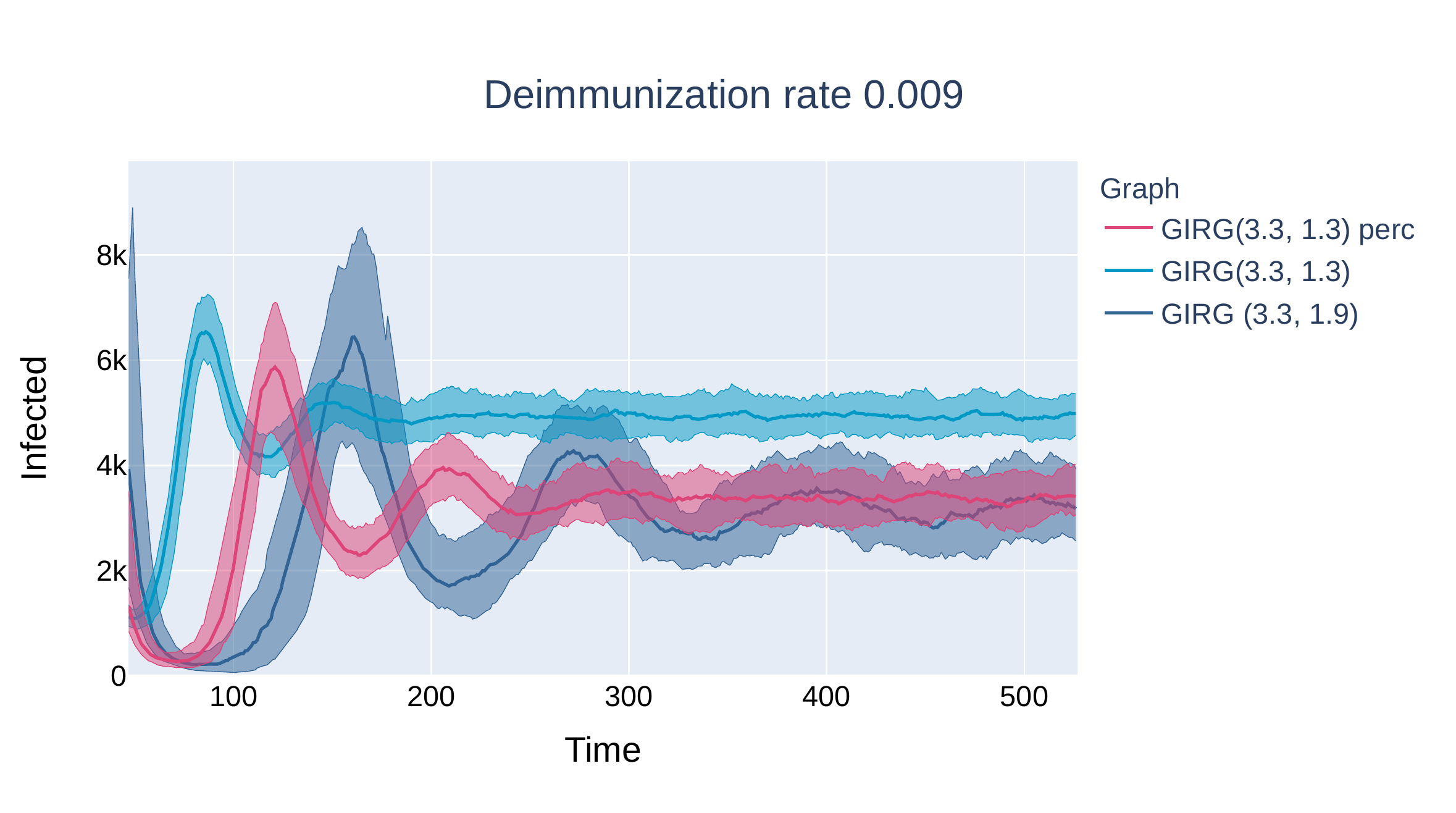}
  		\includegraphics[width=0.9\textwidth]{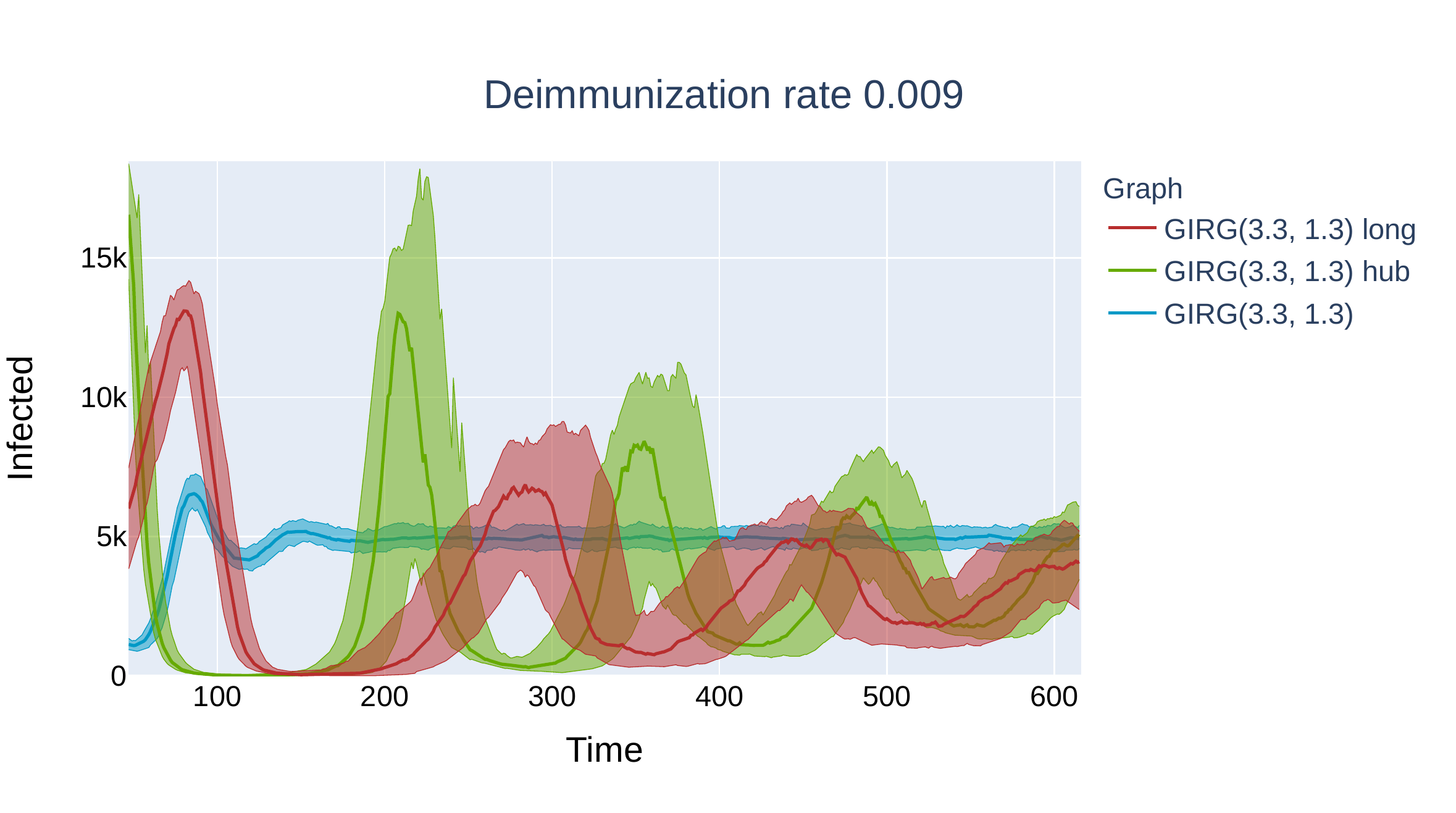}
  		\caption{Later effects of interventions on $G_2$: \emph{Top}: Social distancing (``GIRG$(3.3, 1.3)$ perc''), as well as the weak no-travel rule (``GIRG$(3.3, 1.9)$'') compared to the original network  (``GIRG$(3.3, 1.3)$''). Second and further peaks are present, equilibrium is reached within $600$ days. \emph{Bottom:} Limiting the maximal number of contact (``GIRG$(3.3, 1.3)$ hub''), and the hard no travel rule (``GIRG$(3.3, 1.3)$ long'"): second and further peaks are more profound, equilibrium is only reached around $1200$ days. $G_2$ is a GIRG with $(\tau,\alpha)=(3.3, 1.3)$ and average degree $8.7$ on $N=160000$ nodes. All networks after intervention have average degree $\approx 4.7$.	 The first peak can be seen on Figure \ref{fig:girg-interventions-firstpeak}. The first peak of the red curve (corresponding to truncating long edges) is the first peak of the epidemic.} \label{fig:girg2-interventions}.
  \end{center}\end{figure}

  \begin{figure}[t]\begin{center}
  		\includegraphics[width=0.9\textwidth]{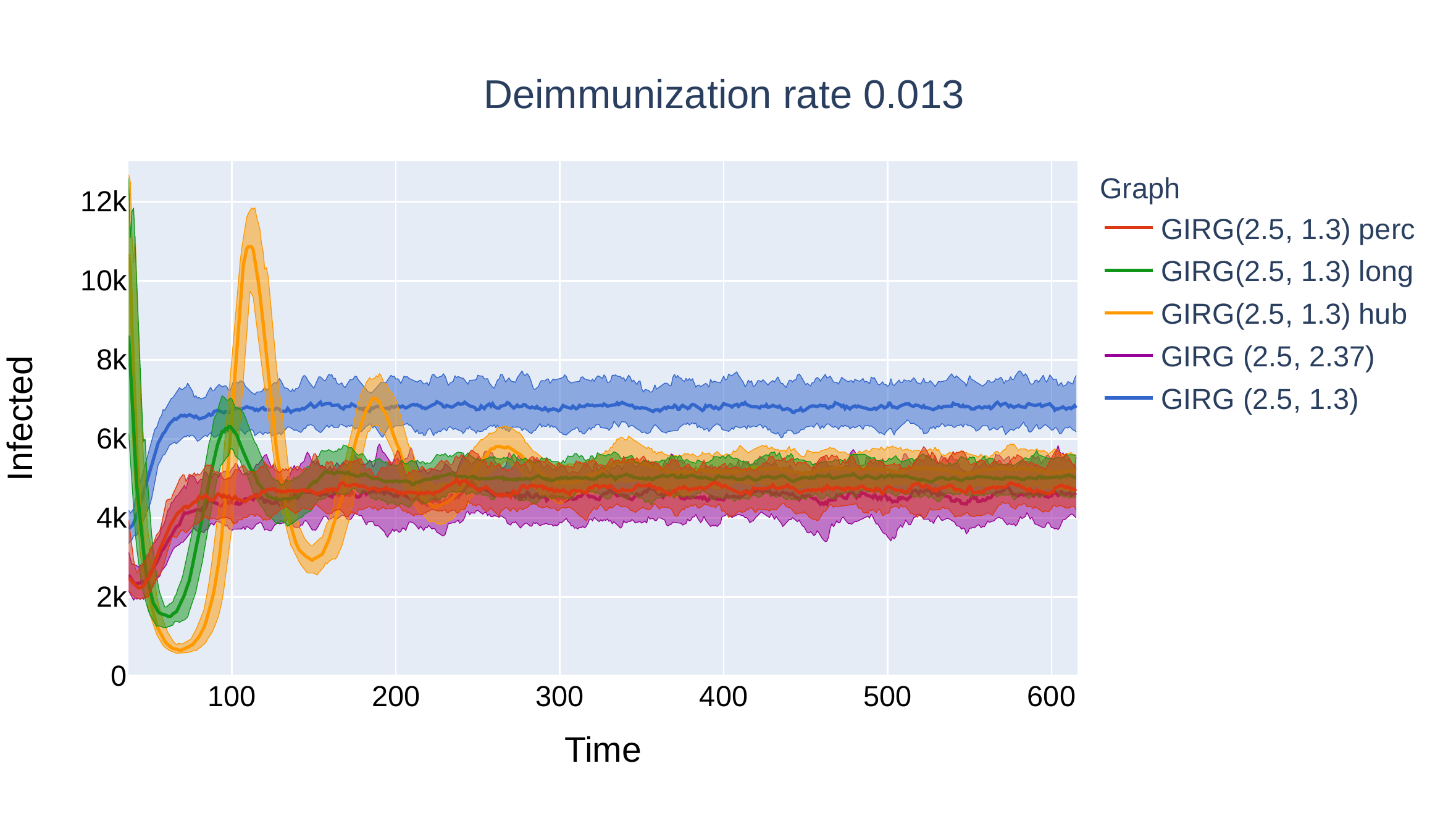}
  		\includegraphics[width=0.9\textwidth]{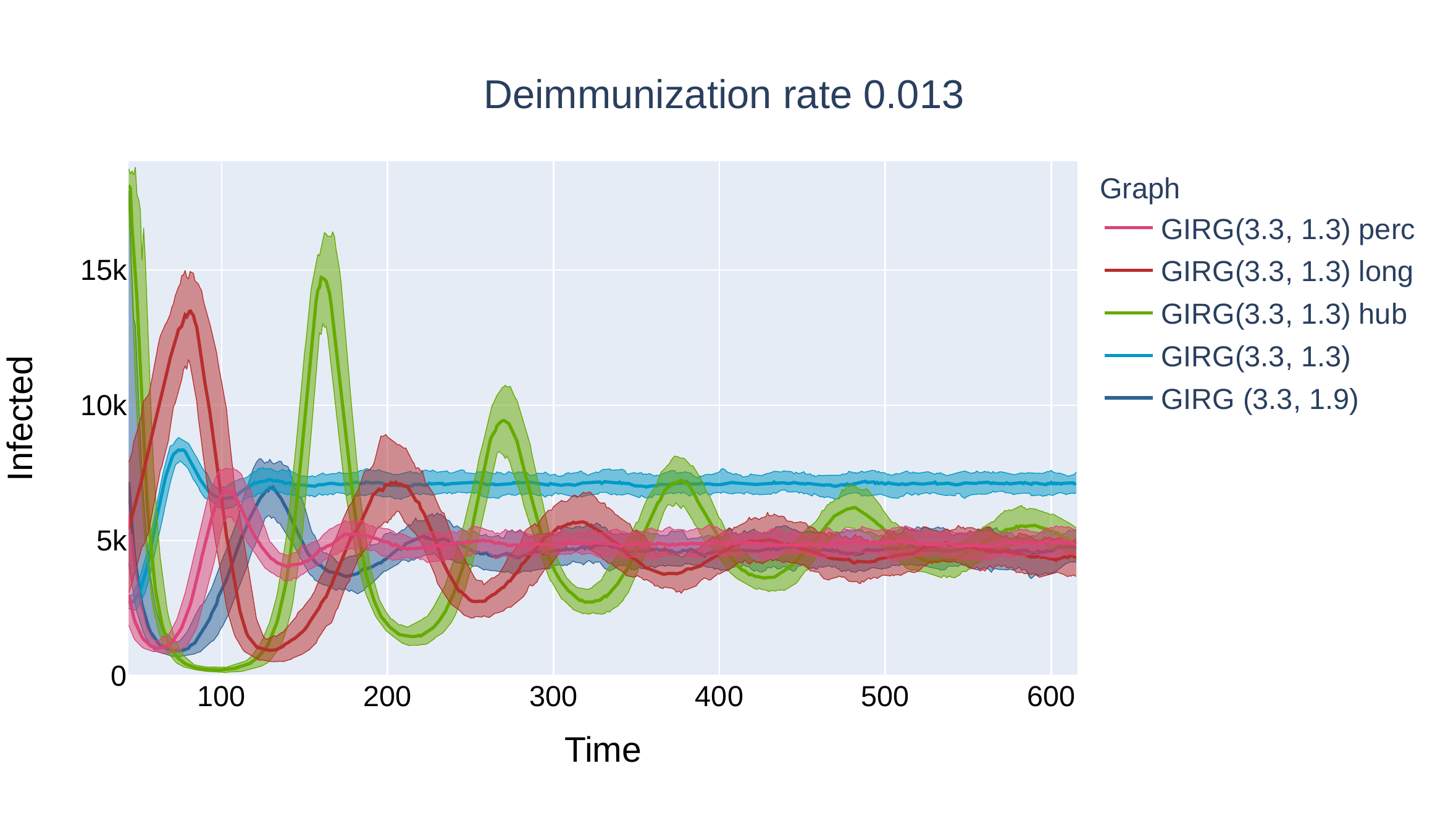}
  		\caption{Later effects of interventions on $G_1$, $G_2$, for $\eta=0.013$: \emph{Top}: $G_1$ and all interventions. \emph{Bottom }  $G_2$ and all interventions. The curves are qualitatively similar to $\eta=0.009$, the period between second and further peaks differ from those on Figures \ref{fig:girg1-interventions} and \ref{fig:girg1-interventions}. The first peak is not affected by the change of $\eta$ (below $2\%$ change in height, that could be also due to statistical error margins, so we omit to plot it. The first peak of the red curve (corresponding to truncating long edges) is the first peak of the epidemic.} \label{fig:girg-interventions-eta13}.
  \end{center}\end{figure}

  \clearpage
  \newpage

  \begin{table}[]
  \resizebox{\textwidth}{!}{%
  \begin{tabular}{@{}cccccccccccccccccccccc@{}}
  \toprule
                         &                                  & \multicolumn{5}{c}{\textbf{First peak $\bm{(\times 10^3)}$}}                        & \multicolumn{5}{c}{\textbf{Time of first peak}}                                     & \multicolumn{5}{c}{\textbf{Second peak $\bm{(\times 10^3)}$}}                       & \multicolumn{5}{c}{\textbf{Time of second peak}}               \\ \midrule
                         & \multicolumn{1}{c|}{$\bm{\eta}$} & \textbf{O} & \textbf{A} & \textbf{Bw} & \textbf{B} & \multicolumn{1}{c|}{\textbf{C}} & \textbf{O} & \textbf{A} & \textbf{Bw} & \textbf{B} & \multicolumn{1}{c|}{\textbf{C}} & \textbf{O} & \textbf{A} & \textbf{Bw} & \textbf{B} & \multicolumn{1}{c|}{\textbf{C}} & \textbf{O} & \textbf{A} & \textbf{Bw} & \textbf{B} & \textbf{C} \\ \midrule
  $\bm{G_1}$ & \multicolumn{1}{c|}{0.001}       & 85        & 51         & 52          & 38         & \multicolumn{1}{c|}{49}         & 8         & 11         & 10          & 25         & \multicolumn{1}{c|}{25}         &           &            &             &            & \multicolumn{1}{c|}{}           &           &            &             &            &            \\
                         & \multicolumn{1}{c|}{0.002}       & 84        & 53         & 52          & 39         & \multicolumn{1}{c|}{49}         & 8         & 10         & 10          & 25         & \multicolumn{1}{c|}{25}         & 1.2       & 1.5        & 1.7         &            & \multicolumn{1}{c|}{}           & 1103      & 630        & 636         &            &            \\
                         & \multicolumn{1}{c|}{0.003}       & 84        & 50         & 51          & 39         & \multicolumn{1}{c|}{48}         & 8         & 11         & 10          & 24         & \multicolumn{1}{c|}{26}         & 1.7       & 1.3        & 1.2         &            & \multicolumn{1}{c|}{}           & 1023      & 1027       & 1126        &            &            \\
                         & \multicolumn{1}{c|}{0.004}       & 84        & 53         & 51          & 39         & \multicolumn{1}{c|}{49}         & 8         & 11         & 11          & 25         & \multicolumn{1}{c|}{26}         & 2.2       & 1.6        & 1.6         & 4.0        & \multicolumn{1}{c|}{}           & 609       & 920        & 634         & 193        &            \\
                         & \multicolumn{1}{c|}{0.005}       & 83        & 51         & 52          & 39         & \multicolumn{1}{c|}{49}         & 9         & 11         & 10          & 25         & \multicolumn{1}{c|}{25}         & 2.8       & 2.0        & 2.0         & 5.0        & \multicolumn{1}{c|}{}           & 1177      & 790        & 697         & 170        &            \\
                         & \multicolumn{1}{c|}{0.006}       & 85        & 53         & 51          & 39         & \multicolumn{1}{c|}{49}         & 8         & 11         & 10          & 25         & \multicolumn{1}{c|}{26}         & 3.3       & 2.4        & 2.3         & 5.1        & \multicolumn{1}{c|}{9.9}        & 772       & 680        & 726         & 151        & 184        \\
                         & \multicolumn{1}{c|}{0.007}       & 83        & 53         & 50          & 39         & \multicolumn{1}{c|}{49}         & 8         & 12         & 10          & 25         & \multicolumn{1}{c|}{25}         & 3.9       & 2.7        & 2.6         & 5.4        & \multicolumn{1}{c|}{10.1}       & 436       & 705        & 708         & 136        & 163        \\
                         & \multicolumn{1}{c|}{0.008}       & 84        & 55         & 50          & 39         & \multicolumn{1}{c|}{49}         & 8         & 10         & 10          & 25         & \multicolumn{1}{c|}{25}         & 4.4       & 3.1        & 3.1         & 5.7        & \multicolumn{1}{c|}{10.5}       & 1184      & 1196       & 460         & 126        & 150        \\
                         & \multicolumn{1}{c|}{0.009}       & 83        & 49         & 49          & 40         & \multicolumn{1}{c|}{50}         & 8         & 12         & 10          & 24         & \multicolumn{1}{c|}{25}         & 4.9       & 3.5        & 3.4         & 5.5        & \multicolumn{1}{c|}{10.4}       & 308       & 895        & 868         & 117        & 140        \\
                         & \multicolumn{1}{c|}{0.010}        & 86        & 52         & 51          & 39         & \multicolumn{1}{c|}{49}         & 8         & 11         & 10          & 25         & \multicolumn{1}{c|}{25}         & 5.4       & 3.9        & 3.8         & 5.8        & \multicolumn{1}{c|}{10.6}       & 439       & 321        & 286         & 110        & 130        \\
                         & \multicolumn{1}{c|}{0.011}       & 83        & 51         & 52          & 40         & \multicolumn{1}{c|}{48}         & 8         & 11         & 10          & 24         & \multicolumn{1}{c|}{25}         & 6.0       & 4.2        & 4.1         & 5.9        & \multicolumn{1}{c|}{10.5}       & 972       & 352        & 244         & 105        & 123        \\
                         & \multicolumn{1}{c|}{0.012}       & 85        & 53         & 52          & 39         & \multicolumn{1}{c|}{48}         & 8         & 11         & 10          & 24         & \multicolumn{1}{c|}{25}         & 6.5       & 4.6        & 4.5         & 6.0        & \multicolumn{1}{c|}{10.6}       & 1071      & 225        & 552         & 102        & 116        \\
                         & \multicolumn{1}{c|}{0.013}       & 86        & 53         & 52          & 39         & \multicolumn{1}{c|}{50}         & 8         & 11         & 10          & 25         & \multicolumn{1}{c|}{25}         & 7.0       & 5.0        & 5.0         & 6.3        & \multicolumn{1}{c|}{10.9}       & 727       & 237        & 246         & 98         & 111        \\ \midrule
  $\bm{G_2}$ & \multicolumn{1}{c|}{0.001}       & 78        & 46         & 32          & 12         & \multicolumn{1}{c|}{37}         & 14        & 21         & 30          & 81         & \multicolumn{1}{c|}{40}         &           &            &             &            & \multicolumn{1}{c|}{}           &           &            &             &            &            \\
                         & \multicolumn{1}{c|}{0.002}       & 79        & 45         & 33          & 13         & \multicolumn{1}{c|}{37}         & 14        & 21         & 31          & 80         & \multicolumn{1}{c|}{40}         &           &            &             &            & \multicolumn{1}{c|}{}           &           &            &             &            &            \\
                         & \multicolumn{1}{c|}{0.003}       & 80        & 45         & 33          & 13         & \multicolumn{1}{c|}{37}         & 14        & 21         & 27          & 79         & \multicolumn{1}{c|}{42}         & 3.3       &            &             &            & \multicolumn{1}{c|}{}           & 154       &            &             &            &            \\
                         & \multicolumn{1}{c|}{0.004}       & 78        & 43         & 33          & 12         & \multicolumn{1}{c|}{38}         & 14        & 21         & 26          & 79         & \multicolumn{1}{c|}{40}         & 4.2       & 3.1        &             &            & \multicolumn{1}{c|}{}           & 130       & 182        &             &            &            \\
                         & \multicolumn{1}{c|}{0.005}       & 79        & 47         & 33          & 12         & \multicolumn{1}{c|}{38}         & 14        & 21         & 31          & 78         & \multicolumn{1}{c|}{40}         & 5.1       & 5.0        & 4.0         &            & \multicolumn{1}{c|}{}           & 110       & 171        & 228         &            &            \\
                         & \multicolumn{1}{c|}{0.006}       & 80        & 44         & 33          & 13         & \multicolumn{1}{c|}{37}         & 14        & 22         & 29          & 79         & \multicolumn{1}{c|}{41}         & 5.3       & 4.8        & 5.4         & 6.2        & \multicolumn{1}{c|}{}           & 101       & 152        & 209         & 332        &            \\
                         & \multicolumn{1}{c|}{0.007}       & 79        & 46         & 33          & 13         & \multicolumn{1}{c|}{37}         & 13        & 21         & 28          & 81         & \multicolumn{1}{c|}{40}         & 5.7       & 5.2        & 5.4         & 6.1        & \multicolumn{1}{c|}{13.5}       & 97        & 143        & 183         & 344        & 242        \\
                         & \multicolumn{1}{c|}{0.008}       & 80        & 46         & 34          & 13         & \multicolumn{1}{c|}{38}         & 14        & 22         & 29          & 82         & \multicolumn{1}{c|}{41}         & 6.1       & 5.5        & 6.0         & 6.7        & \multicolumn{1}{c|}{12.5}       & 89        & 129        & 175         & 314        & 233        \\
                         & \multicolumn{1}{c|}{0.009}       & 78        & 45         & 34          & 13         & \multicolumn{1}{c|}{38}         & 14        & 21         & 29          & 80         & \multicolumn{1}{c|}{39}         & 6.5       & 5.9        & 6.4         & 6.9        & \multicolumn{1}{c|}{13.0}       & 86        & 121        & 161         & 284        & 208        \\
                         & \multicolumn{1}{c|}{0.010}        & 78        & 47         & 33          & 13         & \multicolumn{1}{c|}{38}         & 14        & 21         & 29          & 79         & \multicolumn{1}{c|}{40}         & 7.0       & 6.0        & 6.2         & 7.0        & \multicolumn{1}{c|}{13.8}       & 81        & 112        & 147         & 265        & 198        \\
                         & \multicolumn{1}{c|}{0.011}       & 80        & 47         & 34          & 13         & \multicolumn{1}{c|}{38}         & 13        & 23         & 29          & 80         & \multicolumn{1}{c|}{41}         & 7.4       & 6.5        & 6.6         & 7.0        & \multicolumn{1}{c|}{14.2}       & 78        & 112        & 142         & 236        & 188        \\
                         & \multicolumn{1}{c|}{0.012}       & 77        & 44         & 34          & 13         & \multicolumn{1}{c|}{38}         & 14        & 21         & 28          & 81         & \multicolumn{1}{c|}{40}         & 7.9       & 6.4        & 6.5         & 7.1        & \multicolumn{1}{c|}{14.7}       & 77        & 107        & 136         & 219        & 174        \\
                         & \multicolumn{1}{c|}{0.013}       & 80        & 45         & 35          & 14         & \multicolumn{1}{c|}{39}                              & 14        & 22         & 26          & 81         & \multicolumn{1}{c|}{40}                              & 8.4       & 6.8        & 7.0         & 7.2        & \multicolumn{1}{c|}{14.7}                            & 77        & 102        & 131         & 205        & 162        \\ \bottomrule
  \end{tabular}%
  }
  \medskip

  \caption{Time and location of the first and second peak under interventions, for all values of $\eta$.
  O: original graph, A: social distancing, Bw: weak no-travel rule (increasing $\alpha$, B: hard no-travel rule (cutting long edges), C: limiting maximal node degree. Where the data is missing, the system produces a single peak only. The large numbers for the time of the second peak on $G_1$ are coming from random oscillations of the equilibriated system when there is no clear second peak. }\label{table:heights}
  \end{table}
\end{document}